\documentclass[twocolumn]{aastex63}

\newcommand{\MGL}{\hyperlink{cite.MacDonald2020}{MGL20}}

\usepackage{graphicx}
\usepackage{amsmath}	
\usepackage{amssymb}
\usepackage{booktabs}
\usepackage{enumerate}
\usepackage{dirtytalk}
\usepackage{float}
\usepackage{enumitem}
\usepackage{booktabs}
\usepackage{lipsum} 
\usepackage{float}
\usepackage{xcolor}
\usepackage{multirow}
\usepackage{makecell}

\graphicspath{{./}{figures/}}

\accepted{May 2022}

\submitjournal{ApJ}

\shorttitle{On Atmospheric Retrievals of Exoplanets with Inhomogeneous Terminators}
\shortauthors{Welbanks \& Madhusudhan}

\begin{document}

\title{On Atmospheric Retrievals of Exoplanets with Inhomogeneous Terminators}

\correspondingauthor{Luis Welbanks} 
\email{luis.welbanks@asu.edu}

\author[0000-0003-0156-4564]{Luis Welbanks}
\thanks{NHFP Sagan Fellow}
\affil{Institute of Astronomy, University of Cambridge, Madingley Road Cambridge CB3 0HA, UK}
\affil{School of Earth and Space Exploration, Arizona State University, Tempe, AZ 85281, USA}

\author[0000-0002-4869-000X]{Nikku Madhusudhan}
\affil{Institute of Astronomy, University of Cambridge, Madingley Road Cambridge CB3 0HA, UK}

\begin{abstract}
The complexity of atmospheric retrieval models is largely data-driven and one-dimensional models have generally been considered adequate with current data quality. However, recent studies have suggested that using 1D models in retrievals can result in anomalously cool terminator temperatures and biased abundance estimates even with existing transmission spectra of hot Jupiters. Motivated by these claims and upcoming high-quality transmission spectra we systematically explore the limitations of 1D models using synthetic and current observations. We use 1D models of varying complexity, both analytic and numerical, to revisit claims of biases when interpreting transmission spectra of hot Jupiters with inhomogeneous terminator compositions. Overall, we find the reported biases to be resulting from specific model assumptions rather than intrinsic limitations of 1D atmospheric models in retrieving current observations of asymmetric terminators. Additionally, we revise atmospheric retrievals of the hot Jupiter WASP-43b ($T_{\rm eq}=1440$~K) and the ultra-hot Jupiter WASP-103b ($T_{\rm eq}=2484$~K) for which previous studies inferred abnormally cool atmospheric temperatures. We retrieve temperatures consistent with expectations. We note, however, that in the limit of extreme terminator inhomogeneities and high data quality some atmospheric inferences may conceivably be biased, although to a lesser extent than previously claimed. To address such cases, we implement a 2D retrieval framework for transmission spectra which allows accurate constraints on average atmospheric properties and provides insights into the spectral ranges where the imprints of atmospheric inhomogeneities are strongest. Our study highlights the need for careful considerations of model assumptions and data quality before attributing biases in retrieved estimates to unaccounted atmospheric inhomogeneities.
\end{abstract}

\keywords{methods: data analysis --- planets and satellites: composition --- planets and satellites: atmospheres}

\section{Introduction}

The last decade has witnessed an upsurge in transmission spectra of transiting exoplanets. Atmospheres of several dozens of exoplanets have been observed using transmission spectroscopy, i.e., observations of the wavelength dependent change in transit depth \citep[e.g.,][]{Seager2000}. Transmission spectra provide important constraints on the chemical compositions \citep[e.g.,][]{Charbonneau2002,Deming2013, Sing2016}, presence of clouds and hazes \citep[e.g.,][]{Pont2008, Lecavelier2008a, Lecavelier2008b, Kreidberg2014a, Benneke2019a}, and other atmospheric properties at the day-night terminator regions of exoplanets \citep[see e.g.,][for a review]{Madhusudhan2019}.

Concurrent to this data flood, atmospheric models of varying degrees of complexity have been developed to provide theoretical expectations for the observations and aid in their interpretation. These models range from relatively simple one-dimensional (1D) atmospheres \citep[e.g.,][]{Seager1998, Brown2001b, DeWit2013, Goyal2018} to more complex three-dimensional (3D) general circulation models \citep[GCM, e.g.,][]{Showman2009,Parmentier2016}. Independent of their dimensionality, atmospheric models have a series of physical processes and assumptions built into them to make their computation achievable and apt for their intended purpose. Such assumptions can range from considering isothermal and isobaric atmospheres \citep[e.g.,][]{Heng2017}, and determining chemical abundances following chemical equilibrium expectations \citep[e.g.,][]{Goyal2018}, to considering clouds, dynamics, and disequilibrium chemistry \citep[e.g.,][]{Moses2011, Venot2012, Moses2013, Parmentier2016, Venot2020, Steinrueck2021}. 

The investigation of multidimensional effects in transmission spectra dates back to first detection of an atmosphere around a transiting exoplanet \citep[][]{Brown2001a, Charbonneau2002}. In order to investigate the apparent disparity between the predicted and observed strength of the Na absorption feature in the hot Jupiter HD~209458b, \citet{Fortney2003} modeled the transmission spectrum for a two-dimensional (2D) atmosphere, considering the angular dependence of Na ionization across the day and night sides of the planet. Several subsequent studies have considered multi-dimensional (2D/3D) models of transmission spectra with varying degrees of sophistication and functionality \citep[e.g.,][]{Iro2005,Burrows2010,Fortney2010}.

When interpreting observations, the complexity of the models employed is largely driven by the quality of the data. If the precision and/or resolution of the data are not sufficient to distinguish the impact of a computationally expensive consideration, this overly complex calculation may not be warranted for the purposes of inferring atmospheric properties from spectra. This is particularly the case for atmospheric  retrievals of exoplanets, which compute $10^5$--$10^6$ atmospheric models spanning a wide range of physical conditions and parameter space to obtain statistical constraints on the atmospheric properties of an exoplanet from an observed spectrum \citep[see e.g.,][for a recent review]{Madhusudhan2018}. Therefore, retrieval frameworks have generally been limited to 1D parametric models with certain assumptions aimed to ease their computational efficiency.

Naturally, a key question arises: What are the key physical processes and model assumptions permitted by the data in order to obtain robust and reliable atmospheric inferences using atmospheric retrievals? Answering this question is paramount, given recent observational advancements enabling high-precision transmission spectra of exoplanets over a broad spectral range \citep[e.g.,][]{Deming2013, Kreidberg2014b, Sing2016, Nikolov2018}. Additionally, the launch of the James Webb Space Telescope (JWST) and the promise of spectroscopic data over a wide wavelength range with unprecedented precision \citep[e.g.,][]{Greene2016}, may mean that previously unwarranted model considerations will now be needed. Indeed, a flourishing area of research has emerged around the investigation of possible limitations of 1D retrieval models, especially within the context of planets with cloud inhomogeneities as well as thermal and chemical inhomogeneities \citep[e.g.,][]{Line2016a, Caldas2019, Pluriel2020, Lacy2020, MacDonald2020, Espinoza2021, Pluriel2021}.

\citet{Line2016a} explored the effect of inhomogeneous clouds on atmospheric retrievals of transmission spectra. Their study suggests that changes in spectral shape, i.e. the wavelength-dependent slope of the transit depth, due to the presence of inhomogeneous clouds can be modeled as the linear combination of a cloudy and non-cloudy model. This modeling approach has been expanded to combined effects of inhomogeneous clouds and hazes in exoplanet atmospheres \citep[e.g.,][]{MacDonald2017a, Welbanks2021} and inhomogeneities in the temperature structure and chemical composition between terminators \citep[e.g.,][]{MacDonald2020, Espinoza2021}. Similar treatment of inhomogeneities has also been made for thermal emission spectra \citep[e.g.,][]{Feng2016,Taylor2020,Feng2020} and for the effects of stellar heterogeneities on transmission spectra \citep[e.g.,][]{Pinhas2018,Iyer2020} of exoplanets.

Other recent studies have investigated the impact of nonuniform day and night sides with thermal and chemical inhomogeneities on retrievals using 1D atmospheric models \citep[e.g.,][]{Caldas2019, Pluriel2020, Lacy2020, Pluriel2021}. \citet{Caldas2019} develop a 3D radiative transfer model, create synthetic JWST observations for a planet with day-night temperature inhomogeneities, and retrieve the properties of the synthetic observations using cloud-free isothermal 1D atmospheric models. Their results suggest that while 1D models can provide a good match to the data, the retrieved atmospheric properties may be biased. Particularly, for planets with large thermal gradients between the day and night sides such as ultra-hot Jupiters, the retrieved temperature for transmission spectra may be higher than the terminator temperature when using 1D, cloud-free, isothermal models. Similar conclusions were reached by \citet{Pluriel2020} who find that 1D isothermal models with homogeneous clouds may retrieve biased molecular abundances if confronted with synthetic JWST observations of planets with day-night chemical heterogeneities.

\begin{figure*}[ht!]
\includegraphics[width=\textwidth]{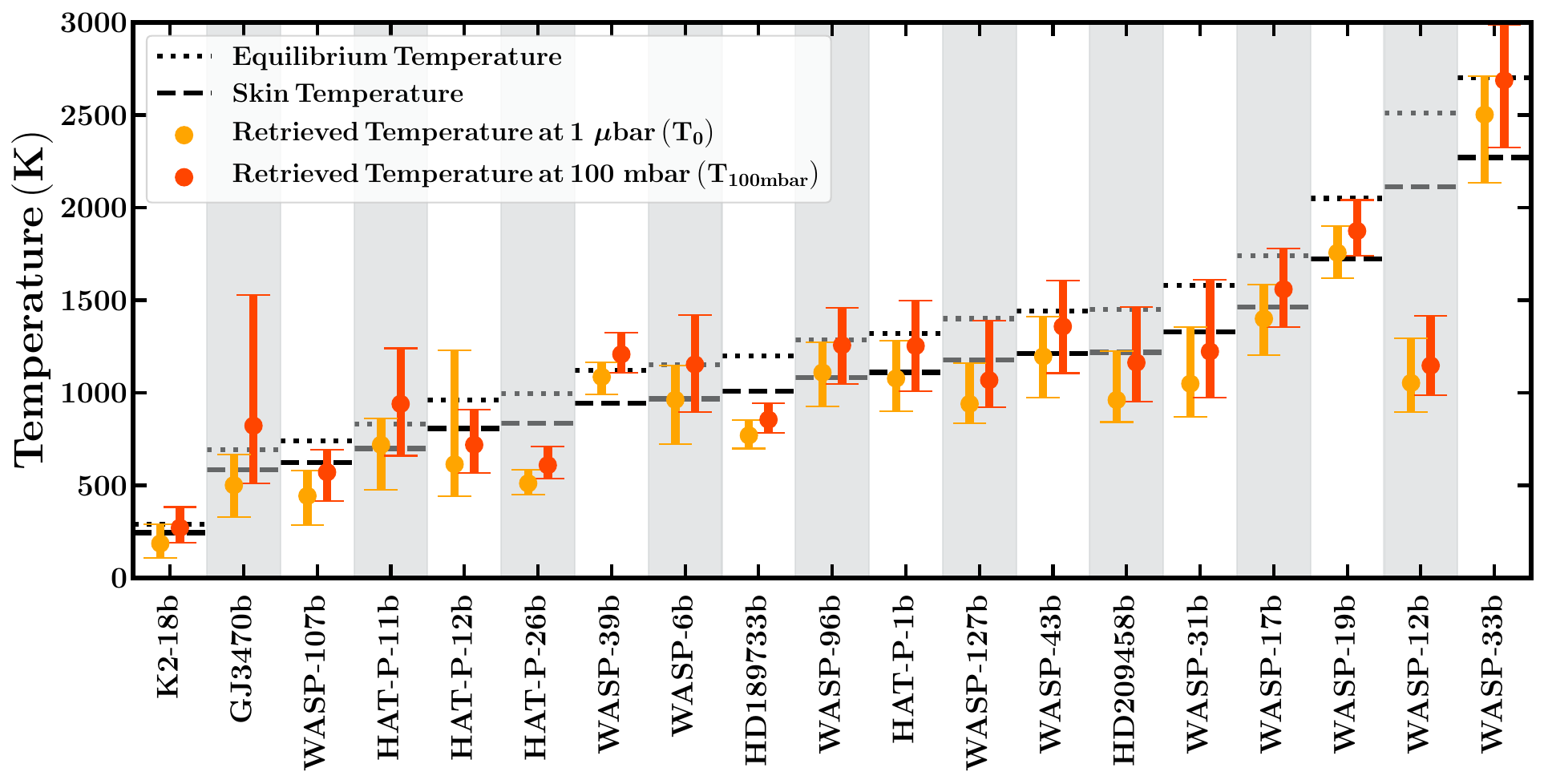}
\centering
\caption[temp_comp]{Equilibrium temperatures ($T_{\rm{eq}}$, dotted line) and skin temperatures ($T_{\rm skin} = 2^{-1/4} \, T_{\rm{eq}}$, dashed line) for the planets in the sample of \citet{Welbanks2019b}. Retrieved temperatures ($T_{\rm{ret}}$) at different pressures and their $1\sigma$ confidence intervals are shown in yellow ($1$~$\mu$bar, $T_0$) and orange ($100$~mbar, $T_{\text{100mbar}}$). Also see Table \ref{tab:litvalues}. \label{fig:temp_comparison}}
\end{figure*}  

\citet{Lacy2020} further explore the impact of day-night temperature gradients on atmospheric retrievals. They propose a parameterization to recover the pressure-temperature ($P$--$T$) profile of an atmosphere with day-night temperature gradients, and investigate the combined effects of day-night temperature gradients and aerosols. Their results, based on synthetic JWST-like observations and assumptions of chemical equilibrium, suggest that day-night temperature gradients may be detectable in the future. Similarly, \citet{Pluriel2021} find that assuming isothermal $P$--$T$ profiles can lead to biased results when interpreting observations of planets with day-night inhomogeneities. Their results, based on cloud free models, suggest that JWST-like data will require accounting for non-isothermal temperature structures, as normally done in retrievals \citep[e.g.,][]{Madhusudhan2009}.

Besides day-night inhomogeneities, exoplanets may also have nonuniform morning and evening terminators. In order to constrain the properties of these morning and evening terminators, \citet{Espinoza2021} present a framework to produce distinct transmission spectra for each planetary limb. Additionally, they modify the retrieval framework CHIMERA \citep{Line2013a} to compute chemically-consistent retrievals (i.e., assuming chemical equilibrium) using the linear combination of two 1D atmospheric models, each for a planet limb, as initially proposed by \citet{Line2016a}. Using real Hubble Space Telescope (HST) data and synthetic JWST data, their results suggest that obtaining separate transmission spectra for each planetary limb and performing a retrieval on them may be a fruitful approach to constrain the atmospheric properties of inhomogeneous planets.

Recent studies have also claimed anomalous retrieved atmospheric temperatures in hot Jupiters with existing observations of transmission spectra. A study by MacDonald, Goyal and Lewis (\citeyear{MacDonald2020}), hereafter \hyperlink{cite.MacDonald2020}{MGL20}, investigated the impact of morning and evening terminator inhomogeneities. Their work claims to find an important anomaly in retrievals of transiting exoplanets using existing observations: the retrieved atmospheric temperatures are notably cooler than expectations from the equilibrium temperature of the planet. They investigate the origin of this bias using analytic derivations and retrieving synthetic HST observations using 1D models. Using cloud free models, they find that the retrieved abundances and temperature profiles for observations of present-day quality can be significantly biased as the result of asymmetric terminators. They conclude that chemical abundances from 1D retrieval techniques are often biased. Such thermal anomalies are also reported by other recent studies performing atmospheric retrievals on ground- and space-based transmission spectra \citep[e.g.,][]{Weaver2020, Kirk2021}. Our present work investigates these anomalies further.

These studies raise compelling concerns regarding the reliability and robustness of 1D atmospheric models. Particularly, the inferred anomalies in the retrieved atmospheric temperatures could mean that 1D retrieval techniques are often biased in their retrieved chemical abundances and associated efforts to constrain planet formation mechanisms from atmospheric abundances. Therefore, there is an urgent need to understand the origin and influence of these thermal biases under different planetary conditions. Moreover, it is imperative to establish whether conclusions derived using isothermal models can be generalised to results from non-isothermal models. Likewise, it is important to ascertain whether the lessons derived from cloud-free models are applicable to models considering the presence of inhomogenous clouds and hazes. Finally, the reliability of state-of-the-art retrieval frameworks with more flexible 1D models considering non-isothermal temperature profiles, inhomogeneous clouds and hazes, and multiple sources of opacity needs to be established.

In the present paper we investigate the reliability of atmospheric temperature estimates from 1D atmospheric models. In particular, we address claims of seemingly anomalous temperatures from retrievals of hot Jupiters from \MGL. Additionally, we implement a two-dimensional (2D) chemically unconstrained retrieval approach for exoplanetary transmission spectra.  In Section \ref{sec:literature_and_analytic} we revisit published retrieval studies of transmission spectra for a wide sample of exoplanets spanning cool mini-Neptunes to ultra-hot Jupiters, and assess whether their retrieved atmospheric temperatures match expectations derived from their equilibrium temperatures and GCM studies. Then, in Section \ref{sec:analytic_solutions} we investigate the impact of asymmetric morning evening terminators on the retrieved atmospheric temperatures using semi-analytic models and retrievals. 

Next, in Section \ref{sec:1D_retrievals} we perform a systematic exploration of retrieved temperature estimates over a wide range of planetary conditions ranging from warm to ultra-hot Jupiters using different families of 1D atmospheric models. We investigate the impact of asymmetric terminators on 1D atmospheric models and their inferred atmospheric properties. We expand our investigation in Section \ref{sec:real_data} by performing retrievals on existing observations of hot/ultra-hot Jupiters for which previous studies \cite[e.g.,][]{Weaver2020,Kirk2021} inferred anomalously cool atmospheric temperatures. We investigate whether these retrieved thermal biases are the result of simplified model considerations or anomalies in 1D models due to asymmetric terminators.

Subsequently, in Section \ref{sec:2D_retrievals} we introduce a 2D atmospheric model in a `free-retrieval' framework to retrieve the atmospheric properties of planets with asymmetric terminators. A similar approach has been implemented recently by \citet{Espinoza2021}, with constraints of chemical equilibrium. This new multidimensional approach retrieves the atmospheric properties of planets with inhomogeneous terminators. Additionally, this framework provides important information about the wavelength ranges most affected by these inhomogeneities as well as the chemical species with non-homogeneous abundances. We summarise our results and key lessons in Section \ref{sec:summary}, and discuss the prospects for future investigations of multidimensional effects in exoplanetary transmission spectra. 

\begin{deluxetable*}{lclcccclclccc}[ht!] 
\tablecaption{Retrieved Temperatures From Exoplanet Transmission Spectra \label{tab:litvalues}}
\tablewidth{0pt}
\tablehead{
\multicolumn{1}{l}{Planet} & \multicolumn{1}{c}{$T_{\rm eq}$}& \multicolumn{1}{c}{$T_{\rm skin}$} & \multicolumn{1}{c}{ $T_{\rm ret}$} & \multicolumn{1}{c}{$T_{\rm ret} - T_{\rm eq}$} & \multicolumn{1}{c}{$\frac{T_{\rm ret}}{T_{\rm skin}}$}  & \multicolumn{1}{c}{ \ldots } & \multicolumn{1}{l}{Planet} & \multicolumn{1}{c}{$T_{\rm eq}$} & \multicolumn{1}{c}{$T_{\rm skin}$} & \multicolumn{1}{c}{ $T_{\rm ret}$} & \multicolumn{1}{c}{$T_{\rm ret} - T_{\rm eq}$} & \multicolumn{1}{c}{$\frac{T_{\rm ret}}{T_{\rm skin}}$} \\ 
 & (K) & (K) & (K) & (K)& & & & (K) &  (K) & (K) & (K) &
}
\startdata \\[-8pt]
    K2-18b & 290 & 244 & $271 ^{+ 112 }_{- 80 }$ &   $-19 ^{+ 112 }_{- 80 }$ & $1.11 ^{+ 0.46 }_{- 0.33 }$     
& & HAT-P-1b & 1320 & 1110 & $1253 ^{+ 246 }_{- 243 }$ &   $-67 ^{+ 246 }_{- 243 }$ & $1.13 ^{+ 0.22 }_{- 0.22 }$  \\
    GJ~3470b & 693 & 583 & $822 ^{+ 705 }_{- 313 }$ &   $129 ^{+ 705 }_{- 313 }$ & $1.41 ^{+ 1.21 }_{- 0.54 }$     
& & WASP-127b & 1400 & 1177 & $1068 ^{+ 323 }_{- 145 }$ &   $-332 ^{+ 323 }_{- 145 }$ & $0.91 ^{+ 0.27 }_{- 0.12 }$  \\
    WASP-107b & 740 & 622 & $571 ^{+ 122 }_{- 156 }$ &   $-169 ^{+ 122 }_{- 156 }$ & $0.92 ^{+ 0.20 }_{- 0.25 }$  
& & WASP-43b & 1440 & 1211 & $1358 ^{+ 247 }_{- 252 }$ &   $-82 ^{+ 247 }_{- 252 }$ & $1.12 ^{+ 0.20 }_{- 0.21 }$ \\
    HAT-P-11b & 831 & 699 & $940 ^{+ 300 }_{- 280 }$ &   $109 ^{+ 300 }_{- 280 }$ & $1.34 ^{+ 0.43 }_{- 0.40 } $
& & HD~209458b & 1450 & 1219 & $1163 ^{+ 299 }_{- 210 }$ &   $-287 ^{+ 299 }_{- 210 }$ & $0.95 ^{+ 0.25 }_{- 0.17 }$ \\
    HAT-P-12b & 960 & 807 & $719 ^{+ 189 }_{- 154 }$ &   $-241 ^{+ 189 }_{- 154 }$ & $0.89 ^{+ 0.23 }_{- 0.19 }$        
& & WASP-31b & 1580 & $1329$ & $1223 ^{+ 386 }_{- 251 }$ &  $ -357 ^{+ 386 }_{- 251 }$ & $0.92 ^{+ 0.29 }_{- 0.19 }$  \\
    HAT-P-26b & 994 & 836 & $609 ^{+ 102 }_{- 72 }$ &  $ -385 ^{+ 102 }_{- 72 }$ & $0.73 ^{+ 0.12 }_{- 0.09 }$           
& &  WASP-17b & 1740 & 1463 & $1559 ^{+ 221 }_{- 204 }$ &   $-181 ^{+ 221 }_{- 204 }$ & $1.07 ^{+ 0.15 }_{- 0.14 }$  \\
    WASP-39b & 1120 & 942 & $1208 ^{+ 115 }_{- 101 }$ &   $88 ^{+ 115 }_{- 101 }$ & $1.28 ^{+ 0.12 }_{- 0.11 }$     
& &   WASP-19b & 2050 & 1724 & $1874 ^{+ 167 }_{- 135 }$ &   $-176 ^{+ 167 }_{- 135 }$ & $1.09 ^{+ 0.10 }_{- 0.08 }$   \\ 
    WASP-6b & 1150 & 967 & $1153 ^{+ 268 }_{- 259 }$ &   $3 ^{+ 268 }_{- 259 }$ & $1.19 ^{+ 0.28 }_{- 0.27 }$      
& &   WASP-12b & 2510 & 2111 & $1147 ^{+ 268 }_{- 161 }$ &   $-1363 ^{+ 268 }_{- 161 }$ & $0.54 ^{+ 0.13 }_{- 0.08 }$  \\ 
    HD~189733b & 1200 & 1009 & $855 ^{+ 88 }_{- 70 }$ &   $-345 ^{+ 88 }_{- 70 }$ & $0.85 ^{+ 0.09 }_{- 0.07 }$   
&& WASP-33b & 2700 & 2270 & $2687 ^{+ 301 }_{- 362 }$ &   $-13 ^{+ 301 }_{- 362 }$ & $1.18 ^{+ 0.13 }_{- 0.16 }$  \\ 
    WASP-96b & 1285 & 1081 & $1257 ^{+ 202 }_{- 211 }$ &   $-28 ^{+ 202 }_{- 211 }$ & $1.16 ^{+ 0.19 }_{- 0.20 }$ 
&&  &  &  &  & &  \\[3pt]
\enddata 
\tablecomments{Equilibrium temperatures ($T_{\rm{eq}}= \left[f\, (1-A_B) \left(\frac{R_{\mathrm{star}}}{2\,a}\right)^2 \right]^\frac{1}{4}T_{\rm eff., \,star}$) assuming no bond albedo ($A_B=0$) and full energy redistribution ($f=1$) from \citet{Welbanks2019b} and reported in \citet{MacDonald2020}. The skin temperature is given by $T_{\rm skin} = 2^{-1/4} \, T_{\rm{eq}}$. $T_{\rm ret}$ is the retrieved temperature at a pressure of 100~mbar, close to the slant photosphere, from \citet{Welbanks2019b}.}
\end{deluxetable*}

\section{On Retrieved Temperature Anomalies from the Literature} \label{sec:literature_and_analytic}

With the growing number of transmission spectra for exoplanet atmospheres, recent studies have pursued systematic explorations of these observations using retrieval frameworks to address population level hypothesis about planet composition and formation \citep[e.g.,][]{Madhusudhan2014a, Barstow2017, Pinhas2019, Welbanks2019b}. A by-product of such studies is a large sample of planetary spectra analysed under similar model assumptions and for which the inferred atmospheric properties can be compared among planets in the sample. One of such studies \citep{Welbanks2019b} was used as a reference by \MGL\, to postulate an anomaly in retrieval studies of transmission spectra. Specifically, \MGL\, state that almost all retrieved temperatures are notably cooler than the planetary equilibrium temperature. Here we seek to investigate the origin of these anomalies. We revisit the inferred atmospheric properties from the complete 19 planets sample from \citet{Welbanks2019b} and compare their retrieved temperatures ($T_{\rm ret}$) to the calculated equilibrium ($T_{\rm eq}$) and skin ($T_{\rm skin}$) temperatures of the planet. We summarise the retrieved atmospheric temperatures from \citet{Welbanks2019b} in Figure \ref{fig:temp_comparison} and Table \ref{tab:litvalues}.

The atmospheric models in \citet{Welbanks2019b} use non-isothermal $P$--$T$ profiles. As a result, the temperature in the atmosphere varies vertically. Due to the lack of thermal inversions in the retrieved $P$--$T$ profiles, the temperature at the top atmosphere will generally be cooler than the temperatures deeper in the atmosphere. Figure \ref{fig:temp_comparison} shows the temperature estimates at the top of the model atmosphere ($T_0$) at a pressure of $10^{-6}$~bar (in yellow), and the temperature estimates near the photosphere (T$_{\text{100mbar}}$) at a pressure of 100~mbar (in orange). The temperature estimates near the photosphere are warmer than the temperature estimates at the top of the atmosphere. This difference in temperature estimates was not considered by \MGL\, who used the cooler temperature estimates at the top of the atmosphere instead of the more representative and warmer retrieved temperature near the photosphere.

Furthermore, when reporting the estimated difference between $T_{\rm ret}$ and $T_{\rm eq}$ or $T_{\rm skin}$, \MGL\, restricted their comparison to the median values only, without considering the uncertainties in the retrieved parameters. Figure \ref{fig:temp_comparison} and Table \ref{tab:litvalues} include the inferred 1$\sigma$ confidence intervals for the estimated temperatures (i.e., the `error' in the estimate). For most planets, $T_{\rm ret}$ and $T_{\rm eq}$ are consistent within 1$\sigma$. Additionally, the ratio between $T_{\rm ret}$ and $T_{\rm skin}$ is consistent with unity for most planets. As such, there is no anomaly in the retrieved temperatures for most planets in the sample. 

Three planets have retrieved temperatures near the photosphere which are inconsistent with their $T_{\rm eq}$ and $T_{\rm skin}$. HAT-P-26b and HD~189733b have inconsistencies of $\lesssim300~$K. Considering that $T_{\rm eq}$ may not be representative of the temperature at the terminator, and the assumption of full redistribution and zero albedo in the computation of the values presented in Table \ref{tab:litvalues}, this apparent discrepancy may be not significant. Finally, WASP-12b has an inconsistency of $\gtrsim1000~$K. This outlier would warrant a detailed inspection of the data and model considerations employed to better understand the origin of this anomaly, as discussed below. \textbf{Overall, when considering the entire sample, a comparison of the median retrieved temperatures at the photosphere with the skin temperature shows no consistent trend of underestimated temperatures.}

It is also important to consider that the photospheric temperature at the terminator of the planet can be lower than $T_{\rm eq}$ and $T_{\rm skin}$. As mentioned above, the calculation of $T_{\rm eq}$ requires making some assumptions about the energy redistribution between the day and night sides of the planet, and the planetary albedo. Therefore, assuming different values for the planetary albedo or energy redistribution can result in cooler estimates. For instance, assuming a bond albedo of 0.1 the equilibrium temperatures reported in Table \ref{tab:litvalues} would decrease by $\sim20$~K for the warm mini-Neptunes and up to $\sim70$~K for the ultra-hot Jupiters. 

On the other hand, $T_{\rm skin} = 2^{-1/4} \, T_{\rm{eq}}$, is the temperature estimate of a hypothetical outer atmospheric layer with low optical depth, transparent to incident stellar radiation, and heated only by outgoing radiation, for a gray atmosphere \citep[][]{Pierrehumbert2010, Parmentier2014}. This temperature estimate would not necessarily represent the photospheric temperature at the terminator. Furthermore, \citet{Parmentier2014} show that non-gray effects can lead to skin temperatures significantly lower than those predicted for the gray case. Although \MGL\, argue for $T_{\rm eq}$ and $T_{\rm skin}$ as the reference values for the terminator temperature expectations, in practice any temperature estimate would be better gauged by GCMs, expected to provide more realistic estimates of these values \citep[][]{Seager2010_book}. Therefore, we inform our expectations using GCM results from the literature below.

GCMs show that the atmospheric temperature of the planet is dependent on a series of factors including the atmospheric composition and the orbital phase of the planet. For instance, for the hot Neptune GJ436b, \citet{Lewis2010} find zonal-mean temperatures at 10 mbar of $T_{\rm 10 \, mbar}\sim500$--$550$~K for a $1\times$ solar composition, estimates cooler than $T_{\rm eq}$ and $T_{\rm skin}$ ($T_{\rm{eq}}=649$~K, $T_{\rm skin} =546$~K). Similarly, for the high eccentricity hot Jupiter HAT-P-2b, \citet{Lewis2014} find temperature averages over latitude and longitude of $T\lesssim1200$~K at $P\sim10^{-3}$--$10^{-1}$~bar at the time of transit, estimates lower than $T_{\rm eq}$ and $T_{\rm skin}$ ($T_{\rm{eq}}=1481$~K, $T_{\rm skin} =1246$~K). 

More recently \citet{Kataria2016} model 9 hot Jupiters common to the sample in \citet{Welbanks2019b}. Their study models HAT-P-12b, WASP-39b, WASP-6b, HD~189733b, HAT-P-1b, HD~209458b, WASP-31b, WASP-17b, and WASP-19b with reported $T_{\rm eq}$ close to those in Table \ref{tab:litvalues}. Their results show that the temperature probed in transmission can be significantly cooler than $T_{\rm eq}$ and $T_{\rm skin}$ of the planet. For instance, the pressures probed in transmission for the hot Jupiter WASP-19b ($T_{\rm{eq}}\sim 2050$~K, $T_{\rm skin}\sim 1724$~K) can correspond to temperatures as cool as $\lesssim 1400$~K cooler than $T_{\rm skin}$ and $T_{\rm eq}$. Similarly, a cooler planet like WASP-39b ($T_{\rm eq}=1120$~K, $T_{\rm skin}=942$~K) exhibits temperatures cooler than $T_{\rm skin}$ and $T_{\rm eq}$ probed in transmission with values as low as $\lesssim900$~K. Similar results are shown for the ultra-hot Jupiter WASP-12b ($T_{\rm{eq}}\sim 2510$~K, $T_{\rm skin}\sim 2111$~K) in recent study by \citet{Arcangeli2021}, where the brightness temperatures at the terminator near photospheric pressures, $T_{\rm 1 \, mbar}\sim1000$--$2000$~K, can be cooler than $T_{\rm eq}$ and $T_{\rm skin}$.

Considering these estimates from the literature, most temperature estimates from \citet{Welbanks2019b} shown in Figure \ref{fig:temp_comparison} and Table \ref{tab:litvalues} are well within the physically plausible range for the photospheric temperature of the terminator. Overall, depending on the atmospheric composition and the pressures probed by the photosphere, the temperature at the terminator of the planet can be significantly cooler than $T_{\rm eq}$ and $T_{\rm skin}$ of the planet. Future studies may investigate how these consideration affect the photospheric temperature near the terminator of the planet to better inform our expectations for ultra-hot Jupiters.

Notwithstanding the fact that we do not find anomalous retrieved temperature estimates in the published results of \citet{Welbanks2019b}, we investigate other claims made in previous studies. For example, \MGL\, claim an analytic justification for substantially cooler temperature estimates. From this analytic solution they state that compositional biases exceeding a factor of 2 result in temperature biases of hundreds of degrees. Additionally, they state that no equivalent temperature can reproduce a 2D spectrum using a 1D model. \MGL\, claim that these thermal biases hold for state-of-the-art retrieval codes and that the chemical abundances derived from 1D retrieval techniques are often biased. Finally, they claim to confirm these biases using atmospheric retrievals with synthetic data. Other studies, have also reported seemingly anomalous temperature estimates with observed transmission spectra of several hot/ultra-hot Jupiters \cite[e.g.,][]{Weaver2020,Kirk2021}. We explore these claims systematically in the sections below.

In Section \ref{sec:analytic_solutions} and in Appendix \ref{app:derivation} we demonstrate analytically that the inferred temperatures from 1D models are the average of the morning and evening terminators. In Section \ref{subsec:analytic_retrieval} we verify our expectations using semi-analytic retrievals and show that 1D models can reproduce 2D spectra. In Section \ref{sec:1D_retrievals} we use the same atmospheric cases as \MGL\, and state-of-the-art retrievals with established $P$--$T$ parameterizations and generalised cloud and haze prescriptions to investigate the claimed thermal biases. Using the same atmospheric models, in Section \ref{sec:real_data} we reanalyse observations from \citet{Weaver2020} and \citet{Kirk2021} to investigate their seemingly anomalous temperature inferences. Finally, in \ref{subsec:retrievals_as_stats} we discuss how several of the simulated retrieval estimates in \MGL\, indeed seem consistent with the true temperature and abundance values if the statistical uncertainties are considered. 

\section{Exploring Semi-Analytic Solutions for Asymmetric Terminators}
\label{sec:analytic_solutions}

Analytic treatments of the transit depth of exoplanets \citep[e.g.,][]{Brown2001b, Lecavelier2008a, Betremieux2017, Heng2017} can be useful to understand the origin of a planetary spectrum and its dependence on the physical parameters such as temperature and chemical composition. However, the analytical tractability of such expressions is based on assumptions of isobaric cross-sections in isothermal atmospheres with assumptions of constant gravity and scale heights \citep[e.g.,][]{Heng2017}. These unphysical assumptions can lead to biased atmospheric interpretations when compared to more physically plausible numerical models \citep[][]{Welbanks2019a}. Nonetheless, these analytic treatments are routinely used to provide a theoretical basis for our understanding of exoplanet atmospheres.

Indeed, \MGL\, follow such a semi-analytic approach to construct a theoretical justification for their inferred biased temperatures. They derive an equivalent 1D temperature for a 2D atmosphere with asymmetric terminators. Here we seek to investigate the expectations for this equivalent 1D temperature using the same semi-analytic approach.

\subsection{Analytic Temperature and Abundance Expectations}\label{subsec:analytic_expectations}

First, we begin with the general wavelength-dependent transit depth

\begin{equation} \label{eq:transit_depth_general}
\Delta_{\lambda} = \frac{\pi R_{\mathrm{p}}^{2} + \displaystyle\int_{0}^{2 \pi} \displaystyle\int_{R_{\mathrm{p}}}^{\infty} \left( 1 - e^{-\tau_{\lambda}(b, \theta)} \right) \, b \, db \, d\theta}{\pi R_{\mathrm{star}}^{2}},
\end{equation} 

\noindent where $R_{\mathrm{p}}$ and $R_{\mathrm{star}}$ are the planetary radius and stellar radius respectively, $b$ is the impact parameter, $\tau_{\lambda}$ is the slant optical depth, and $\theta$ is the azimuthal angle. We can split the contribution of the optical depth in Equation \ref{eq:transit_depth_general} into two separate integrals, as proposed by \MGL, each representing half a planetary hemisphere, a morning and an evening one as 

\begin{align}
\label{eq:transit_depth_2D}
\resizebox{.99\hsize}{!}{$
\Delta_{\lambda, 2D} = \frac { R_{\mathrm{p}}^{2} + \displaystyle\int\limits_{R_{\mathrm{p}}}^{\infty} b \left( 1 - e^{-\tau_{\lambda, M}(b)} \right) db + \displaystyle\int\limits_{R_{\mathrm{p}}}^{\infty} b \left( 1 - e^{-\tau_{\lambda, E}(b)} \right) db }{R_{\mathrm{star}}^{2} }$}.
\end{align}

\noindent This result is equivalent to taking the well known expression for 1D atmospheres in transmission geometry \citep[e.g.,][]{MacDonald2017a, Welbanks2021} and reformulating the contribution of the atmosphere as a linear combination of two distinct 1D atmospheres, one for the evening `E' and other for the morning `M'. 

Then, and in order to easily evaluate the integrals of the above expressions, one can assume isobaric and isothermal conditions, a single chemical absorber, constant gravity and scale height ($H=kT/\mu g$), as previously performed in numerous works \citep[e.g.,][]{Lecavelier2008a, DeWit2013, Betremieux2017, Heng2017, Sing2018}. With the above assumptions, considering two separate semi-hemispheres (morning and evening) with distinct isothermal temperatures and associated scale heights ($H_E=kT_E/\mu g$ for the evening and $H_M=kT_M/\mu g$ for the morning), Equation \ref{eq:transit_depth_2D} simplifies to 

\begin{equation} \label{eq:transit_depth_2D_analytic}
\resizebox{.99\hsize}{!}{$
\Delta_{\lambda, 2D} = \frac{R_{\mathrm{p}}^{2} + R_{\mathrm{p}} H_{M} (\gamma + \ln{\tau_{0, \lambda, M}}) + R_{\mathrm{p}} H_{E} (\gamma + \ln{\tau_{0, \lambda, E}}) }{R_{\mathrm{star}}^{2}}$}
\end{equation} 

\noindent with $\tau_{0, \lambda, i}$ given under the assumption of isobaric cross sections and hydrostatic equilibrium as

\begin{equation} \label{eq:analytic_tau}
    \tau_{0, \lambda, i}= \frac{P_0}{kT_i}\sqrt{2\pi R_{\rm p} H_i}X \sigma_{\lambda},
\end{equation}

\noindent where the index $i$ is used to represent the evening (E) or morning (M) parameters, and $X$ and $\sigma$ are the single chemical absorber's volume mixing ratio and cross section. Similarly, the 1D semi-analytic transmission spectrum is

\begin{equation} \label{eq:transit_depth_1D_analytic}
\Delta_{\lambda, 1D} = \frac{R_{\mathrm{p}}^{2} + 2 R_{\mathrm{p}} H_{1D} (\gamma + \ln{\tau_{0, \lambda, 1D}})}{R_{\mathrm{star}}^{2}}
\end{equation} 

\noindent with only one isothermal temperature $T_{1D}$, its associated scale height $H_{1D}=kT_{1D}/\mu g$, and $\tau_{0, \lambda, 1D}$ given by Equation \ref{eq:analytic_tau} with the 1D temperature, scale height, and abundances. 

\MGL\, investigate the resulting atmospheric properties from using a 1D model to interpret the transmission spectrum of a 2D model atmosphere, using the semi-analytic expressions in Equations \ref{eq:transit_depth_2D_analytic} and \ref{eq:transit_depth_1D_analytic}. They investigate the isothermal temperature of an equivalent 1D model to produce the same transit depth as a 2D model at a given wavelength, i.e., $\Delta_{\lambda, 2D}=\Delta_{\lambda, 1D}$. According to \MGL\,, while one may expect the 1D model to obtain an average temperature (i.e. $T_{1D}=\bar{T} = \frac{1}{2}(T_E+T_M)$) they find $T_{1D}$ to be lower than $\bar{T}$. This result was used by \MGL\, to explain the erroneously cold temperatures presumably derived using 1D retrievals of hot Jupiters, with implications also to retrieved abundance estimates.

We revisit the analytic formulation in Appendix \ref{app:derivation} to assess this result. \MGL \, equate the transit depths of the 1D and 2D models at a single wavelength, implicitly assuming that the atmospheric properties can be inferred from a single transit depth measurement, which is not feasible. In practice, it is the shape of a transmission spectrum, i.e. the gradient of the transit depth with wavelength, that provides constraints on the atmospheric properties in transmission geometry \citep{Lecavelier2008a,Benneke2012,Line2016a,Sing2018}. Therefore, we equate the spectral shapes between the 1D and 2D models to determine the relation between the 1D and 2D atmospheric properties.

In contrast to \MGL, as shown in Appendix \ref{app:derivation}, we find that an equivalent 1D model can indeed recover the 1D temperature as the average of the morning and evening temperatures in the 2D model, i.e., $T_{1D}=\bar{T}$. Our analysis also recovers the well known degeneracy due to the unphysical assumptions of this semi-analytic formalism between the chemical abundances and the reference pressure of the planet \citep[see e.g.,][]{Lecavelier2008a, Heng2017, Welbanks2019a}. Furthermore, the atmospheric properties of the 1D model as a function of the properties of the 2D model are related as

\begin{multline}\label{eq:1d_properties}
   P_{0, \mathrm{1D}} \,  X_{\mathrm{1D}} \, \sigma_{\bar{T}} = \bar{T}^\frac{1}{2} \left(P_{0, \mathrm{M}}\, X_{\mathrm{M}} \, \sigma_{\rm M} \right) ^ \frac{\beta}{2}
     \left(P_{0, \mathrm{E}}\, X_{\mathrm{E}} \, \sigma_{\rm E}\right) ^ \frac{\alpha}{2}
   \\ \left(\bar{T}-\Delta T \right) ^{-\frac{\beta}{4}}
    \left(\bar{T}+\Delta T \right)  ^{-\frac{\alpha}{4}},
\end{multline}

\noindent where for each morning (M), evening (E) or 1D model (1D), $X_i$ is the chemical abundance, $\sigma_i$ is the temperature dependent cross section, and $P_{0\,i}$ is the reference pressure; $\alpha= 1+\frac{\Delta T}{\bar{T}}$, $\beta = 1-\frac{\Delta T}{\bar{T}}$, $\bar{T}=\frac{T_{\rm E}+T_{\rm M}}{2}$, and $\Delta T=\frac{T_{\rm E}-T_{\rm M}}{2}$.

In order to derive a 1D chemical abundance in this analytic framework, one must assume a reference pressure, or vice versa \citep[e.g.,][]{Lecavelier2008a}. Additionally, Equation \ref{eq:1d_properties} shows that the 1D chemical abundance is not expected to be the average of the morning and evening abundances as assumed by \MGL. We test the predictions from our derivation using an atmospheric retrieval framework below.

\subsection{Atmospheric Retrieval Using Semi-Analytic Models}
\label{subsec:analytic_retrieval}

We generate a synthetic HST-WFC3 transmission spectrum following the 2D semi-analytic atmospheric model in Equation \ref{eq:transit_depth_2D_analytic}, and then retrieve the atmospheric properties using the 1D semi-analytic model in Equation \ref{eq:transit_depth_1D_analytic}. The atmospheric model producing the synthetic observations and in the atmospheric retrieval considers the system properties of the canonical hot Jupiter HD~209458b: $R_{\rm p} = 1.359 R_{\rm J}$ with $R_{\rm p,\, ref} = 1.33182 R_{\rm J}$ as the assumed reference radius following \MGL, $M_{\rm p} = 0.685 M_{\rm J}$, and $R_{\mathrm{star}}=1.155 R_\odot $ \citep{Torres2008}. To satisfy the assumptions of the semi-analytic treatments, we employ isothermal atmospheres with a single absorber, H$_2$O. The assumption of isobaric cross sections from this semi-analytic treatment is satisfied by calculating the H$_2$O cross sections at $10$~bar. The mean molecular weight is fixed to $2.3$ amu. 

Similarly to section 2.1 of \citet{Welbanks2019a}, we replace the numerical atmospheric model in the retrieval framework of \citet{Welbanks2021} with the semi-analytic model in Equation \ref{eq:transit_depth_1D_analytic}. The parameter estimation is performed using the nested sampling algorithm MultiNest \citep{Feroz2009, Feroz2013} through the implementation PyMultiNest \citep[][]{Buchner2014}. The priors used are explained in Appendix \ref{app:priors}. The synthetic observations follow the resolution ($R=60$) and precision ($50$ ppm) chosen by \MGL\, for HST-WFC3, representative of current HST observations \citep[e.g.,][]{Sing2016}. The model spectra are generated by sampling the high-resolution opacities at their native resolution ($0.1$~cm$^{-1}$). The high resolution models are binned to the instrumental resolution to perform the model-data comparison \citep[see e.g.,][and Section \ref{subsec:generating_data}]{Pinhas2019}. The synthetic data do not include Gaussian scatter following the approach of \MGL. 

The 2D semi-analytic model that produced the synthetic observations assumes a warm evening terminator with an isothermal temperature $T_E=2000$~K, H$_2$O volume mixing ratio of $\log_{10}(X_{\text{H}_2\text{O}})_E=-6.0$, and a reference pressure of 10~bar. The cooler morning terminator assumes an isothermal temperature $T_M=1000$~K and $\log_{10}(X_{\text{H}_2\text{O}})_M=-3.0$. The reference pressure for the morning terminator is derived in Appendix \ref{app:semi_analytic_values} to be $\sim7$~mbar, assuming that the reference radius in the deep atmosphere is the same between the morning and evening terminator and that both reference points correspond to the same optical depth surface. The chosen input values are only illustrative and selected as an extreme case of thermal and chemical inhomogeneities.

\begin{figure}
\includegraphics[width=0.45\textwidth]{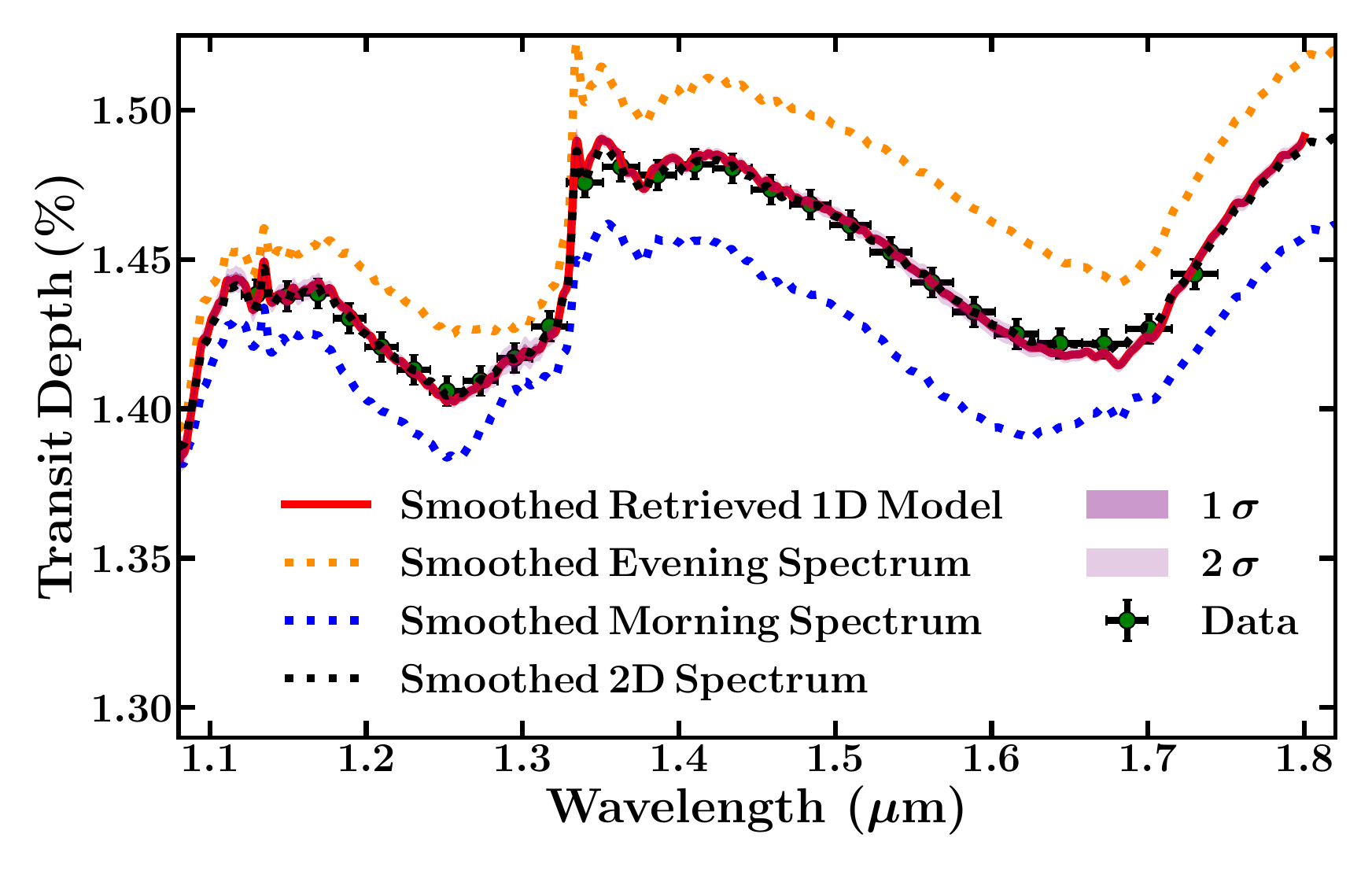}
\includegraphics[width=0.45\textwidth]{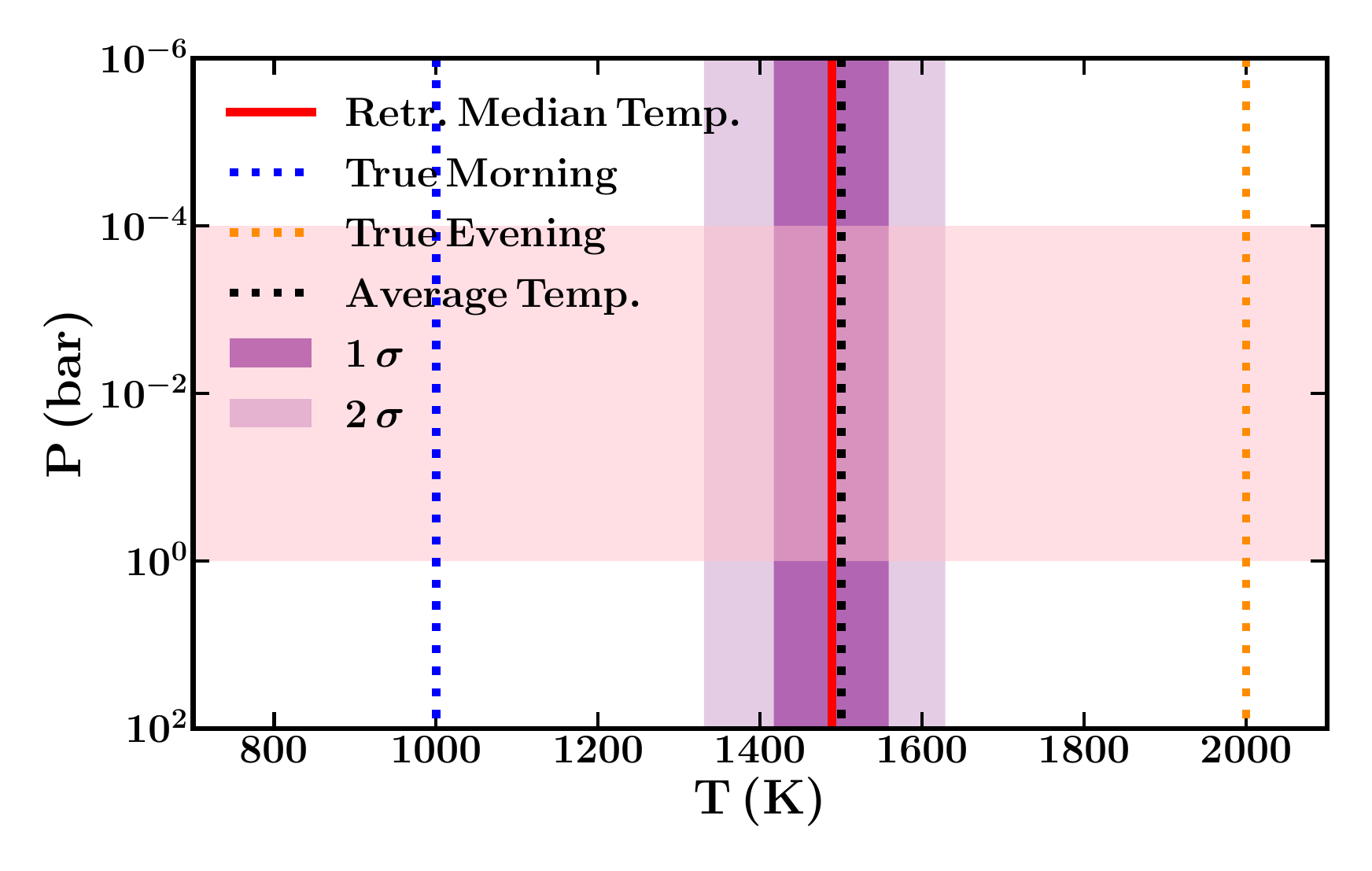}
\centering
\caption[]{Retrieved results using a semi-analytic 1D model on a semi-analytic 2D synthetic dataset. Top: Retrieved 1D transmission spectrum (red line) and synthetic observations (black error bars) from a 2D transmission spectrum (black dotted line). Coloured dotted lines show the morning (blue) and evening (orange) spectra. Bottom: Red line shows the retrieved median 1D temperature. Dotted lines show the morning (blue) and evening (orange) input temperatures, and their average (black). Pink horizontal shaded region shows the estimated photospheric pressure range from \MGL. Retrieved 1$\sigma$ and 2$\sigma$ confidence intervals in the transmission spectrum and $P$--$T$ profile are shown using purple shaded regions. Confidence intervals on the transmission spectrum are not visible due to their narrow range. \label{fig:analytic_spectra_pt_1} }  
\end{figure}

\begin{figure}
\includegraphics[width=0.45\textwidth]{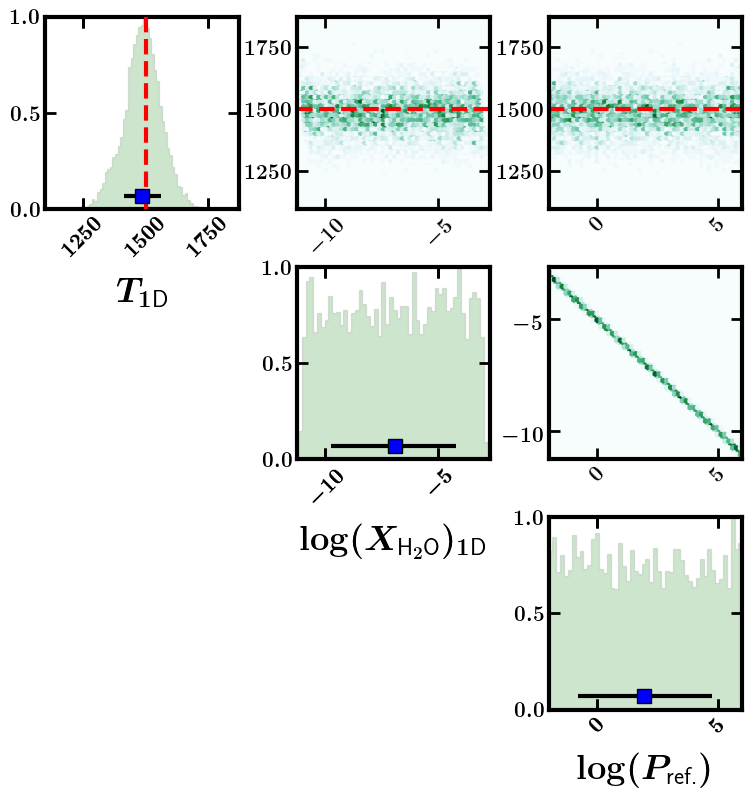}
\centering
\caption[]{Posterior distributions for the retrieval using a semi-analytic 1D model on a semi-analytic 2D synthetic dataset without assuming a reference pressure. The green histograms show the retrieved 1D isothermal temperature, 1D H$_2$O mixing ratio, and 1D reference pressure. The expectation (red line) from our semi-analytic derivation is $T_{1D}=\bar{T}=1500$~K. Without assumptions about the reference pressure in the 1D model, the semi-analytic treatment expects a degeneracy between the chemical abundance and reference pressure. The retrieval matches the expectations from our semi-analytic treatment. \label{fig:analytic_corner_1} }
\end{figure}

Using the above input values, the expected retrieved isothermal temperature for the 1D model would be $T_{1D}=\frac{1}{2}(T_E+T_M)=\frac{1}{2}(2000+1000)=1500$~K. As dictated by Equation \ref{eq:1d_properties}, without any assumptions about the reference pressure of the 1D model we expect a degeneracy between the chemical abundance and reference pressure. First, we perform a retrieval for which we retrieve the 1D isothermal temperature, 1D H$_2$O mixing ratio, and 1D reference pressure for a radius of $R_{\rm p,\, ref} = 1.33182 R_{\rm J}$. Figures \ref{fig:analytic_spectra_pt_1} and \ref{fig:analytic_corner_1} show the results from this retrieval. 

In agreement with the expectations from our semi-analytic derivation, our retrieval finds a 1D isothermal temperature of $T_{\rm 1D}=1487 ^{+ 72 }_{- 73 }$ consistent with the average of the input morning and evening temperatures. The abundance and reference pressure are not constrained and degenerate with each other. In order to break this degeneracy in semi-analytic models and derive a single chemical abundance, one must assume a reference pressure \citep[][]{Lecavelier2008a}. For illustration, we perform a second retrieval assuming an arbitrary reference pressure of 0.1~bar for a radius of $R_{\rm p,\, ref} = 1.33182 R_{\rm J}$.

\begin{figure}
\includegraphics[width=0.45\textwidth]{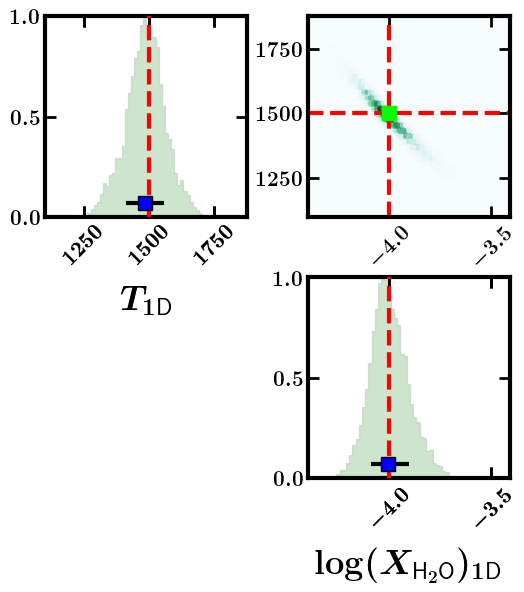}
\centering
\caption[]{As Figure \ref{fig:analytic_corner_1} but for the retrieval assuming a reference pressure in the 1D model of 0.1~bar. The expectations (red lines) from our semi-analytic derivation are $T_{1D}=\bar{T}=1500$~K, and $\log_{10}(X_{\text{H}_2\text{O}})=-4$. The retrieval matches the expected values. \label{fig:analytic_corner_2} }
\end{figure} 

Figure \ref{fig:analytic_corner_2} shows the posterior distributions from the retrieval assuming a reference pressure of 10~bar. With these input values, the 1D chemical abundance from Equation \ref{eq:1d_properties} is expected to be $\log_{10}(X_{\text{H}_2\text{O}})=-4$, as derived in Appendix \ref{app:semi_analytic_values}. Our retrieval finds $T_{\rm 1D}=1485 ^{+ 71 }_{- 75 }$ and $\log_{10}(X_{\text{H}_2\text{O}})=-4.0 ^{+ 0.1 }_{- 0.1 }$, consistent with expectations. The retrieved transmission spectrum and $P$--$T$ profile is virtually unchanged from the ones shown in Figure \ref{fig:analytic_spectra_pt_1}.

Overall, these semi-analytic retrievals confirm the intuition derived from the semi-analytic derivation included in Appendix \ref{app:derivation}. When equating semi-analytic 1D and 2D models, the temperature from the 1D model is the average temperature of the two terminators in the 2D model. Our results contrast with the properties of the analytic expression from \MGL, particularly three of their important takeaways. First, \MGL\, claim that compositional differences exceeding a factor of 2 (i.e., $\Delta\log_{10}(X)>0.3$) like the one in our example ($\Delta\log_{10}(X)=3.0$) result in biases many hundreds of degrees cooler than the average temperature of the two terminators. We find that, even in this case where the compositional difference is a factor of 1000, we retrieve a temperature consistent with the average temperature of the morning and evening terminators. Second, \MGL\, claim that the wavelength dependency in their derived analytic expression implies that no one equivalent temperature can perfectly reproduce a 2D spectrum using a 1D model. Our results find that using this semi-analytic approach, a 1D model can reproduce a 2D spectrum. Third, \MGL\, highlight that their semi-analytic derivation predicts $T_{1D}<\frac{1}{2}(T_E+T_M)$ under the assumption that the retrieved 1D mixing ratio is the average of the morning and evening abundances. Our results and analytic derivation indicate show that the 1D volume mixing ratio in the semi-analytic treatment is not expected to be the average of the morning and evening mixing ratios. Finally, our results recover the well-known degeneracy between abundances and reference pressure from simplistic semi-analytic models. 

Semi-analytic models and retrievals remain powerful tools to help build our understanding of transmission spectra. Nonetheless, when looking for more physically realistic models that incorporate multidimensional thermal and compositional inhomogeneities, the results of semi-analytic formalisms can be misleading. As comprehensively discussed in \citet{Welbanks2019a}, simplified semi-analytic models are inadequate for reliable interpretations of transmission spectra. Instead, numerical models that relax some of these unphysical assumptions can be better tools in our analysis of exoplanet atmospheres. As such, we now investigate the retrieved atmospheric temperatures for 2D spectra using 1D models within a numerical atmospheric retrieval framework.  

\section{1D Atmospheric Retrievals with Synthetic Observations}
\label{sec:1D_retrievals}

We explore the possibility of thermal and compositional biases resulting from retrieving the atmospheric properties of an exoplanet with asymmetric terminators (i.e., a 2D transmission spectra) using 1D atmospheric models within a numerical Bayesian atmospheric retrieval framework. Using synthetic observations, we appraise the performance of three different model families: (1) models as in \MGL\, using a modified $P$--$T$ parameterization and assuming a clear atmosphere; (2) models using the $P$--$T$ parameterization of \citet{Madhusudhan2009} and assuming a clear atmosphere; and (3) models using the $P$--$T$ parameterization of \citet{Madhusudhan2009} and including the presence of inhomogenous clouds and hazes as described in \citet{Welbanks2021}, equivalent to the current state-of-the art in the field. We select three diverse cases with thermal and compositional differences ranging from warm Jupiters ($T_{\rm eq}\sim 1000$~K) to ultra-hot Jupiters ($T_{\rm eq}\sim2500$~K). To allow for a direct comparison with \MGL\, and their assumptions, we adopt their proposed atmospheric case studies summarized in Table \ref{tab:case_studies}.  

\begin{deluxetable*}{|l|c|c|c|c|c|c|}[ht!] 
\tablecaption{Morning And Evening Chemical And Thermal Parameters For Selected Case Studies. \label{tab:case_studies}}
\tablewidth{0pt}
\tablehead{
Atmospheric Case & WJ Morning & WJ Evening & HJ Morning & HJ Evening & UHJ Morning & UHJ Evening
}
\startdata
$\log_{10}(X_{\text{H}_2\text{O}})$ & $-3.3$  & $-3.3$  & $-3.3$  & $-3.3$  & $-3.3$  & $-3.3$  \\
$\log_{10}(X_{\text{Na}})$          & $-8.0$  & $-6.0$  & $-6.0$  & $-6.0$  & $-6.0$  & $-6.0$  \\
$\log_{10}(X_{\text{K}})$           & $-9.0$  & $-7.0$  & $-7.0$  & $-7.0$  & $-7.0$  & $-7.0$  \\
$\log_{10}(X_{\text{CH}_4})$        & $-4.0$  & $-6.0$  & N/A   & N/A   & N/A   & N/A   \\
$\log_{10}(X_{\text{TiO}})$         & N/A  & N/A    & N/A   & $-7.0$  & N/A   & N/A   \\
$\log_{10}(X_{\text{VO}})$          & N/A  & N/A    & N/A   & $-8.0$  & N/A   & N/A   \\
$\log_{10}(X_{\text{H}^-})$         & N/A  & N/A    & N/A   & N/A   & N/A   & $-8.0$  \\
$\alpha_{1}$                        & 0.6  & 0.7    & 0.6   & 0.7   & 0.5   & 0.7   \\
$\alpha_{2}$                        & 0.5  & 0.6    & 0.5   & 0.6   & 0.4   & 0.6   \\
$\log_{10}(P_{1})$                  & $-2.0$ & $-2.0$   & $-2.0$  & $-2.0$  & $-2.0$  & $-2.0$  \\
$\log_{10}(P_{2})$                  & $-5.0$ & $-5.0$   & $-5.0$  & $-5.0$  & $-5.0$  & $-5.0$  \\
$\log_{10}(P_{3})$                  & 1.0  & 1.0    &  1.0  &  1.0  & 1.0   &  1.0  \\
$T_{\rm deep}$                      & 1600 & 1600   & 2200  & 2200 & 3000   & 3000  \\
\enddata 
\tablecomments{The warm Jupiter (WJ), hot Jupiter (HJ), and ultra-hot Jupiter (UHJ) selected cases are constructed with the system and bulk properties of the prototypical hot Jupiter HD~209458b, and assume a reference radius of $R_{\rm p, 10bar} = 1.33182 R_{\rm J}$ at a reference pressure of $P_{\rm ref}=10$~bar. N/A means that the parameter is not included in the model by design. Models parameters adopted from \MGL. }
\end{deluxetable*}

\subsection{Generating Synthetic Observations}\label{subsec:generating_data}

We generate synthetic HST-STIS and HST-WFC3 observations assuming spectral resolutions and precisions comparable to current observations \citep[e.g.,][]{Sing2016}. To avoid possible biases due to different data generation procedures, we adopt the same parameters and procedure as \MGL. First, we generate a higher resolution ($R\sim2000$) spectrum from $0.3$--$2.0$~$\mu$m solving line-by-line radiative transfer in transmission geometry in a plane-parallel atmosphere for each terminator (morning and evening). The model atmosphere for each terminator is discretized into 81 pressure layers uniformly spaced in $\log_{10}(P)$ from $10^{-6}$ to $10^2$ bar under hydrostatic equilibrium. The high-resolution spectra are linearly combined to construct a 2D transmission spectrum. Each 2D spectrum is then convolved using the point spread function of the HST STIS G430/G750 gratings and HST-WFC3 G141 grisms and integrated over the sensitivity function of each instrument (see binning strategy in Section 2.1.6 in \citealt{Pinhas2018}). The assumed spectral resolutions and precisions are $R_{\rm HST-STIS}=20$ and $100$~ppm, respectively, for HST-STIS, and $R_{\rm HST-WFC3}=60$ and $50$~ppm for HST-WFC3. The observations do not include Gaussian scatter following \MGL. 

\subsection{Retrieval Setup}
\label{subsec:1d_retrieval_setup}

The atmospheric retrievals are performed using Aurora \citep[][]{Welbanks2021}. The atmospheric model computes line-by-line radiative transfer in transmission geometry for a plane-parallel planet atmosphere under hydrostatic equilibrium \citep[see e.g.,][]{Pinhas2018, Welbanks2021}. The parameter estimation is performed using the nested sampling algorithm MultiNest \citep[][]{Feroz2009, Feroz2013}, through PyMultiNest \citep{Buchner2014}. Each retrieval is performed using $4000$ nested sampling live points. We consider three model configurations: 

\begin{itemize}
    \item MGL20: These models follow the procedure of \MGL\, where the $P$--$T$ profile is parameterized using a modification to the parameterization of \citet{Madhusudhan2009}. That is, instead of retrieving the temperature at the top of the atmosphere ($T_0$) as \citet{Madhusudhan2009}, \MGL\, retrieve the temperature at 10 bar ($T_{\rm deep}$). These models retrieve the reference planetary radius at a reference pressure of 10 bar ($R_{\rm p, \, 10 \, bar}$), and assume a clear atmosphere.
    \item Clear atmosphere: These models follow the standard retrieval procedure explained in \citet{Welbanks2021} for a clear atmosphere. The $P$--$T$ parameterization follows the prescription of \citet{Madhusudhan2009}. The reference pressure is retrieved for the reference radius used in the morning and evening models.
    \item Cloudy atmosphere: These models follow the same procedure as the clear atmosphere models, but include the possibility of clouds and hazes using the generalised prescription of \citet{Welbanks2021}. The parameterization considers the possibility of four different sectors: 1) a clear sector, 2) a sector with hazes only, 3) a sector with clouds only, and 4) a sector with clouds and hazes. Clouds are included by a pressure parameter ($P_{\mathrm{cloud}}$) that determines the pressure level at which the atmosphere becomes optically thick due to the presence of a cloud deck. Hazes are included as a deviation from H$_2$ Rayleigh scattering using a parametric cross section $\sigma_{\mathrm{hazes}}=a\sigma_0(\lambda/\lambda_0)^\gamma$, where $\gamma$ is the scattering slope, $a$ is the Rayleigh-enhancement factor, and $\sigma_0$ is the H$_2$ Rayleigh-scattering cross section ($5.31\times10^{-31}$~m$^2$) at a reference wavelength $\lambda_0$ ($350$~nm). The fractional cover of each of the four sectors above is a free parameter and follows the unit-sum constraint (i.e., $\phi_{\mathrm{clear}}=1-\phi_{\mathrm{hazes}}- \phi_{\mathrm{clouds}}-\phi_{\mathrm{clouds+hazes}}$).
\end{itemize}

To avoid biases due to different pressure grids between models, we maintain the choice of 81 pressure levels from $10^{-6}$ to $10^2$ bar of \MGL\, for all model configurations. The retrievals include as free parameters the abundances of the chemical species used as input in each of the atmospheric cases (i.e., warm Jupiter: H$_2$O, Na, K, and CH$_4$; hot Jupiter: H$_2$O, Na, K, TiO, and VO; ultra-hot Jupiter: H$_2$O, Na, K, H$^-$). The sources of opacity included are H$_2$-H$_2$ and H$_2$-He CIA \citep[][]{Richard2012}, H$_2$O \citep{Rothman2010}, CH$_4$ \citep{Yurchenko2013, Yurchenko2014}, Na \citep{Allard2019}, K \citep{Allard2016}, TiO \citep{Kurucz1992, Schwenke1998, Hill2013}, VO \citep{McKemmish2016}, and bound-free H$^-$ \citep{John1988} with their computation following the methods and descriptions in \citet{Gandhi2017, Gandhi2018, Gandhi2020a, Gandhi2020b} and \citet{Welbanks2019b}. The priors, summarized in Appendix \ref{app:priors} Table \ref{table:priors}, are generally standardised and follow the description of \MGL\, for the MGL20 models, and \citet{Welbanks2021} for the clear and cloudy models. We choose to retrieve the reference pressure instead of the reference radius as the planetary radius is generally the observable quantity.

\subsection{Retrieval Results}
\label{subsec:1d_results}

\begin{figure*}[ht!]
    \centering
    \includegraphics[width=0.94\textwidth]{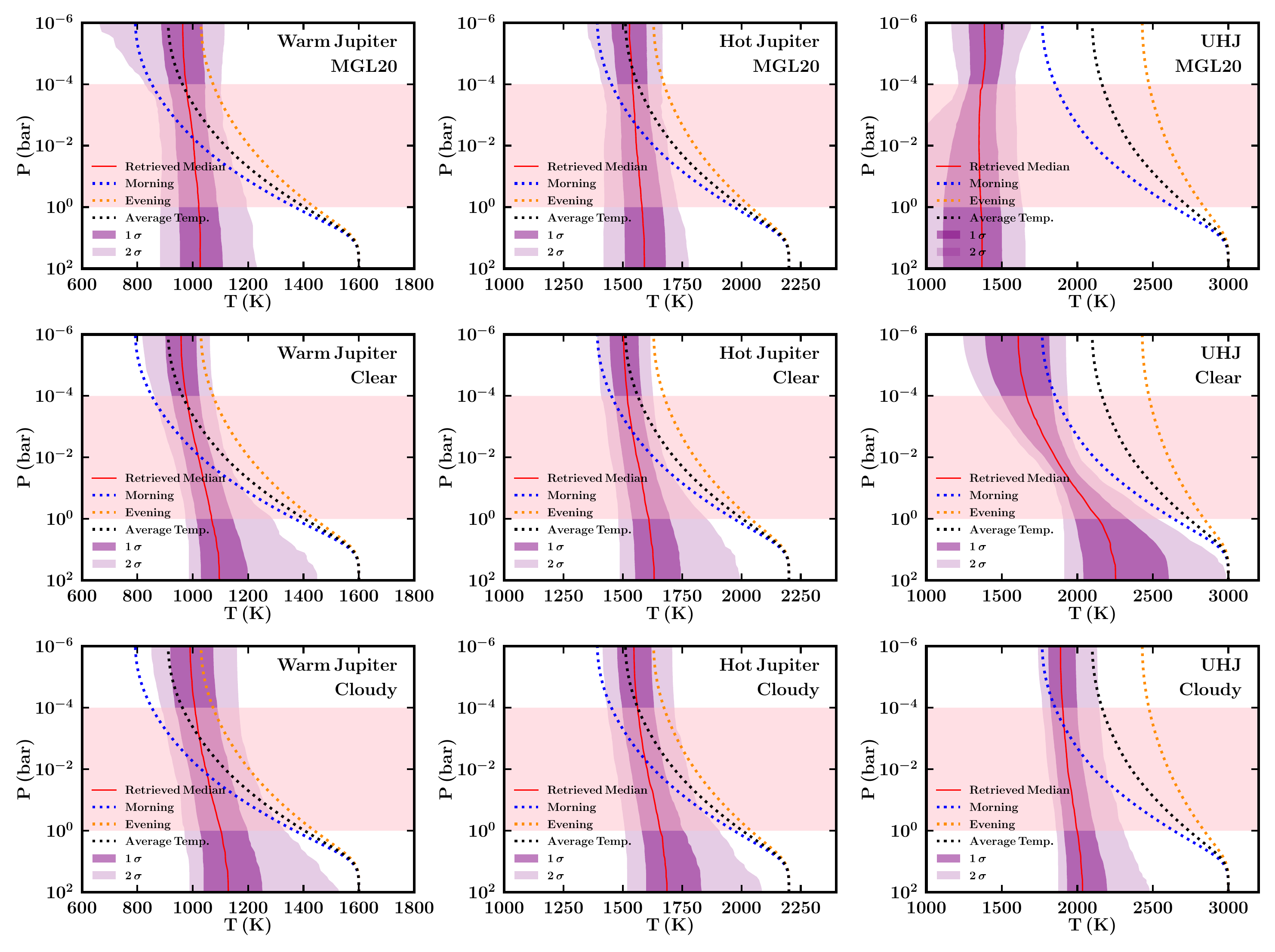}
    \caption{Retrieved $P$--$T$ profiles from our numerical retrievals using 1D models on synthetic data from a 2D model. Each column corresponds to an atmospheric case adopted after \MGL: a warm Jupiter ($\bar{T}\sim 1000$~K), a hot Jupiter ($\bar{T}\sim 1600$\,K), and an ultra-hot Jupiter ($\bar{T}\sim 2200$\,K). Each row corresponds to a different model strategy: a clear atmospheric model with the $P$--$T$ parameterization from \MGL\, (top), a cloud-free model using the $P$--$T$ parameterization from \citet{Madhusudhan2009} (middle), and a model with inhomogeneous clouds and hazes and the $P$--$T$ parameterization from \citet{Madhusudhan2009} (bottom). The true morning (blue) and evening (orange) $P$--$T$ profiles used to generate the 2D model and synthetic data, along with their average (black), are shown using dotted lines. The retrieved median (red line) and $1\sigma$ and $2\sigma$ confidence intervals (purple shaded regions) are shown. Pink horizontal shaded region shows the estimated photospheric pressure range from \MGL.}
    \label{fig:pt_grid}
\end{figure*}

The retrieved $P$--$T$ profiles from our retrievals are shown in Figure \ref{fig:pt_grid}, while the retrieved spectra and posterior distributions are included in Appendix \ref{app:1D_retrieval_figures}, Figures \ref{fig:spectrum_grid} and \ref{fig:posterior_grid}. First, our reproduction of the results from \MGL\, are consistent with their findings. When using the $P$--$T$ parameterization from \MGL, the retrieved photospheric temperatures ($T_{\rm 1 \, bar}$ to $T_{\rm 0.1 \,mbar}$) are generally cooler than the average of their morning and evening terminator temperatures. For the warm Jupiter case, the average temperature is consistent within the 2$\sigma$ confidence contour of the retrieved estimates for $P\lesssim0.01$~bar. The retrieved temperature estimates for the deeper atmosphere are inconsistent with the average temperature with discrepancies of $\sim 200$~K at $P\sim1$~bar. The hot Jupiter case is similar, with a similar discrepancy of $\sim 200$~K between the retrieved $2\sigma$ confidence contour and the average temperature at $P\sim 1$~bar. Finally, the largest bias is found in the ultra-hot Jupiter scenario, for which the retrieved temperature and associated confidence intervals are inconsistent with the average temperature. The $2\sigma$ confidence interval and the average temperature are inconsistent at $P\sim1$~bar to $\gtrsim1000$~K.

Our reproduction of \MGL\, also obtains the compositional biases they highlight and the relatively poor constraints of Na and K for the hot and ultra-hot Jupiter scenarios. Although the spectral fits (top row, Figure \ref{fig:spectrum_grid}) are generally a good fit to the data, the retrieved abundances (top row, Figure \ref{fig:posterior_grid}) can be inconsistent with the expectation of average chemical abundances to within 1$\sigma$ for the warm and hot Jupiter scenarios. On the other hand, the ultra-hot Jupiter results show a worse spectral fit and the retrieved abundances of H$_2$O and H$^-$ can be inconsistent with the assumption of average chemical assumptions above $3\sigma$. 

Next, we perform a retrieval using our standard setup for clear atmospheres using the $P$--$T$ parameterization from \citet{Madhusudhan2009}. The results are shown in the middle row of Figure \ref{fig:pt_grid} and Figures \ref{fig:spectrum_grid} and \ref{fig:posterior_grid} in Appendix \ref{app:1D_retrieval_figures}. The retrieved $P$--$T$ profiles for this model are somewhat less biased than those obtained using the \MGL\, $P$--$T$ parameterization. The discrepancy between the $2\sigma$ confidence interval and the average temperature at $P\sim1$~bar is $\sim100$~K for the warm and hot Jupiter cases. Similarly, the retrieved temperatures at the slant photosphere \citep[$P\sim10^{-3}$--$10^{-1}$~bar, e.g.,][]{Welbanks2019a} are generally consistent with the average temperature of the morning and evening terminators within the 2$\sigma$ confidence intervals to within 100~K. 

On the other hand, the ultra-hot Jupiter temperature estimates at the transmission spectroscopy photosphere remain inconsistent, although to a lesser degree. The difference between the $2\sigma$ confidence interval and the average temperature for the ultra-hot Jupiter is $\lesssim350$~K at most pressures. The retrieved transmission spectra (middle panel of Figure~\ref{fig:spectrum_grid}) for the clear model shows a reasonably good fit to the synthetic observations, with a worse fit for the ultra-hot Jupiter case in the optical wavelengths. The retrieved chemical abundances (middle panel of Figure~\ref{fig:posterior_grid}) remain mostly unchanged from those retrieved with the \MGL\, models, with the exception of H$^-$ in the ultra-hot Jupiter case which is consistent with the average of the morning and evening terminators. The retrieved Na and K abundances are relatively unconstrained in the hot and ultra-hot Jupiter cases.

The results from our models with inhomogeneous clouds and hazes are shown in the bottom row of Figures \ref{fig:pt_grid}, \ref{fig:spectrum_grid}, and \ref{fig:posterior_grid}. The retrieved $P$--$T$ profile estimates from the cloudy models for the warm and hot Jupiter cases are largely consistent with the average temperature of the two terminators near the slant photosphere ($P\sim10^{-3}$--$10^{-1}$~bar). The average temperature is marginally consistent with the retrieved 2$\sigma$ contour at $P\sim1$~bar to within $\lesssim100$~K for the warm and hot Jupiter cases. The ultra-hot Jupiter temperature estimates are consistent with the average temperature for $P\lesssim0.1$~mbar, and inconsistent at $\lesssim450$~K for higher pressures up to $\sim1$~bar.  

The cloudy models retrieved spectra that is in good agreement with the synthetic data for all planet scenarios. Although our models allow for the possibility of clouds and hazes, the retrieved solutions correctly indicate that the input data corresponds to a cloud-free atmosphere (i.e., $\phi_{\mathrm{hazes}}\sim 0$, $\phi_{\mathrm{clouds}}\sim 0$, $\phi_{\mathrm{clouds+hazes}}\sim 0$). Furthermore, the retrieved chemical abundances are consistent with the average of the abundances of the morning and evening terminators within $\sim1\sigma$ for the warm and hot Jupiter cases, even though the retrieved abundances may not strictly be expected to be the average of the two terminators. The retrieved H$_2$O and H$^-$ abundances for the ultra-hot Jupiter scenario are consistent with the average of the two terminators at $\sim3\sigma$. Importantly, the retrieved posterior distributions for some chemical abundances begin to show multi-modal behaviour. Observing such modes in a posterior distribution would indicate us to be careful with our interpretation of the retrieved median abundance, and to not treat that value as a reliable indication of the overall abundance of the planet \citep[see e.g., WASP-39b in][]{Welbanks2019b}.

Even for the most extreme case of an ultra-hot Jupiter, our photospheric temperature discrepancy of $\lesssim450$ K from the true average at the $2\sigma$ boundary is still significantly lower than that of \MGL. Nevertheless, it is important to establish the cause of any such discrepancy. An inspection of the model behind the synthetic observations for this case, following \MGL\, offers some insights. The third column in Figure \ref{fig:spectrum_grid} shows the evening and morning spectra that formed the 2D input spectrum via linear combination. The evening terminator has almost no visible spectral features in the optical due to the high H$^-$ abundance assumed in the case study, resulting in an almost flat spectrum. On the other hand, the morning terminator spectrum has clear Na, K and H$_2$O features. Therefore the ultra-hot Jupiter scenario is a pathological scenario due to its almost featureless evening spectrum. 

It is also important to note that parts of both terminators face the same sides, i.e. dayside or nightside. Compositional and temperature inhomogeneities may be expected to be stronger across day-night boundary than the morning-evening boundary as considered here. The biases uncovered by this pathological case may, therefore, be representative of only the most extreme of cases and not of the broader hot Jupiter population. Nevertheless, we still address this extreme case with a new retrieval framework in Section~\ref{sec:2D_retrievals}. 

Overall, our results find that the modified $P$--$T$ profile used by \MGL\, result in cooler temperature estimates by a factor of $\sim2$ or more compared to the other cases we consider. On the other hand, the models using the $P$--$T$ parameterization from \citet{Madhusudhan2009} find retrieved temperatures near the photosphere for transmission spectroscopy largely consistent with the average of the two terminators for the warm and hot Jupiter scenarios and a smaller thermal bias for the ultra-hot Jupiter. Finally, the inclusion of inhomogeneous clouds and hazes in the atmospheric models results in chemical abundances that are less biased than those obtained with the cloud-free models. 

\section{1D Atmospheric Retrievals with Existing Observations}
\label{sec:real_data}

We extend our investigation into the origin of thermal anomalies by revisiting published transmission spectra for which previous studies have inferred anomalously cool atmospheric temperatures. We choose two planets with widely different temperatures and for which broad optical to infrared observations are available. Particularly, we revisit recent inferences on the atmospheric properties of the hot Jupiter WASP-43b \citep[][]{Hellier2011} and ultra-hot Jupiter WASP-103b \citep[][]{Gillon2014} made by \citet{Weaver2020} and \citet{Kirk2021}, respectively. We revisit their reported inferences, reproduce them using their model assumptions where possible, and perform additional retrievals using the full model set-up discussed above to resolve biases.

\subsection{The Hot Jupiter WASP-43b}

\citet{Weaver2020} performed an atmospheric retrieval of the hot Jupiter WASP-43b using a ground-based optical transmission spectrum ($\lambda=0.53$--$0.9$~$\mu$m) obtained as part of the ACCESS Survey on Magellan/IMACS,in combination with HST-WFC3 observations from \citet{Kreidberg2014b}. Their atmospheric retrieval framework, adapted from the work of \citet{Espinoza2019}, assumes an isothermal and isobaric atmosphere following semi-analytic treatments \citep[e.g.,][]{Betremieux2017, Heng2017}. Their results find no evidence for chemical absorbers such as Na and K in the optical, but confirm the presence of H$_2$O in the atmosphere of the planet with abundances consistent with previous results \citep[e.g.,][]{Kreidberg2014b}. Furthermore, their results suggest the presence of star spots on the surface of WASP-43 impacting the transmission spectrum of the planet. Nonetheless, they retrieve an isothermal temperature for the atmosphere of the planet of $T_{\rm iso.}=352.91^{+206.14}_{-125.08}$~K, an estimate considerably cooler than the equilibrium temperature of the planet $T_{\rm eq}=1440$~K \citep[e.g.,][]{Welbanks2021} and expectations from GCM studies \citep[e.g.,][]{Kataria2015}.

\subsubsection{Retrieval Setup} \label{subsubsec:wasp43_setup}

\citet{Weaver2020} employ simplified semi-analytic models with isothermal and isobaric assumptions. The limitations of these models, and their impact on retrieved atmospheric properties, has been extensively discussed in \citet{Welbanks2019a}. Although their general retrieval setup is presented, some key aspects of their modelling strategy are not self-evident (e.g., the complete list of chemical species considered and whether collision-induced absorption was included). As a result, we do not perform a direct reproduction of their results using semi-analytic models, nor are we able to incorporate their exact modelling assumptions into our numerical 1D models. Instead, we perform a numerical retrieval relaxing the isothermal and isobaric assumptions of their semi-analytic model. We employ the $P$--$T$ parameterization of \citet{Madhusudhan2009} with the same priors as \citet{Welbanks2019b} (e.g., as shown in Table \ref{table:priors}, except for $T_0$ which follows a uniform prior between $800$~K and $1540$~K). 

Our retrieval for WASP-43b follows the general description presented in Section \ref{subsec:1d_retrieval_setup}. The parameter estimation is performed using PyMultiNest \citep[][]{Buchner2014} with 2000 nested sampling live points. Our model considers free parameters for the volume mixing ratios of different chemical species, the possibility of inhomogeneous clouds and hazes, and a parameterization for the $P$--$T$ profile of the planet. 

The retrieval in \citet{Weaver2020} does not consider non-uniform cloud cover. Instead, they interpret their retrieved reference pressure as a cloud-top pressure that indicates the transition between an optically thin and an optically thick atmosphere due to clouds. Additionally, \citet{Weaver2020} include the presence of hazes using the parameterization explained in Section \ref{subsec:1d_retrieval_setup} and in \citet{Lecavelier2008a, MacDonald2017a, Pinhas2018, Welbanks2019a} and \citet{Welbanks2021}. Our retrieval considers the presence of clouds and hazes following the description in Section \ref{subsec:1d_retrieval_setup}, with the full four-sector generalised parameterization of \citet{Welbanks2021} for non-uniform cloud/haze cover.

\citet{Weaver2020} explore the impact of stellar heterogeneities by following the formalism described in \citet{Rackham2018, Rackham2019}, and \citet{Pinhas2018}. We do the same using the functionalities in Aurora inherited from AURA \citep{Pinhas2018}. We adopt a uniform prior for the spot fraction between $0$ and $0.5$, a uniform prior in the spot temperatures between $0.6$ and $1.2$ times the stellar effective temperature \citep[i.e., $T_{\rm eff.}=4520$~K, e.g.,][]{Gillon2012}, and a Gaussian prior for the photospheric temperature with a mean of $T_{\rm eff.}=4520$~K and standard deviation of $120$~K informed by \citet{Gillon2012}.

In agreement with \citet{Weaver2020}, we also include a free parameter to retrieve an instrumental shift between the optical and infrared observations. We use a Gaussian prior with a standard deviation of $1000$~ppm, similar to \citet{Weaver2020}, and a mean of $0$~ppm. We consider absorption due to H$_2$O, Na, K, CH$_4$, NH$_3$, CO, CO$_2$, TiO and VO, with a log-uniform prior on their volume mixing ratios from $-16$ to $-1$. Our retrieval uses the optical data from \citet{Weaver2020}, the HST-WFC3 observations from \citet{Kreidberg2014b}, the Spitzer transit depths at 3.6~$\mu$m and 4.5~$\mu$m from \citet{Stevenson2017}, and input system parameters from \citet{Gillon2012}. 

\subsubsection{Retrieval Results}

\begin{figure}
    \centering
    \includegraphics[width=0.45\textwidth]{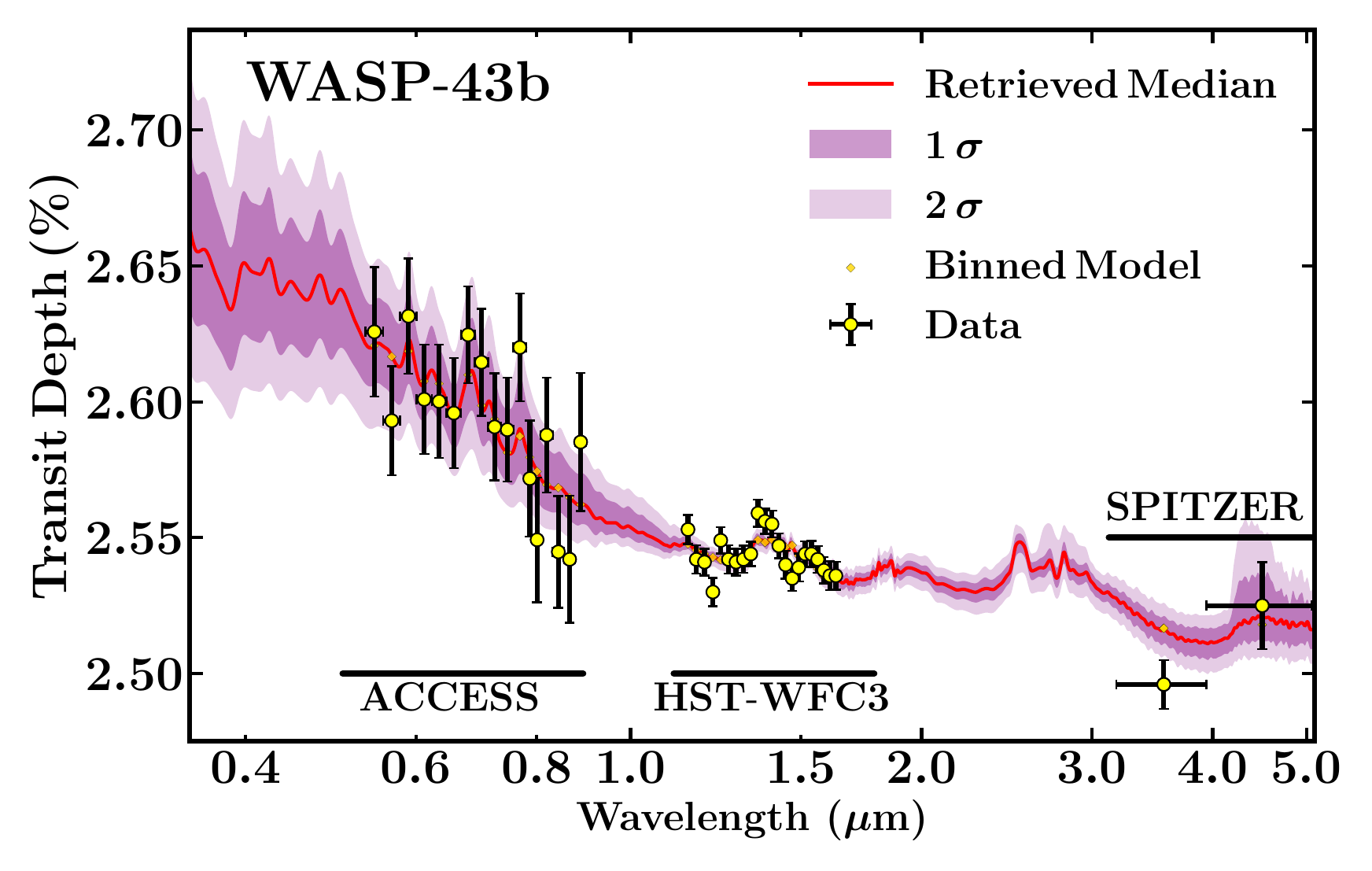}
    \includegraphics[width=0.45\textwidth]{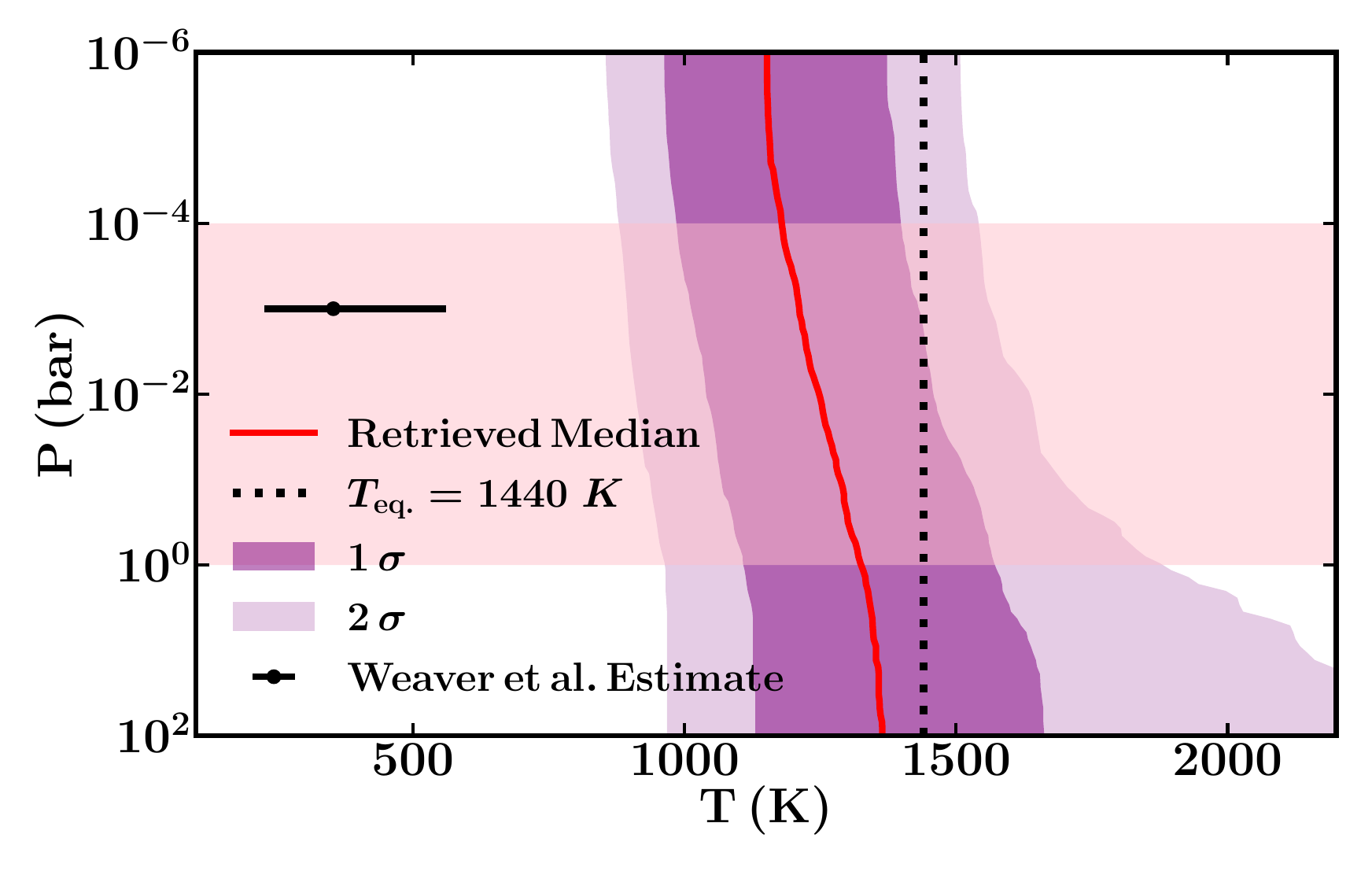}
    \caption{Results for the retrieval of ACCESS, HST-WFC3, and Spitzer WASP-43b data from \citet{Weaver2020, Kreidberg2014b} and \citet{Stevenson2017}. Top: Retrieved median transmission spectrum (red line) and 1$\sigma$ and 2$\sigma$ confidence intervals (purple shaded regions) are shown alongside the observations (error bars) and binned median model (diamonds). Bottom: Retrieved median $P$--$T$ profile (red line) and $1\sigma$ and $2\sigma$ confidence intervals (purple shaded regions). Pink horizontal shaded region shows the estimated photospheric pressure range from \MGL. Horizontal error bar shows the median and 1$\sigma$ retrieved temperature from \citet{Weaver2020}. Our retrieved temperature is consistent with the equilibrium temperature of the planet ($T_{\rm eq}=1440$~K, dotted line). }
    \label{fig:results_wasp43b}
\end{figure}

Our retrieval of the optical and infrared transmission spectrum of the hot Jupiter WASP-43b suggest the presence of H$_2$O ($\sim 2 \sigma$), a strong preference for an offset between the optical and infrared observations ($\sim 7 \sigma$), and evidence for stellar contamination in the optical transmission spectrum ($\sim 4 \sigma$). Figure \ref{fig:results_wasp43b} shows the retrieved transmission spectrum and the retrieved $P$--$T$ profile. Our retrieved transmission spectrum is in good agreement with the observations. Furthermore, our retrieved $P$--$T$ profile has a temperature near the slant photosphere ($P\sim10^{-3}$--$10^{-1}$~bar) consistent with the equilibrium temperature of the planet ($T_{\rm eq}=1440$~K) within $\sim1\sigma$.

Our retrieved H$_2$O volume mixing ratio of $\log_{10}(X_{\text{H}_2\text{O}})=-5.27 ^{+ 0.61 }_{- 0.46 }$ is consistent within $\sim 1 \sigma$ with previous estimates using infrared observations only \citep[e.g.,][]{Kreidberg2014b, Welbanks2019b}, and lower than the estimate from \citet{Weaver2020} not using the Spitzer transit depths (i.e., $\log_{10}(X_{\text{H}_2\text{O}})=-2.78 ^{+ 1.38 }_{- 1.47}$). Our retrieval suggest the presence of stellar heterogeneities in the transmission spectrum as the result of spots with an average temperature of $T_{\rm het.}=4102.84 ^{+ 86.99 }_{- 99.56 }$, $\sim 300$~K cooler than the retrieved stellar photosphere $T_{\rm phot.}=4484.76 ^{+ 82.72 }_{- 72.46 }$, with a cover fraction of $\delta=0.11 ^{+ 0.05 }_{- 0.03 }$. These estimates are generally consistent with the retrieved parameters of \citet{Weaver2020}. 

Additionally, we retrieve an offset of $f_{\rm offset}=847 ^{+133}_{-128}$~ppm between the optical and infrared observations. \citet{Weaver2020} retrieve an offset of $f_{\rm offset}=25749.40^{+101.54}_{-105.70}$~ppm. Considering the mean white light curve depth of $25071$~ppm derived by \citet{Weaver2020}, their retrieved offset translates to a value of $f_{\rm offset}=678.40^{+101.54}_{-105.70}$~ppm, which is consistent with our estimate to within $1\sigma$. We note, however, that our retrieval includes both HST-WFC3 and Spitzer observations whereas \citet{Weaver2020} use only HST-WFC3 in the infrared.

Our retrieved $P$--$T$ profile is largely consistent with the equilibrium temperature of the planet $T_{\rm eq}=1440$~K \citep[e.g.,][]{Welbanks2021}. The retrieved temperature at 100~mbar, near the photosphere, is $T_{\rm 100~mbar}=1257^{+235}_{-214}$~K and consistent with the equilibrium temperature of the planet within 1$\sigma$. Additionally, the retrieved temperature at 100~mbar falls within the range of temperatures at $\sim100$~mbar found for the same planet by \citet{Kataria2015} using GCMs. Figure \ref{fig:results_wasp43b} also shows the retrieved isothermal temperature of $T_{\rm iso.}=352.91^{+206.14}_{-125.08}$~K from \citet{Weaver2020}. Our retrieved median temperature at 100~mbar is $\sim 900$~K warmer than the median retrieved temperature of \citet{Weaver2020}.

We further assess whether our models preferentially explain the optical data using models that consider the impact of stellar heterogeneities or the presence of inhomogeneous clouds and hazes. \citet{Weaver2020} find that given current data quality, it is not possible to fully distinguish between the impact on the spectrum due to hazes and the contamination from stellar heterogeneities. We find no preference for models with clouds and hazes relative to models without. Our retrieved cloud and haze properties are not constrained and largely suggestive of an atmosphere with low cloud and haze cover (i.e., $\phi_{\rm clouds}\approx\phi_{\rm hazes} \approx \phi_{\rm clouds+hazes} \approx 0$). These results suggest that the data is preferentially explained by contamination from a non-homogenous stellar photosphere instead of non-homogenous cloud and haze cover. Future observations can better inform these poor constraints on the cloud and haze properties of WASP-43b.

Our results indicate that the use of semi-analytic models with isothermal and isobaric assumptions\footnote{Failing to consider collision-induced absorption will additionally impact any atmospheric inferences \citep[see e.g.,][]{Welbanks2019a}. If that is the case for the semi-analytic models of \citet{Weaver2020}, this assumption would have exacerbated any retrieved biases.} by \citet{Weaver2020} may have biased their atmospheric temperature estimates, leading to cooler temperatures inconsistent with expectations for the atmospheric temperature of the planet. On the other hand, our numerical models with height-dependent gravity, a parametric $P$--$T$ profile, and inhomogeneous clouds and hazes result in temperature estimates consistent with the equilibrium temperature of the planet and GCM simulations. Our retrieval using the full numerical model suggests that the retrieved temperature biases from \citet{Weaver2020} for WASP-43b may likely be due to the use of simplified semi-analytic models rather than due to biases in 1D retrievals applied to  asymmetric terminators in hot Jupiters. 

\subsection{The Ultra-Hot Jupiter WASP-103b}

\citet{Kirk2021} use a ground-based optical transmission spectrum ($\lambda=0.390$--$0.945$~$\mu$m) from the ACCESS (Magellan/IMACS) and LRG-BEAST (William Herschel Telescope-ACAM) surveys, as well as archival data from Gemini-GMOS and VLT-FORS2, in combination with infrared HST-WFC3 and Spitzer data from \citet{Kreidberg2018} to infer the atmospheric properties of the ultra-hot Jupiter WASP-103b. Their atmospheric retrievals are performed using the frameworks petitRADTRANS \citep[][]{Molliere2019} for the optical transmission spectrum, and POSEIDON \citep{MacDonald2017a} for the combined optical and infrared transmission spectrum. Their results with POSEIDON using the combined transmission spectrum find weak to non-detections (i.e., $\lesssim2.0\sigma$) of H$_2$O, TiO, and HCN. However, their results indicate a $\sim 4.0 \sigma$ preference for models considering contamination in the spectrum from unnocculted stellar heterogeneities. However, the retrieved isothermal atmospheric temperature for the full model using POSEIDON of $T_{\rm iso.}=782^{+283}_{-231}$~K is lower than the equilibrium temperature of this ultra-hot Jupiter \citep[e.g., $T_{\rm eq}=2484$~K][]{Delrez2018}, and estimates from phase-curve observations and GCM simulations \citep[e.g.,][]{Kreidberg2018}.

\subsubsection{Retrieval Setup}

We perform two numerical atmospheric retrievals for our analysis of WASP-103b using ground- and space-base observations. First, we perform a reproduction of the `full' POSEIDON retrieval performed by \citet{Kirk2021}. We use their same free parameters and priors, with the exception of their choice to retrieve a reference radius at 10~bar. Instead, we retrieve the reference pressure ($P_{\rm ref}$) at the planetary radius uncorrected for asphericity ($R_{\rm p}=1.623~R_{\rm J}$) used by \citet{Kirk2021}. In total, our reproduction retrieval has 22 free parameters: 12 for the volume mixing ratios\footnote{The sources of opacity remain as described in Section \ref{subsec:1d_retrieval_setup} with the addition of \citet{Patrascu2015} for AlO, \citet{Bauschlicher2001} for CrH, \citet{Rothman2010} for CO and CO$_2$, \citet{Barber2014} for HCN, and \citet{Dulick2003, Wende2010} and \citet{Hargreaves2010} for FeH.} of Na, K, H$^-$, TiO, VO, AlO, CrH, FeH, H$_2$O, CO, CO$_2$, and HCN; 1 for an isothermal $P$--$T$ profile; 1 parameter for the reference pressure $P_{\rm ref}$ at a reference radius; 4 parameters for the presence of inhomogenous clouds and hazes with the assumption of a single sector capturing their joint contribution \citep[e.g., as in][]{MacDonald2017a}; 3 parameters for contributions from stellar heterogeneities as described above and in \citet{Pinhas2018}; and 1 parameter for an instrumental shift between the optical and infrared observations. 

Then, we perform a second retrieval of WASP-103b relaxing some of the assumptions made by \citet{Weaver2020}. First, instead of using an isothermal $P$--$T$ profile, we use the parameterization of \citet{Madhusudhan2009} with priors as in Table \ref{table:priors}, except for $T_0$ which follows a uniform prior between $800$~K and $2584$~K. Second, instead of using an unphysical prior on the volume mixing ratio of H$^-$ (i.e., log-uniform from -16 to -1), we adopt a physically motivated prior where unrealistically high H$^-$ abundances \citep[i.e., $\log_{10}(X_{\rm H^-})>-7.0$ e.g.,][]{Parmentier2018} are not allowed, following a log-uniform distribution from -16 to -7. Third, we use as our reference radius the planetary radius corrected for asphericity ($R_{\rm p}=1.681~R_{\rm J}$) reported by \citet{Delrez2018}. Fourth, we do not restrict the presence of clouds and hazes to a single sector, and instead use the more generalised parameterization for inhomogenous clouds and hazes of \citet{Welbanks2021} as explained in Section \ref{subsec:1d_retrieval_setup}. All other parameters and priors remain as in the reproduction retrieval described above. For both retrievals, the parameter estimation is performed using PyMultiNest \citep[][]{Buchner2014} with 2000 nested sampling live points.

\subsubsection{Retrieval Results}

\begin{figure}[ht!]
    \centering
    \includegraphics[width=0.45\textwidth]{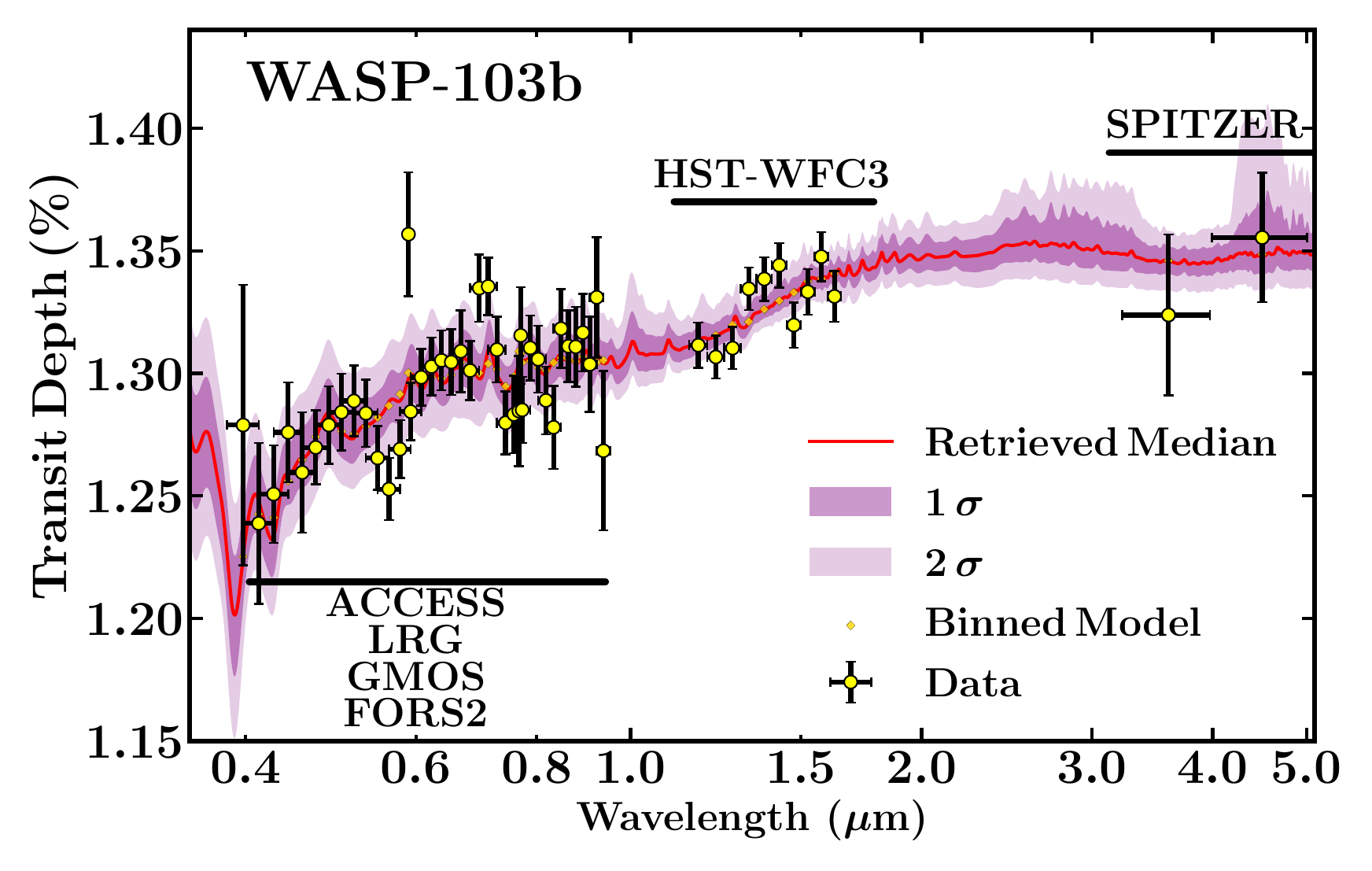}
    \includegraphics[width=0.45\textwidth]{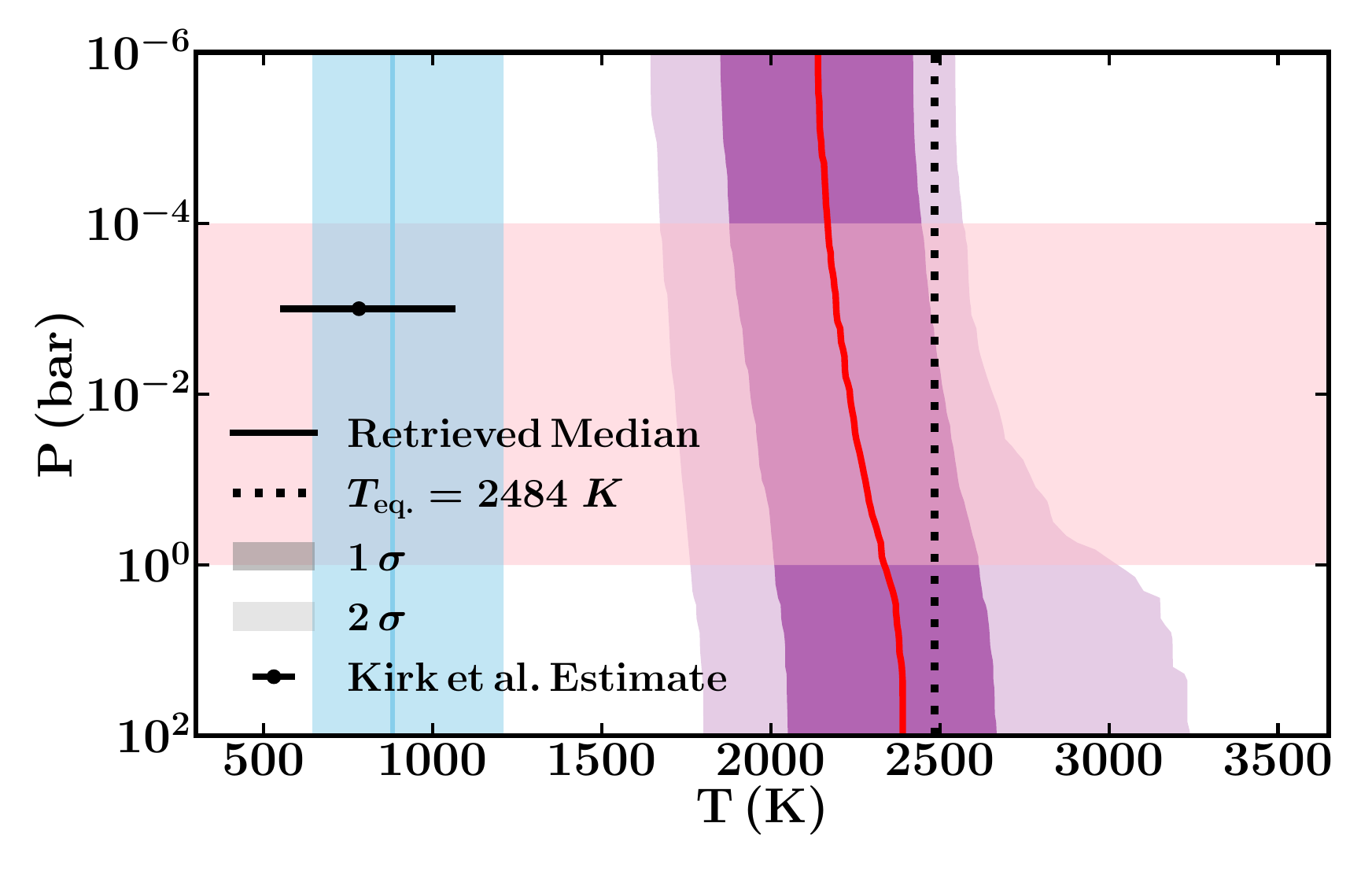}
    \caption{Results for the retrieval of ACCESS, LRG, Gemini-GMOS, VLT-FORS2, HST-WFC3, and Spitzer WASP-103b data from \citet{Kirk2021}. Top: Retrieved transmission spectrum for our numerical retrieval of the planet, relaxing the assumptions from \citet{Kirk2021}. Median transmission spectrum (red line) and 1$\sigma$ and 2$\sigma$ confidence intervals (purple shaded regions) are shown alongside the observations (error bars) and binned median model (diamonds). Bottom: Retrieved $P$--$T$ profiles. Reproduction from \citet{Kirk2021} shown using blue. Retrieval following our modelling procedure is shown in red. Median value is shown with the solid line, and confidence intervals with vertical shaded regions. Pink horizontal shaded region shows the estimated photospheric pressure range from \MGL. Horizontal error bar shows the median and 1$\sigma$ retrieved temperature from \citet{Kirk2021}. Our retrieved temperature is consistent with the equilibrium temperature of the planet ($T_{\rm eq}=2484$~K, vertical dotted line).}
    \label{fig:results_wasp103b}
\end{figure}

Figure \ref{fig:results_wasp103b} shows the retrieved transmission spectrum from our full retrieval in the top panel. The bottom panel of Figure \ref{fig:results_wasp103b} shows the retrieved $P$--$T$ profile from both the reproduction of the `full' retrieval with POSEIDON from \citet{Kirk2021} and our full reanalysis. The black horizontal error bar in Figure \ref{fig:results_wasp103b} shows the isothermal temperature estimate from \citet{Kirk2021}. 

Our reproduction of \citet{Kirk2021} finds poor constraints on the chemical abundances of most species considered. The retrieved abundances of $\log_{10}(X_{\text{H}_2\text{O}})=-3.85 ^{+ 1.21 }_{- 4.21 }$ and $\log_{10}(X_{\text{TiO}})=-8.20 ^{+ 1.56 }_{- 1.61 }$ are consistent with the reported abundances by \citet{Kirk2021}. Their associated detection significances are 1.8$\sigma$ and 1.9$\sigma$ for H$_2$O and TiO, respectively. HCN is not detected (1.0$\sigma$ preference) and its abundance is poorly constrained with a retrieved volume mixing ratio of $\log_{10}(X_{\text{HCN}})=-6.58 ^{+ 3.57 }_{- 5.61 }$, consistent with the retrieved value from \citet{Kirk2021}.

Similarly, the reproduction retrieval exhibits a strong preference for models considering stellar heterogeneity to a 4.6$\sigma$ level. The retrieved faculae have a temperature of $T_{\rm het}=6498.43 ^{+ 202.80 }_{- 176.63 }$~K and a cover fraction $\delta= 0.25 ^{+ 0.11 }_{- 0.09 }$. The retrieved photospheric temperature is $T_{\rm phot.}=6187.70 ^{+ 111.74 }_{- 132.05 }$~K. The retrieved parameters associated with stellar contamination are consistent with those retrieved by \citet{Kirk2021}. Additionally, we retrieve an offset between the optical and infrared observations of $f_{\rm shift}=226^{+81}_{-95}$~ppm, consistent with the estimate from POSEIDON. 

In agreement with \citet{Kirk2021}, the retrieved isothermal temperature from our reproduction is cooler than the equilibrium temperature of the planet. The retrieved isothermal temperature $T_{\rm iso.}=862.70 ^{+ 292.98 }_{- 217.61 }$~K has a median value $\sim1600$~K cooler than the equilibrium temperature of $T_{\rm eq}=2484$~K. Our reproduction shows a general agreement with the results of \citet{Kirk2021} indicating a good agreement in our implementation of their model assumptions. 

Next, we present the results from our full reanalysis with different model assumptions. Figure \ref{fig:results_wasp103b} shows that the retrieved spectrum is in agreement with the observations. Additionally, the bottom panel shows the agreement between the retrieved $P$--$T$ profile and the equilibrium temperature of the planet. Our results find weak to no-detections of TiO and H$_2$O with detection significances at the 1.3$\sigma$ and 1.4$\sigma$ levels, respectively. On the other hand, absorption due to HCN is not preferred by our models. In contrast, the retrieval indicates a strong preference for stellar heterogeneities in the spectrum at a 4.8$\sigma$ level. 

The retrieved chemical abundances are poorly constrained. The retrieved volume mixing ratios of $\log_{10}(X_{\text{H}_2\text{O}})=-7.65 ^{+ 2.18 }_{- 5.14 }$, $\log_{10}(X_{\text{TiO}})=-9.84 ^{+ 0.87 }_{- 1.04 }$, and $\log_{10}(X_{\text{HCN}})=-10.82 ^{+ 3.22 }_{- 3.17 }$ are consistent with the estimates of \citet{Kirk2021}. Furthermore, the retrieved stellar contamination parameters suggest the presence of faculae with temperatures of $T_{\rm het.}=6653.62 ^{+ 313.47 }_{- 251.27 }$~K covering $\delta=0.20 ^{+ 0.10 }_{- 0.08 }$, $\sim400$~K hotter than the retrieved photospheric temperature $T_{\rm phot.}=6205.64 ^{+ 119.50 }_{- 128.52 }$~K, consistent with those from \citet{Kirk2021}.

Our reanalysis finds a temperature near the photosphere of $T_{\rm 100~mbar}=2313^{+279}_{-308}$, consistent with the equilibrium temperature ($T_{\rm eq}=2484$~K) within 1$\sigma$. Additionally, this retrieved temperature is  consistent with the photospheric temperature estimates inferred from phase curve observations of the same planet \citep[e.g.,][]{Kreidberg2018}. This non-isothermal model finds warmer temperatures than the isothermal temperature of $T_{\rm iso.}=782^{+283}_{-231}$~K inferred by \citet{Kirk2021} and shown in Figure \ref{fig:results_wasp103b}. Overall, our results suggest that the previously retrieved cooler temperatures for this ultra-hot Jupiter are likely the result of assumptions in the models employed instead of asymmetric terminators affecting 1D models.

\section{A 2D Retrieval Framework for Asymmetric Terminators}\label{sec:2D_retrievals}

An important next step in the development of atmospheric models is the development of models capable of capturing the properties of planets with asymmetric terminators. To this effect, we demonstrate a multidimensional (1D+1D), referred to here as 2D, retrieval framework for exoplanetary transmission spectra. Our model explores the atmospheric properties of the morning and evening terminators of an exoplanet using a linear combination of atmospheric models. This is, to the best of our knowledge, the first implementation of a `free-retrieval' (i.e., not restricted by chemical equilibrium assumptions) for exoplanet transmission spectra. We note, however, that the use of linear combinations to model exoplanetary inhomogeneities was first introduced by \citet{Line2016a}. Additionally, this approach to model the morning and evening terminators of exoplanets was recently implemented by \citet{Espinoza2021} using the retrieval framework CHIMERA \citep{Line2013a}, although with constraints of chemical equilibrium and forcing thermal inhomogeneities between terminators. Unlike our general approach described below, \citet{Espinoza2021} restrict their models to the cases in which the contribution of each terminator to the final spectrum is 1/2. Our implementation uses the existing Aurora framework \citep{Welbanks2021} due to its modular and multidimensional functionalities. In what follows we describe our 2D implementation and assess its performance on the synthetic observations of asymmetric terminators generated above. 

\begin{figure*}[ht!]
  \centering
  \includegraphics[width=0.94\textwidth]{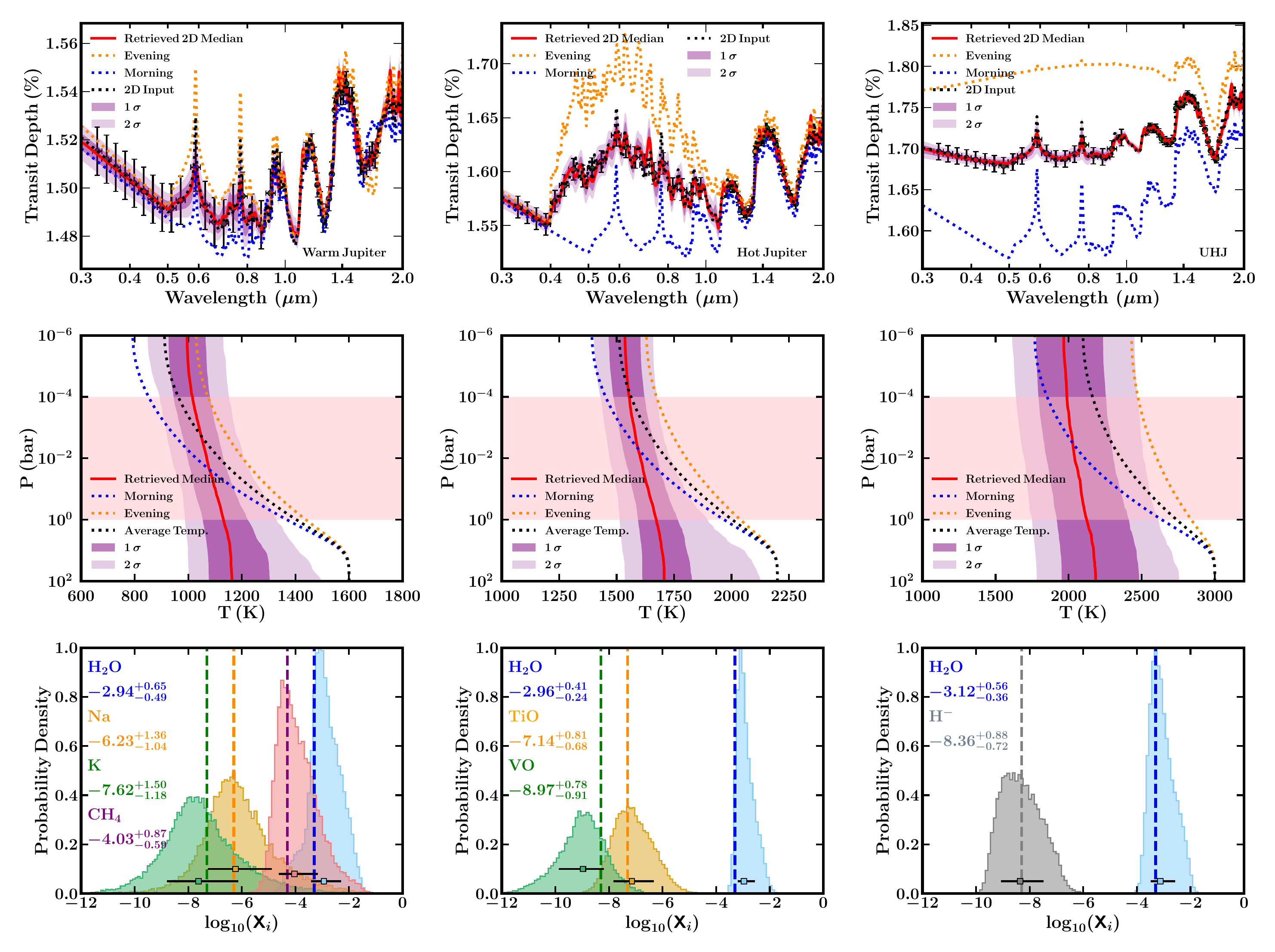}
  \caption{Results from our 2D retrievals on synthetic data from a 2D model. Each column corresponds to an atmospheric case explained in Section \ref{sec:1D_retrievals} adopted after \MGL: a warm Jupiter ($\bar{T}\sim 1000$~K), a hot Jupiter ($\bar{T}\sim 1600$~K), and an ultra-hot Jupiter ($\bar{T}\sim 2200$~K). Top: A morning (blue) and evening (orange) spectra are linearly combined to produce a 2D model (black) to generate HST-STIS and HST-WFC3 synthetic observations (black error bars). The retrieved median 2D spectrum (red) and $1\sigma$ and $2\sigma$ confidence intervals (purple shaded regions) from the 2D retrieval are shown. Middle: The true morning (blue) and evening (orange) $P$--$T$ profiles used to generate the 2D model and synthetic data, along with their average (black), are shown using dotted lines. The retrieved 2D $P$--$T$ profile is shown using a red line (median) with purple shaded regions showing the $1\sigma$ and $2\sigma$ confidence intervals. Pink horizontal shaded region shows the estimated photospheric pressure range from \MGL. Bottom: Posterior distributions corresponding to the average of the morning and evening abundances and error bars indicating the median and $1\sigma$ intervals (numerical value shown in the labels). The averages of the input morning and evening abundances are shown using vertical dashed lines.}
  \label{fig:2d_grid}
\end{figure*}

\subsection{A 2D Model for Asymmetric Terminators}
\label{subsubsec:2d_explanation}

We expand the atmospheric models in Aurora to include 1D+1D models for asymmetric terminators, which we refer to as 2D models. These 2D models are the result of solving numerically for the transit depth of each terminator and calculating their linear combination. The transit depth for each terminator $T_i$ is given by 

\begin{align}
\resizebox{.99\hsize}{!}{$
\Delta_{\lambda, \, T_i} = \frac{R_{\rm p}^{2} +  \displaystyle\int\limits_{R_{\rm p}}^{R_{\rm p}+ H_{A, \, T_i}} 2 b \left(1 - e^{-\tau_{\lambda, \, T_i}(b)} \right) db -  \displaystyle\int\limits_{0}^{R_{\rm p}} 2 b e^{-\tau_{\lambda, \, T_i}(b)} db}{R_{\text{star}}^2}$ },
\label{eq:td}
\end{align}

\noindent where ${R_{\rm p}}$ is the planetary radius, $R_{\text{star}}$ is the stellar radius, $H_{A, \, Ti}$ and $\tau_{\lambda, \, T_i}$ are the maximum height of the planetary atmosphere and the slant optical depth at terminator $T_i$, respectively. We maintain the choice of \citet{Welbanks2021} to show Equation \ref{eq:td} as a three part integral to emphasise that the chosen $R_{\rm p}$ may not correspond to an optically thick part of the atmosphere. The choice of $R_{\rm p}$ and associated $P_{\rm ref}$ can be different for each terminator. However, for this work we perform an initial appraisal for which the reference pressure $P_{\rm ref}$ and reference radius $R_{\rm p}$ pair is a shared anchor point for the pressure and radial grids of both atmospheric models. 

The calculation of each transit depth requires solving line-by-line radiative transfer in a transmission geometry in a plane-parallel atmosphere. Each terminator atmosphere has its own set chemical compositions, $P$--$T$ profile, and cloud/haze properties. Each model atmosphere has a pressure grid with a number of layers uniformly spaced in $\log_{10}(P)$, and a corresponding radial grid determined by hydrostatic equilibrium. As with the 1D models described in Section \ref{sec:1D_retrievals}, the calculation of hydrostatic equilibrium considers altitude-dependent gravity and an atmospheric mean molecular weight determined by the model's chemical composition.

The resulting two dimensional model is given by 

\begin{equation}
\Delta_{\lambda, \, 2D} =\Phi_{T_1} \Delta_{\lambda, \, T_1} + \Phi_{T_2 } \Delta_{\lambda, \, T_2} 
\label{eq:1d+1d}
\end{equation}

\noindent where $\Phi_{T_i}$ is the fraction of the of the total planetary atmosphere that terminator $T_i$ represents. The sum of the terminator fractions follows the unit sum constraint. For the case of a homogeneous atmosphere, the total fraction of one of the terminators becomes unity and we recover the transit depth for a homogeneous 1D planet atmosphere \citep[e.g.,][]{Welbanks2021}.

\subsection{2D Retrieval Setup}
\label{subsubsec:2d_setup}

The atmospheric retrieval setup follows the description in Section \ref{subsec:1d_retrieval_setup} for the parameter estimation scheme, number of nested sampling live points, sources of opacity, and number of pressure layers. Although the prescription in Section \ref{subsubsec:2d_explanation} can incorporate the effects of inhomogeneous clouds and hazes in each terminator, we limit our models to clear atmospheres only to focus on the effects of compositional and thermal inhomogeneities. Future work will explore the effects of including inhomogeneous clouds and hazes in these models. Additionally, we do not retrieve the terminator fraction for each model and instead assume that each $\Phi_{T_i}=0.5$. Due to the symmetry of the problem, the retrieved posterior distributions for the parameters of each terminator are commutable and representative of either. 

We explore the three cases considered in Section \ref{sec:1D_retrievals} of the warm, hot, and ultra-hot Jupiters. The priors for these models are the same as those used for the clear models in Section \ref{subsec:1d_retrieval_setup} shown in Table \ref{table:priors} in Appendix \ref{app:priors}. The models have a total of 21 parameters for the warm Jupiter and ultra-hot Jupiter cases (8 chemical abundances, 12 $P$--$T$ parameters, and 1 for $R_{\rm p}$ at $P_{\rm ref}=10$~bar), and 23 parameters for the hot Jupiter case (10 chemical abundances, 12 $P$--$T$ parameters, and 1 for $R_{\rm p}$ at $P_{\rm ref}=10$~bar).

\subsection{2D Retrieval Results}
\label{subsubsec:2d_results}

Figure \ref{fig:2d_grid} shows the results of our 2D retrieval on the synthetic observations of a planet with asymmetric evening and morning terminators. The top row of Figure \ref{fig:2d_grid} shows the retrieved 2D spectrum, the middle row shows the 2D $P$--$T$ profile, and the bottom row shows the average of the morning and evening abundances. In Appendix \ref{app:2D_retrieval_figures} we show the retrieve spectra, $P$--$T$ profiles, and abundances for each morning and evening terminator. 

For all three cases, the retrieved 2D transmission spectrum is in good agreement with the synthetic observations. Furthermore, the retrieved spectra for each of the terminators (e.g., top row in Figures \ref{fig:wj_2d_grid}, \ref{fig:hj_2d_grid}, and \ref{fig:uhj_2d_grid}) are consistent with the input morning and evening spectra used in the linear combination. While the retrieved 2D transmission spectra provide a fit to the observations, the retrieved 1D spectra for each terminator provides important information about the family of models contributing to the 2D spectra. Additionally, by inspecting the retrieved confidence contours of the morning and evening spectra we can appreciate which wavelength ranges are affected by the inhomogeneities in the asymmetric terminators. For example, the retrieved transmission spectrum for each terminator in the hot Jupiter case seen in Figure \ref{fig:hj_2d_grid} shows that the confidence intervals are wider in the HST-STIS wavelengths ($\lesssim1$~$\mu$m) where the TiO and VO abundance inhomogeneities have a strong impact, compared to the HST-WFC3 wavelengths ($\sim1.1$--$1.8$~$\mu$m) with strong contributions from a homogeneous H$_2$O abundance.

The retrieved 2D $P$--$T$ profiles in the photosphere for transmission spectra ($P\sim10^{-3}$--$10^{-1}$~bar) are all consistent with the average of the morning and evening terminator temperatures. The retrieved $P$--$T$ profiles for each terminator have wide 1$\sigma$ and 2$\sigma$ confidence contours consistent with the input $P$--$T$ profiles (e.g., middle row in Figures \ref{fig:wj_2d_grid}, \ref{fig:hj_2d_grid}, and \ref{fig:uhj_2d_grid}). Overall, the use of this 2D retrieval results in no significant biases in the thermal inferences at the photosphere for these input models with asymmetric terminators.

Similarly, the average retrieved chemical abundances are consistent with the average of the input abundances within $\sim1\sigma$ for all atmospheric cases. The retrieved constraints on the average chemical abundances are comparable to those obtained with existing observations and models with inhomogenous clouds and hazes \citep[$\lesssim1$~dex, e.g.,][]{Welbanks2019b}. Furthermore, the posterior distributions for the retrieved chemical abundances of each terminator (e.g., bottom row in Figures \ref{fig:wj_2d_grid}, \ref{fig:hj_2d_grid}, and \ref{fig:uhj_2d_grid}) are consistent with the input values within $\sim1\sigma$. The posterior distributions for the chemical species with strong inhomogeneities (e.g., present in one terminator and absent in the other) are multi-modal and suggestive of their presence in one terminator and absence in the other. 

Finally, we compare these 2D models with the 1D models explored in Section \ref{sec:1D_retrievals}. We find that, for the synthetic observations employed, there is no significant preference for the 2D models over the 1D models in the warm and hot Jupiter scenarios. However, our 2D model is preferred over all 1D models at a $\gtrsim4\sigma$ level for the ultra-hot Jupiter scenario. This suggests that our 2D atmospheric model is better than its 1D counterparts at explaining the spectroscopic features of asymmetric terminators, especially those with strong chemical and thermal inhomogeneities, as expected.

In summary, the implementation of a 2D retrieval results in no distinguishable biases in the inferred $P$--$T$ profile, atmospheric chemical composition, or transmission spectra for the atmospheric cases investigated. Additionally, this 1D+1D procedure provides us with meaningful inferences that show the families of atmospheric models that can explain the observed spectra, as well as the individual properties of each terminator. This new atmospheric model enables initial inferences about chemical and thermal inhomogeneities in exoplanet atmospheres and is ready to be applied to current and upcoming ground and space-based observations.

\section{Summary and Discussion}\label{sec:summary}

We investigated the origin of previous inferences of thermal anomalies in 1D retrievals of transmission spectra. We revisited results from homogeneous retrieval studies of a large sample of exoplanets and informed their thermal expectations using GCM studies. Then, we derived expectations for an equivalent 1D temperature from a 2D spectrum using a semi-analytic formalism and confirmed them with atmospheric retrievals. Additionally, we perform 1D atmospheric retrievals on synthetic observations of a 2D spectrum with asymmetric terminators and on existing observations of the hot Jupiter WASP-43b and ultra-hot Jupiter WASP-103b. Finally, we introduce a multidimensional 2D atmospheric retrieval for planets with asymmetric terminators. Here, we summarise our conclusions on atmospheric temperature estimates from exoplanetary transmission spectra.

\begin{itemize}
    \item Existing atmospheric temperature estimates from the population study of \citet{Welbanks2019b} are generally consistent with equilibrium and skin temperature expectations and fall within the physically plausible range of the terminator photospheric temperature estimates from GCM studies. 
    \item Our semi-analytic derivation finds that the 1D equivalent temperature from a 2D spectrum with asymmetric terminators is the average of the temperatures of the morning and evening terminators. Our retrieval with semi-analytic models confirms this expectation. Furthermore, the 1D equivalent volume mixing-ratio is not expected to be the average of the morning and evening abundances under the semi-analytic assumptions. 
    \item Our numerical retrievals using synthetic observations find that previous inferences of temperature biases were affected by their choice of $P$--$T$ parameterization and resulted in lower temperatures estimates by a factor of 2 or more. On the other hand, models using well validated $P$--$T$ parameterizations result in no significant biases for warm and hot Jupiters, and comparatively smaller biases for the chosen ultra-hot Jupiter scenario. 
    \item Using synthetic data we find that models with inhomogeneous clouds and hazes, inherently multidimensional, find chemical abundances consistent with expectations for most cases. Additionally, the retrieved posterior distributions for chemical species with strong abundance inhomogeneities between the morning and evening terminators can be multi-modal suggesting that the median retrieved abundance is not a reliable indicator of the overall abundance of the atmosphere.
    \item Our reanalysis of published transmission spectra for the hot Jupiter WASP-43b and ultra-hot Jupiter WASP-103b finds that previous inferences of low atmospheric temperatures are likely the result of model assumptions such as isothermal $P$--$T$ profiles, and not due to limitations in 1D models due to asymmetric terminators. When these model assumptions are relaxed, our atmospheric models retrieve warmer atmospheric temperatures generally consistent with equilibrium temperature estimates, GCM studies, and phase curve observations.
    \item We introduce the first 2D atmospheric free retrieval for transiting exoplanets. This multidimensional scheme allows for the retrieval of the spectrum, thermal properties, and chemical abundances of asymmetric terminators. Furthermore, this framework helps identify the wavelength range most affected by the inhomogeneities among terminators. This multi-dimensional framework is ready to be confronted with the high-precision and broad wavelength observations expected from JWST.
\end{itemize}

Additionally, we find that a series of factors likely affected previous studies inferring anomalously cool temperatures in atmospheric retrievals of transmission spectra, starting with \MGL. These include:
\begin{itemize}
    \item Failing to consider the uncertainties in the retrieved temperature estimates from the literature.
    \item Using the retrieved temperature at the top of the atmosphere ($10^{-6}$~bar) as a representative temperature for the photosphere in transmission spectra. 
    \item Assuming that the temperatures probed with transmission spectroscopy cannot be cooler than the skin temperature for a gray atmosphere.
    \item Choosing inappropriate $P$--$T$ profile parameterizations.
    \item Generalising findings derived from non-cloudy models and extreme atmospheric scenarios that may not be representative of the general population of exoplanets.
    \item Assuming that the atmospheric properties at the terminator can be inferred from a single transit depth measurement at an arbitrary wavelength.
\end{itemize}

We discuss additional considerations for future works as we move towards the characterisation of transiting exoplanets with multidimensional models.

\subsection{Understanding the Statistical Nature of Retrievals}\label{subsec:retrievals_as_stats}

Our study highlights the importance of interpreting the inferred atmospheric properties as statistical estimates with associated uncertainties. After all, the product of an atmospheric retrieval are the posterior distributions of the model parameters given the assumptions of the model. We emphasise the importance of interpreting the retrieved median results in the context of the posterior distribution they come from. As explained above, if the posterior distribution is multi-modal, the median value may not be representative of the overall parameter recovered. Additionally, choosing the most probable mode and ignoring others can skew the interpretation of the results. Similarly, it is important to consider the $1\sigma$ and $2\sigma$ confidence intervals in the retrieved $P$--$T$ profiles, especially for transmission spectroscopy where temperature constraints can have precisions of a few $100$~K at the photosphere. 

For example, \MGL\, compute transmission spectra for the hot Jupiter WASP-17b and the ultra-hot Jupiter WASP-12b using self-consistent atmospheric profiles and retrieve them using 1D models. Their retrieval results are consistent with the input models for WASP-17b. However, the retrieved solutions for WASP-12b, multi-modal in nature, are interpreted as inconsistent and biased. Their interpretation focuses on the most probable mode and does not explore the properties of the other modes. Indeed, their sub-dominant mode is consistent with the input parameters and is indicative of the true input answer. Similarly, their suggested biases in the retrieved photospheric temperatures for the same models may be less severe when considering that the retrieved $1\sigma$ and $2\sigma$ confidence intervals shown in their Figure 4 are largely consistent with the input profiles, especially near the slant photosphere.

Next, there is the consideration of observations using white noise (i.e., independently distributed Gaussian errors). For instance, \citet[][]{Feng2018} highlight the possibility of biases in retrieved properties due to different noise instances. This argument is the motivation for \MGL\, to ignore a Gaussian scatter in their synthetic transit depths. However, the likelihood function in the retrieval framework employed in their study assumes that the observations have independently distributed Gaussian errors \citep{MacDonald2017a}. As a result, utilising observations that do not have independently distributed Gaussian errors invalidates the assumption in their likelihood function. Future work may look into the effects of invalidating the likelihood function and how it affects the retrieved inferences and their uncertainties.

Finally, it is important to highlight that any atmospheric inference is only as good as the assumptions of the models employed. For example, clear atmospheric models can result in more precise abundance estimates than models with inhomogeneous clouds and hazes \citep[e.g.,][]{Welbanks2019a}. Failing to consider the limitations of these seemingly more precise models, and extrapolating the lessons learned from them to models with \textit{different} model assumptions would be erroneous. As such, it is important to contextualise the inferences made by studies using simplified models by their limiting assumptions and avoid generalising their conclusions to a broader family of models.

As we move towards the characterisation of exoplanet atmospheres with multidimensional models, an investigation of the model and data limitations will be paramount. Future studies can investigate the uncertainties in the retrieved atmospheric properties as a function of model complexity. Moreover, future studies can also investigate the degeneracies introduced by building these multidimensional frameworks. Overall, it will be important to keep in mind the statistical nature of retrievals and their output as we inform inferences made with current and upcoming facilities.

\subsection{Obtaining Temperature Constraints from Transmission Spectroscopy}

Finally, we note that the constraints obtained for the $P$--$T$ profiles in this work, and in the literature (see e.g., Table \ref{tab:litvalues}), are of the order of a few $100$~K in precision. This is largely due to the fact that transmission spectroscopy is limited in constraining the vertical temperature structure of the planet, as the main temperature dependence of the spectrum is encoded through the scale height and not directly through thermal emission. Nevertheless, these temperature constraints are still important to obtain information about the nominal conditions at the terminator and are the only means for planets which may not be conducive for emission spectroscopy. But where possible, better constraints on the atmospheric temperature structure can be obtained through thermal emission spectra and phase curves. For instance, emission spectroscopy encodes in its observed spectrum the brightness temperature of the planet at different wavelengths allowing for robust constraints on the $P$--$T$ profile in the atmosphere \citep[see e.g.,][]{Madhusudhan2019}.

\subsection{Concluding Remarks}

Our study highlights the importance of considering the statistical nature of retrievals and contextualising any atmospheric inferences by the model assumptions inevitably employed in deriving them. As we incorporate previously overlooked physical considerations, we must systematically explore the limitations of our models and the impact of their assumptions on the inferred atmospheric properties and their uncertainties. We caution against encompassing all possible biases in 1D retrievals under the scope of inhomogeneous terminators/hemispheres without considering the impact of chemical equilibrium assumptions, simplified semi-analytic models and isothermal atmospheres \citep[e.g.,][]{Welbanks2019a}, inhomogeneities in the presence of clouds and hazes \citep[e.g.,][]{Line2016a, Barstow2020a, Welbanks2021}, and other model considerations such as the choice of $P$--$T$ parameterization \citep[e.g.,][]{Blecic2017}. By jointly considering the limitations of our data and models, we can provide robust and reliable atmospheric inferences. 

The future of transmission spectroscopy is bright. For current observations, our results suggest that state-of-the-art 1D models are overall unaffected by terminators with compositional and thermal inhomogeneties in most cases. Only the most extreme terminator asymmetries result in biases for 1D models assuming homogeneous conditions, as expected. For these exotic cases, we have extended our models and relaxed the assumption of symmetric planet terminators. These 2D models successfully capture the average atmospheric properties of these planets, and provide an insightful window into the individual properties of each terminator. This more advanced framework not only provides a fit to the observed spectra, but also reveals the wavelength ranges most affected by planetary inhomogeneities. Interpreting upcoming ground- and space-based observations, including those with JWST, with our new multidimensional retrieval framework promise to unveil a new dimension for understanding exoplanet atmospheres.

\acknowledgments
{We thank the anonymous referee for their helpful comments. L.W. thanks the Gates Cambridge Trust for support toward his doctoral studies. Support for this work was provided by NASA through the NASA Hubble Fellowship grant \#HST-HF2-51496.001-A awarded by the Space Telescope Science Institute, which is operated by the Association of Universities for Research in Astronomy, Inc., for NASA, under contract NAS5-26555. This work was performed using resources provided by the Cambridge Service for Data Driven Discovery (CSD3) operated by the University of Cambridge Research Computing Service (\url{www.csd3.cam.ac.uk}), provided by Dell EMC and Intel using Tier-2 funding from the Engineering and Physical Sciences Research Council (capital grant EP/P020259/1), and DiRAC funding from the Science and Technology Facilities Council (\url{www.dirac.ac.uk}). L.W. thanks Siddharth Gandhi for providing the H$^-$ opacities used in this work. L.W. thanks Savvas Constantinou, Michael Zhang, Vivien Parmentier, and Michael Line for fruitful conversations regarding semi-analytic models of transmission spectra.}

\appendix
\restartappendixnumbering
\twocolumngrid 

\section{Temperature and Mixing Ratio Equivalence Between a 1D and a 2D Model}\label{app:derivation}

Here we derive thermal and chemical expectations from equating a 2D transmission spectrum, resulting from the linear combination of two 1D transmission spectra, to a 1D spectrum. For this, we use semi-analytic models of transmission spectra which assume isobaric and isothermal atmospheres. While these models have simplified and often unphysical assumptions, their analytic tractability makes them useful tools to help build initial intuition about some of the atmospheric properties. The 1D transmission spectrum under the assumptions of an isobaric, isothermal atmosphere, a single chemical absorber, hydrostatic equilibrium, and ideal gas law is given by \citep[e.g.,][]{Lecavelier2008a,Lecavelier2008b, DeWit2013, Betremieux2017,Heng2017, Sing2018} 

\begin{equation}\label{eq:appendix_1d}
    \Delta_{\lambda, 1D} = \frac{R_{\mathrm{p\, 1D}}^{2} + 2 R_{\mathrm{p\, 1D}} H_{\rm 1D} (\gamma + \ln{\tau_{\rm 1D}})}{R_{\mathrm{star}}^{2}}
\end{equation}

\noindent where $H_{\rm 1D}=k T_{\rm 1D} (\mu g)^{-1}$ is the 1D scale height, $R_{\mathrm{p\, 1D}}$ is the reference radius at a reference pressure $P_0$, $\gamma$ is the Euler-Mascheroni constant, and $\tau_{1D}$ is the wavelength dependent optical depth at the reference planetary radius and reference pressure 

\begin{equation} \label{eq:appendix_tau}
    \tau_{\rm 1D}= \frac{P_0}{k T_{\rm 1D}}\sqrt{2\pi R_{\mathrm{p\, 1D}} H_{\mathrm{1D}} } X_{\mathrm{1D}} \, \sigma_{\lambda}(T_{\rm 1D}),
\end{equation}

\noindent assuming a single absorber with abundance $X_{\mathrm{1D}}$ and isobaric, temperature dependent cross section $\sigma_{\lambda}(T_{\rm 1D})$.

Next, let the 2D transmission spectra be the result of a linear combination of two 1D spectra each corresponding to a planetary terminator, M for morning and E for evening, 

\begin{equation}
    \Delta_{\lambda, 2D}= \frac{\Delta_{\lambda, M} +\Delta_{\lambda, E} }{2}.
\end{equation}

\noindent The 1D spectrum for each terminator is given by Equation \ref{eq:appendix_1d}, where instead of the 1D parameters (i.e., $H_{\rm 1D}$, $R_{\mathrm{p\, 1D}}$, $\tau_{\rm 1D}$) we use the parameters for each of the terminators (i.e., $H_{\rm E}$, $H_{\rm M}$, $\tau_{\rm E}$, $\tau_{\rm M}$) and a reference radius $R_{\mathrm{p\, 2D}}$

\begin{equation} \label{eq:appendix_2d}
\resizebox{.99\hsize}{!}{$
    \Delta_{\lambda, 2D} = \frac{ R_{\mathrm{p\, 2D}}^2 + R_{\mathrm{p\, 2D}} H_{\rm M} (\gamma + \ln{\tau_{\rm M}}) + R_{\mathrm{p\, 2D}} H_{\rm E} (\gamma + \ln{\tau_{\rm E} ) } } { {R_{\mathrm{star}}^2}}$}.
\end{equation}
Our objective is to find the relationship between the atmospheric properties (temperature and abundance) of the 1D model and those of the evening and morning terminators in the 2D models. 

We begin by requiring that the 1D model correctly reproduces the transmission spectrum of the 2D model.  We assume that the shape of the spectral features in transit transmission spectra encode the information about the exoplanetary atmospheric properties, as shown by \citet{Lecavelier2008a,Benneke2012} and \citet{Line2016a}. We equate the shape of the spectral features resulting from both models by considering the difference between transit depths at two wavelengths $\lambda_1$ and $\lambda_2$ for each model. The difference in transit depth for the 1D model in Equation \ref{eq:appendix_1d} is then

\begin{align}
\begin{split} \label{eq:1d_transit_change}
 \delta  \Delta_{\lambda_{2-1}, 1D} &= \Delta_{\lambda_2, 1D}- \Delta_{\lambda_1, 1D} 
 \\ &= 2 \frac{R_{\mathrm{p\, 1D}}}{R_{\mathrm{star}}^{2}}H_{\rm 1D} \ln\left( \frac{\sigma_{\lambda_2}(T_{\rm 1D})}{\sigma_{\lambda_1}(T_{\rm 1D})} \right).
\end{split}
\end{align}

\noindent This dependence of the relative transit depth on the cross sections is well known \citep[e.g.,][]{Sing2018}, and is equivalent to the definition of spectral shape in equation 5 of \citet{Line2016a}. The corresponding result for the 2D model in Equation \ref{eq:appendix_2d} is

\begin{align}
\begin{split} \label{eq:2d_transit_change}
& \delta  \Delta_{\lambda_{2-1}, 2D} = \Delta_{\lambda_2, 2D}- \Delta_{\lambda_1, 2D} 
 \\& = \frac{R_{\mathrm{p\, 2D}}}{R_{\mathrm{star}}^{2}}\left[ H_{\rm M} \ln\left( \frac{\sigma_{\lambda_2}(T_{\rm M})}{\sigma_{\lambda_1}(T_{\rm M})} \right)
 +
 H_{\rm E} \ln\left( \frac{\sigma_{\lambda_2}(T_{\rm E})}{\sigma_{\lambda_1}(T_{\rm E})}
 \right) \right].
\end{split}
\end{align}

Equating \ref{eq:1d_transit_change} and \ref{eq:2d_transit_change} leads to

\begin{align}
\begin{split}
&\frac{R_{\mathrm{p\, 2D}}}{R_{\mathrm{star}}^{2}}\left[ H_{\rm M} \ln\left( \frac{\sigma_{\lambda_2}(T_{\rm M})}{\sigma_{\lambda_1}(T_{\rm M})} \right) + H_{\rm E} \ln\left( \frac{\sigma_{\lambda_2}(T_{\rm E})}{\sigma_{\lambda_1}(T_{\rm E})}
 \right) \right]
 \\ &= 2\frac{R_{\mathrm{p\, 1D}}}{R_{\mathrm{star}}^{2}}H_{\rm 1D} \ln\left( \frac{\sigma_{\lambda_2}(T_{\rm 1D})}{\sigma_{\lambda_1}(T_{\rm 1D})} \right).
\end{split}
\end{align}

\noindent Following \MGL, we consider the case where the reference pressure is deep enough that it corresponds to a high optical depth and the reference radii are the same (i.e., $R_{\mathrm{p\, 2D}}=R_{\mathrm{p\, 1D}}$). This leads to

\begin{align}
\begin{split}\label{eq:scaleheight_result}
&H_{\rm M} \ln\left( \frac{\sigma_{\lambda_2}(T_{\rm M})}{\sigma_{\lambda_1}(T_{\rm M})} \right)+ H_{\rm E} \ln\left( \frac{\sigma_{\lambda_2}(T_{\rm E})}{\sigma_{\lambda_1}(T_{\rm E})}
 \right)
\\&= 2 H_{\rm 1D} \ln\left( \frac{\sigma_{\lambda_2}(T_{\rm 1D})}{\sigma_{\lambda_1}(T_{\rm 1D})}\right).
\end{split}
\end{align}

If the cross sections are allowed to vary weakly with $T$ (i.e.,  $ \frac{\sigma_{\lambda_2}(T_{\rm 1D})}{\sigma_{\lambda_1}(T_{\rm 1D})} \approx \frac{\sigma_{\lambda_2}(T_{\rm E})}{\sigma_{\lambda_1}(T_{\rm E})}
 \approx \frac{\sigma_{\lambda_2}(T_{\rm M})}{\sigma_{\lambda_1}(T_{\rm M})}$), as in \MGL, we find
 
 \begin{equation} \label{eq:transit_change_equality}
H_{\rm 1D} = \frac{H_{\rm E} + H_{\rm M}} {2}.
\end{equation}
 
Finally, under the assumptions of constant gravity and constant mean molecular weight part of this semi-analytic treatment by construction, Equation \ref{eq:transit_change_equality} readily implies that 

\begin{equation}\label{eq:1d_temperature}
 T_{\rm 1D} = \frac{T_{\rm E} + T_{\rm M}} {2},
\end{equation}
i.e., the isothermal temperature derived using the equivalent 1D model is the average of the two isotherms in the 2D model.

We next assess the chemical abundances in the 1D model. We begin by equating the 1D transit depth to the 2D transit dept, that is $\Delta_{\lambda, 2D}=\Delta_{\lambda, 1D}$, 

\begin{align}
\begin{split}
    &\frac{R_{\mathrm{p\, 1D}}^{2} + 2 R_{\mathrm{p\, 1D}} H_{\rm 1D} (\gamma + \ln{\tau_{\rm 1D}})}{R_{\mathrm{star}}^{2}}
    \\&= \frac{ R_{\mathrm{p\, 2D}}^2 + R_{\mathrm{p\, 2D}} H_{\rm M} (\gamma + \ln{\tau_{\rm M}}) + R_{\mathrm{p\, 2D}} H_{\rm E} (\gamma + \ln{\tau_{\rm E} ) } } { {R_{\mathrm{star}}^2}}.
\end{split}
\end{align}

\noindent Assuming $R_{\mathrm{p\, 2D}}=R_{\mathrm{p\, 1D}}$, as above, and rearranging the terms leads to 

\begin{equation} \label{eq:transit_depth_equality}
 \gamma (2 H_{\rm 1D}- H_{\rm E} - H_{\rm M}) = H_{\rm M}\ln{\tau_{\rm M}} + H_{\rm E}  \ln{\tau_{\rm E}} - 2 H_{\rm 1D}\ln{\tau_{\rm 1D}}.
\end{equation}

\noindent Equation \ref{eq:transit_change_equality} implies that the left-hand side of Equation \ref{eq:transit_depth_equality} is zero. As a result, 

\begin{equation}
    \ln{\tau_{\rm 1D}} = \frac{H_{\rm M}} {H_{\rm E} + H_{\rm M}} \ln{\tau_{\rm M}} + \frac{H_{\rm E}}{H_{\rm E} + H_{\rm M}}  \ln{\tau_{\rm E}}.
\end{equation}

\noindent Again, under the assumptions of this semi-analytic treatment the gravity and mean molecular weight are constants which further simplify the expression

\begin{equation} \label{eq:tau_equality}
    \ln{\tau_{\rm 1D}} = \frac{T_{\rm M}}{T_{\rm E} + T_{\rm M}} \ln{\tau_{\rm M}} + \frac{T_{\rm E}}{T_{\rm E} + T_{\rm M}}  \ln{\tau_{\rm E}},
\end{equation}

Next, we rearrange Equation \ref{eq:tau_equality} and use the definition of the optical depth in Equation \ref{eq:appendix_tau} to obtain

\begin{align}
\begin{split}\label{eq:general}
    &\frac{P_{0, \mathrm{1D}}}{k T_{\rm 1D}}\sqrt{2\pi  R_{\mathrm{p\, 1D}} H_{\mathrm{1D}} } X_{\mathrm{1D}} \, \sigma_{\rm 1D} 
    \\ = & \left( \frac{P_{0, \mathrm{M}}}{k T_{\rm M}}\sqrt{2\pi R_{\mathrm{p\, 2D}} H_{\mathrm{M}} } X_{\mathrm{M}} \, \sigma_{\rm M} \right) ^ \frac{T_{\rm M}}{T_{\rm E} + T_{\rm M}}
     \\& \left( \frac{P_{0, \mathrm{E}}}{k T_{\rm E}}\sqrt{2\pi  R_{\mathrm{p\, 2D}} H_{\mathrm{E}} } X_{\mathrm{E}} \, \sigma_{\rm E} \right) ^ \frac{T_{\rm E}}{T_{\rm E} + T_{\rm M}}.
\end{split}
\end{align}

\noindent where we have expressed the wavelength-dependent cross sections evaluated at the corresponding temperature $\sigma_{\lambda}(T_{i})=\sigma_{i}$ for simplicity. Substituting our result from Equation \ref{eq:1d_temperature} and assuming $R_{\mathrm{p\, 1D}}=R_{\mathrm{p\, 2D}}$ gives 

\begin{align}
\begin{split}\label{eq:abundance_result1}
   P_{0, \mathrm{1D}} \, X_{\mathrm{1D}} \, \sigma_{\mathrm{1D}}  =  & \sqrt{\frac{T_{\rm E} + T_{\rm M}}{2 }} \left(P_{0, \mathrm{M}}\, X_{\mathrm{M}} \, \sigma_{\rm M} \right) ^ \frac{T_{\rm M}}{T_{\rm E} + T_{\rm M}}
    \\& \left(P_{0, \mathrm{E}}\, X_{\mathrm{E}} \, \sigma_{\rm E}\right) ^ \frac{T_{\rm E}}{T_{\rm E} + T_{\rm M}}
  \\ & T_{\rm M}  ^{-\frac{T_{\rm M}}{2(T_{\rm E} + T_{\rm M}) } }
    T_{\rm E}  ^{-\frac{T_{\rm E}}{2(T_{\rm E} + T_{\rm M}) } },
\end{split}
\end{align}
\noindent which is the expression that represents the 1D atmospheric properties (i.e., reference pressure, single absorber abundance, and cross section) as a function of the properties of the morning and evening terminators. 

We can further recast Equation \ref{eq:abundance_result1}, following \MGL, in terms of the average temperature of the terminator $\bar{T}=\frac{T_{\rm E}+T_{\rm M}}{2}$ and the deviation $\Delta T=\frac{T_{\rm E}-T_{\rm M}}{2}$, i.e., $T_{\rm E}=\bar{T}+\Delta T$ and $T_{\rm M}=\bar{T}-\Delta T$. We can also define 
\begin{align}
\alpha= \frac{T_{\rm E}}{\bar{T}}= 1+\frac{\Delta T}{\bar{T}}, && \beta = \frac{T_{\rm M}}{\bar{T}}= 1-\frac{\Delta T}{\bar{T}}.
\end{align}

\noindent Using these quantities we find  
\begin{align}
\begin{split}\label{eq:abundance_result2}
   P_{0, \mathrm{1D}} \,  X_{\mathrm{1D}} \, \sigma_{\bar{T}} =& \bar{T}^\frac{1}{2} \left(P_{0, \mathrm{M}}\, X_{\mathrm{M}} \, \sigma_{\rm M} \right) ^ \frac{\beta}{2}
     \left(P_{0, \mathrm{E}}\, X_{\mathrm{E}} \, \sigma_{\rm E}\right) ^ \frac{\alpha}{2}
    \\ &\left(\bar{T}-\Delta T \right) ^{-\frac{\beta}{4}}
    \left(\bar{T}+\Delta T \right)  ^{-\frac{\alpha}{4}}.
\end{split}
\end{align}

\subsection{Interpreting the Semi-Analytic Result}\label{app:semi_analytic_values}

First, Equation \ref{eq:abundance_result2} shows that under the assumptions of this semi-analytic treatment, the 1D chemical abundance is not the algebraic average of the abundances of the morning and evening terminators. Second, Equation \ref{eq:abundance_result2} shows the well known degeneracy between the reference pressure and abundance that results from this semi-analytic treatment \citep[see e.g.,][]{Lecavelier2008a, Heng2017, Welbanks2019a}. If the wavelength dependence of the cross section is known, this result suggests that one needs to assume an abundance to derive the reference pressure, or assume a reference pressure to derive an abundance \citep[e.g.,][]{Lecavelier2008a}. Therefore, without any assumptions about the reference pressure of the 1D model atmosphere, the transit depth at a single wavelength cannot provide enough information to obtain a single 1D chemical abundance under this semi-analytic treatment. 

As an illustration, let us consider the atmospheric scenario that serves as input conditions for Section \ref{subsec:analytic_retrieval}. Let there be a hot Jupiter with a hot evening terminator of temperature $T_{\mathrm{E}}=2000$~K, H$_2$O abundance of $\log_{10}(X_{\mathrm{H}_2\mathrm{O}, \, \mathrm{E}})=-6.0$, and a reference pressure of $P_{0, \mathrm{E}}=10$~bar. The cooler morning terminator has a $T_{\mathrm{M}}=1000$~K and H$_2$O abundance of $\log_{10}(X_{\mathrm{H}_2\mathrm{O}, \, \mathrm{M}})=-3.0$. In order to derive the reference pressure for the morning terminator, we assume that the reference point in both terminators corresponds to the same reference radius and optical depth surface. We can use this assumption and Equation \ref{eq:appendix_tau} to derive an expression for the reference pressure of the morning terminator 

\begin{equation} 
    P_{0, \mathrm{M}}= P_{0, \mathrm{E}}  \frac{X_{\mathrm{H}_2\mathrm{O}, \, \mathrm{E}}}{X_{\mathrm{H}_2\mathrm{O}, \, \mathrm{M}}} \frac{\sigma_{\rm E}}{\sigma_{\rm M}}\sqrt{\frac{T_{\mathrm{M}}}{T_{\mathrm{E}}}}.
\end{equation}

\noindent If we assume, for simplicity, cross sections with small temperature variations (i.e., $\sigma_{1500}\approx\sigma_{1000}\approx\sigma_{2000}$) we find $P_{0, \mathrm{M}}\approx 7$~mbar. Using the above input values, we obtain $\Delta T /\bar{T}=1/3$, $\alpha=4/3$, and $\beta=2/3$.

Substituting the above values into Equation \ref{eq:abundance_result2} shows that for any chosen value of $ X_{\mathrm{1D}}$, one can choose a different value of $P_{0, \mathrm{1D}}$ to satisfy the equality. Indeed, in Section \ref{subsec:analytic_retrieval} we perform a retrieval on a synthetic 2D spectrum with the above inputs using a 1D semi-analytic model and find the expected degeneracy between the abundance and the reference pressure. 

If on the other hand we assume a value for the reference pressure of the 1D model, we can derive a 1D abundance. We can choose an arbitrary pressure of $P_{0, \mathrm{1D}}=0.1$~bar for illustration. Then $X_{\mathrm{1D}}=10^{-4}$. In Section \ref{subsec:analytic_retrieval} we perform a second semi-analytic retrieval on the same synthetic data, but this time we assume a reference pressure of $0.1$~bar in the 1D model. This retrieval finds a 1D abundance consistent with the expected value of $\log_{10}(X_{\text{H}_2\text{O}})=-4$.

\subsection{Analytic Comparison with MGL20}\label{app:comparison_w_MGL}

 The derivation in \MGL\, considers the equivalent transit depth between a 1D and a 2D model at a single wavelength. This approach assumes that the atmospheric properties can be deduced from a transit depth measurement at a single wavelength, which is not feasible. Inference of atmospheric properties in transmission geometry requires measurements of the transit depth at two or more wavelengths, i.e., a transmission spectrum \citep{Lecavelier2008a,Benneke2012,Line2016a}. Therefore, our derivation shown in Section \ref{app:derivation} uses the spectral shape of the transmission spectrum, i.e., the transit depth at two or more wavelengths, to find the equivalent atmospheric properties between a 1D and 2D model. 

 Implementing our procedure on the derivation of \MGL\, results in the same outcome as the derivation shown in this work. That can be demonstrated by starting with equation A22 of \MGL, which shows the transit depth equivalence condition, and differentiating it with respect to $\lambda$ to obtain the equivalence of the spectral shape between the two models. Following the assumption in \MGL\, where the reference pressure is deep enough that it corresponds to a high optical depth and that the reference radii are the same (i.e., $R_{\mathrm{p\, 1D}}=R_{\mathrm{p\, 2D}}$), leads to 

 \begin{equation}\label{eq:lam_derivative}
     H_{\rm 1D} \left( \frac{d\ln(\sigma_\lambda)}{d\lambda} \right) =  \bar{H} \left( \frac{d\ln(\sigma_\lambda)}{d\lambda} \right),
 \end{equation}

 \noindent where $\bar{H}=\frac{H_{\rm E} + H_{\rm M}} {2}$ and $\sigma_\lambda$ is the wavelength dependent cross section assumed to vary weakly with temperature. Equation \ref{eq:lam_derivative} is equivalent to Equation \ref{eq:scaleheight_result} and readily implies that $ H_{\rm 1D}=\frac{H_{\rm E} + H_{\rm M}} {2}$ and $T_{\rm 1D} = \frac{T_{\rm E} + T_{\rm M}} {2}$.

\section{Priors for Retrievals in this Work }\label{app:priors}

The priors for the semi-analytic models in Section \ref{subsec:analytic_retrieval} are uniform for $T_{\rm 1D}$ between 300~K and 3000~K, Log-uniform for $X_{\text{H}_2\text{O}}$ between $10^{-12}$ and $10^{-0.3}$, and Log-uniform for $P_{\rm ref}$ between $10^{-2}$~bar and $10^{6}$~bar.

The priors adopted for the 1D models in Section \ref{sec:1D_retrievals} are shown in Table \ref{table:priors}. The priors are mostly standardised and follow the description of \MGL\, for the MGL20 models, and \citet{Welbanks2019b} and \citet{Welbanks2021} for the clear and cloudy models. The prior for the temperature parameter at the top of the atmosphere ($T_0$) was chosen so that it could span the entire range of possible temperatures in an analogous form to the chosen prior for $T_{\rm deep}$ by \MGL. The priors for 2D models in Section \ref{sec:2D_retrievals} are the same as the clear atmosphere model column in Table \ref{table:priors}.

\begin{deluxetable*}{c|c|c|c|c}
\tablecaption{Parameters and Priors for Retrievals in this Work
\label{table:priors}}
\tablecolumns{5}
\tablewidth{\columnwidth}
\tablehead{
\colhead{Parameter} & \colhead{Prior Distribution} & \colhead{MGL20} & \colhead{Clear Atmosphere} & \colhead{Cloudy Atmosphere}
}
\startdata
$X_i$ & Log-uniform & $10^{-12}$--$10^{-0.3}$ &$10^{-12}$--$10^{-0.3}$ & $10^{-12}$--$10^{-0.3}$\\ \hline
$T_{\rm deep}$ & Uniform & $400$--$3000$ K & N/A & N/A    \\ \hline
$T_0$ & Uniform &  N/A  & $400$--$4000$ K &  $400$--$4000$ K \\ \hline
$\alpha_{1,2}$  &  Uniform & $0.02$--$2.00$ K$^{-1/2}$& $0.02$--$2.00$ K$^{-1/2}$& $0.02$--$2.00$ K$^{-1/2}$\\ \hline
$P_{1,2,\text{cloud},\text{ref}}$ & Log-uniform & $10^{-6}$--$10^{2}$ bar& $10^{-6}$--$10^{2}$ bar& $10^{-6}$--$10^{2}$ bar\\ \hline
$P_{3 }$ & Log-uniform & $10^{-2}$--$10^{2}$ bar & $10^{-2}$--$10^{2}$ bar & $10^{-2}$--$10^{2}$ bar \\ \hline
$R_{\rm p, \, 10 \, bar}$ & Uniform & $0.85$--$1.15$~$R_{\rm p}$ & N/A & N/A  \\ \hline
$P_{\rm ref. }$ & Log-uniform & N/A  & $10^{-6}$--$10^{2}$ bar & $10^{-6}$--$10^{2}$ bar \\ \hline
$a$ & Log-uniform & N/A & N/A  & $10^{-4}$--$10^{10}$\\ \hline
$\gamma$ & Uniform & N/A & N/A & $-20$--$2$\\ \hline
 \makecell{$\phi_{\mathrm{clouds}}$\\$\phi_{\mathrm{hazes}}$\\$\phi_{\mathrm{clouds+hazes}}$} & Uniform & N/A & N/A & $0$--$1$
\enddata 
\tablecomments{N/A means that the parameter was not considered in the model by construction.}
\end{deluxetable*}

\section{1D Retrievals with Synthetic Observations}\label{app:1D_retrieval_figures}

\begin{figure*}
\includegraphics[width=1.0\textwidth]{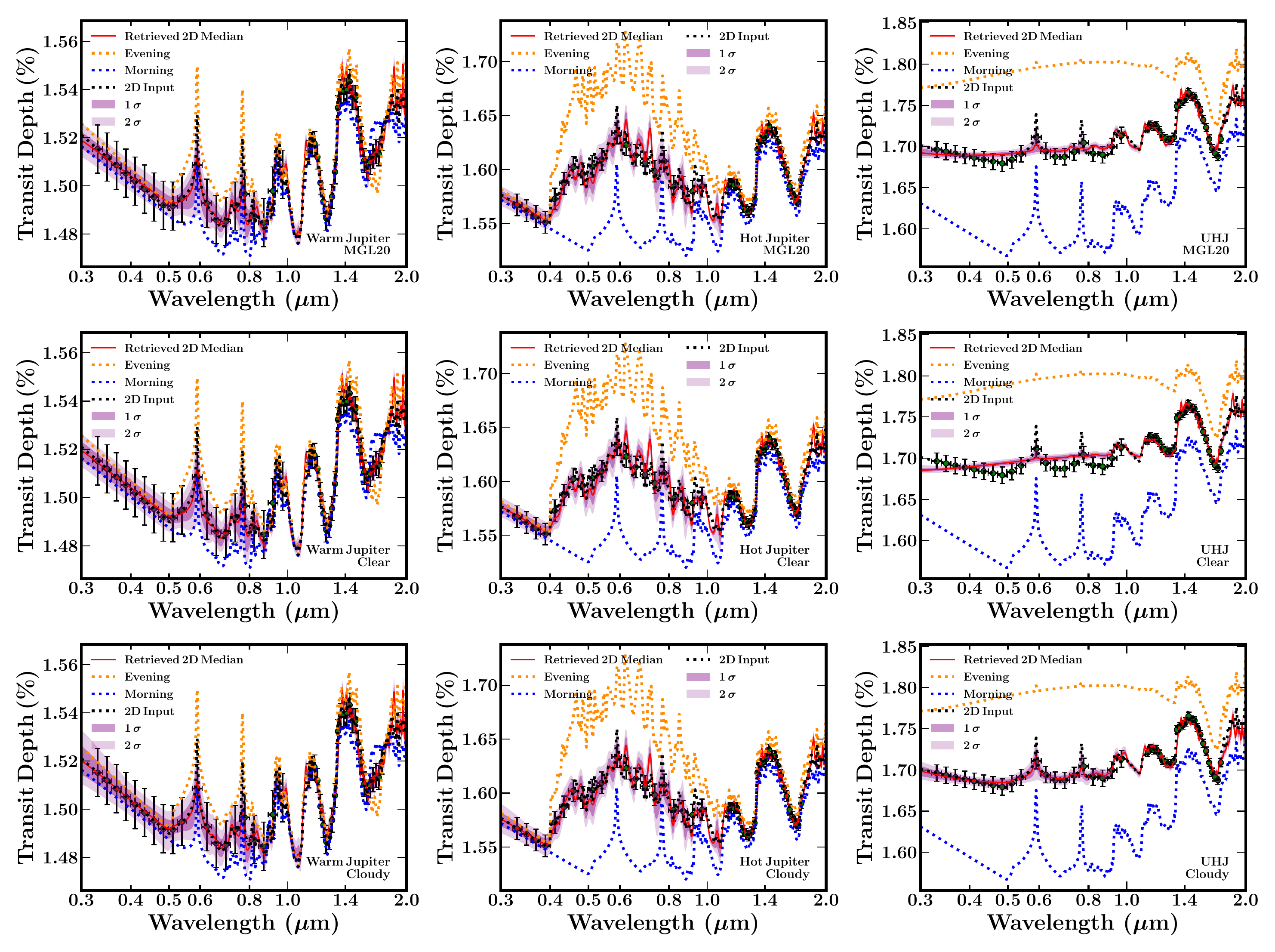}
\caption[]{Retrieved transmission spectra for our numerical retrievals using 1D models on synthetic data from a 2D model in Section \ref{sec:1D_retrievals}. Each column corresponds to an atmospheric case explained in Section \ref{sec:1D_retrievals} adopted after \MGL: a warm Jupiter ($\bar{T}\sim 1000$~K), a hot Jupiter ($\bar{T}\sim 1600$\,K), and an ultra-hot Jupiter ($\bar{T}\sim 2200$\,K). Each row corresponds to a different model strategy: a clear atmospheric model with the $P$--$T$ parameterization from \MGL\, (top), a cloud-free model with the $P$--$T$ parameterization from \citet{Madhusudhan2009} (middle), and a model with inhomogeneous clouds and hazes from \citet{Welbanks2021} and the $P$--$T$ parameterization from \citet{Madhusudhan2009} (bottom). Each panel shows the morning (blue) and evening (orange) spectra that are linearly combined to produce a 2D model (black) and HST-STIS/HST-WFC3 synthetic observations (error bars). The retrieved median spectrum (red) and $1\sigma$ and $2\sigma$ confidence intervals (purple shaded regions) from the 1D retrieval are shown. } \label{fig:spectrum_grid}
\end{figure*}

\begin{figure*}
\includegraphics[width=1.0\textwidth]{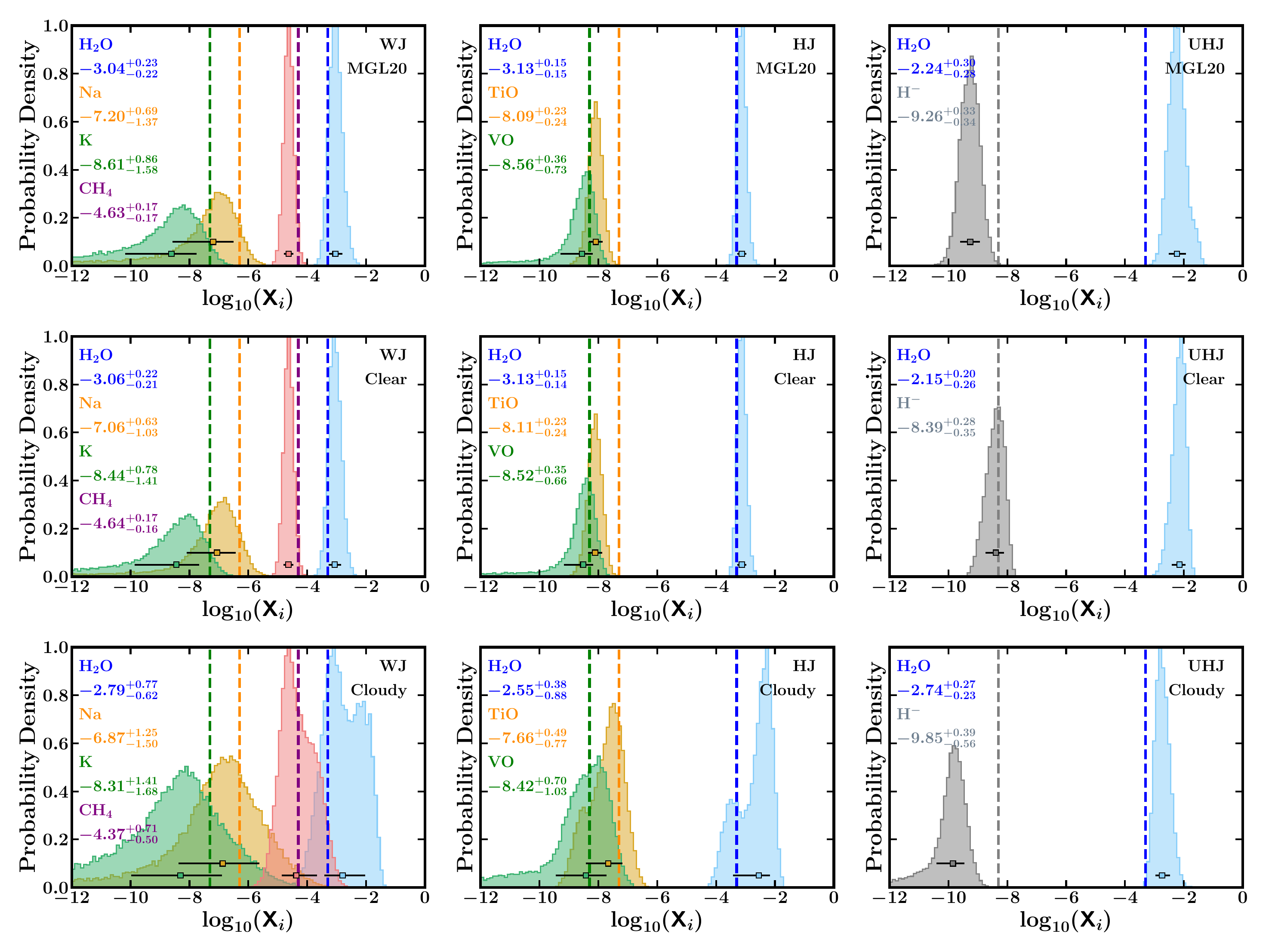}
\caption[]{Retrieved posterior distributions for the chemical abundances in our numerical retrievals using 1D models on synthetic data from a 2D model in Section \ref{sec:1D_retrievals}. Each column corresponds to an atmospheric case explained in Section \ref{sec:1D_retrievals} adopted after \MGL: a warm Jupiter ($\bar{T}\sim 1000$~K), a hot Jupiter ($\bar{T}\sim 1600$\,K), and an ultra-hot Jupiter ($\bar{T}\sim 2200$\,K). Each row corresponds to a different model strategy: a clear atmospheric model with the $P$--$T$ parameterization from \MGL \, (top), a cloud-free model with the $P$--$T$ parameterization from \citet{Madhusudhan2009} (middle), and a model with inhomogeneous clouds and hazes from \citet{Welbanks2021} and the $P$--$T$ parameterization from \citet{Madhusudhan2009} (bottom). Each panel shows the retrieved mixing ratio posteriors and error bars indicating the median and $1\sigma$ intervals (numerical value shown in the labels). The averages of the morning and evening abundances are shown using vertical dashed lines.} \label{fig:posterior_grid}
\end{figure*}

Here we present the retrieved transmission spectra and posterior distributions for the 1D retrievals in Section \ref{sec:1D_retrievals}. Figure \ref{fig:spectrum_grid} shows the input transmission spectra (morning, evening, and their linear combination), the synthetic observations, and the retrieved median model and confidence intervals. Figure \ref{fig:posterior_grid} shows the posterior distributions for the retrieved volume mixing ratios of the chemical species of interest for the warm Jupiter, hot Jupiter, and ultra-hot Jupiter atmospheric scenarios.

\section{2D Retrievals with Synthetic Observations}\label{app:2D_retrieval_figures}

\begin{figure*}
\includegraphics[width=1.0\textwidth]{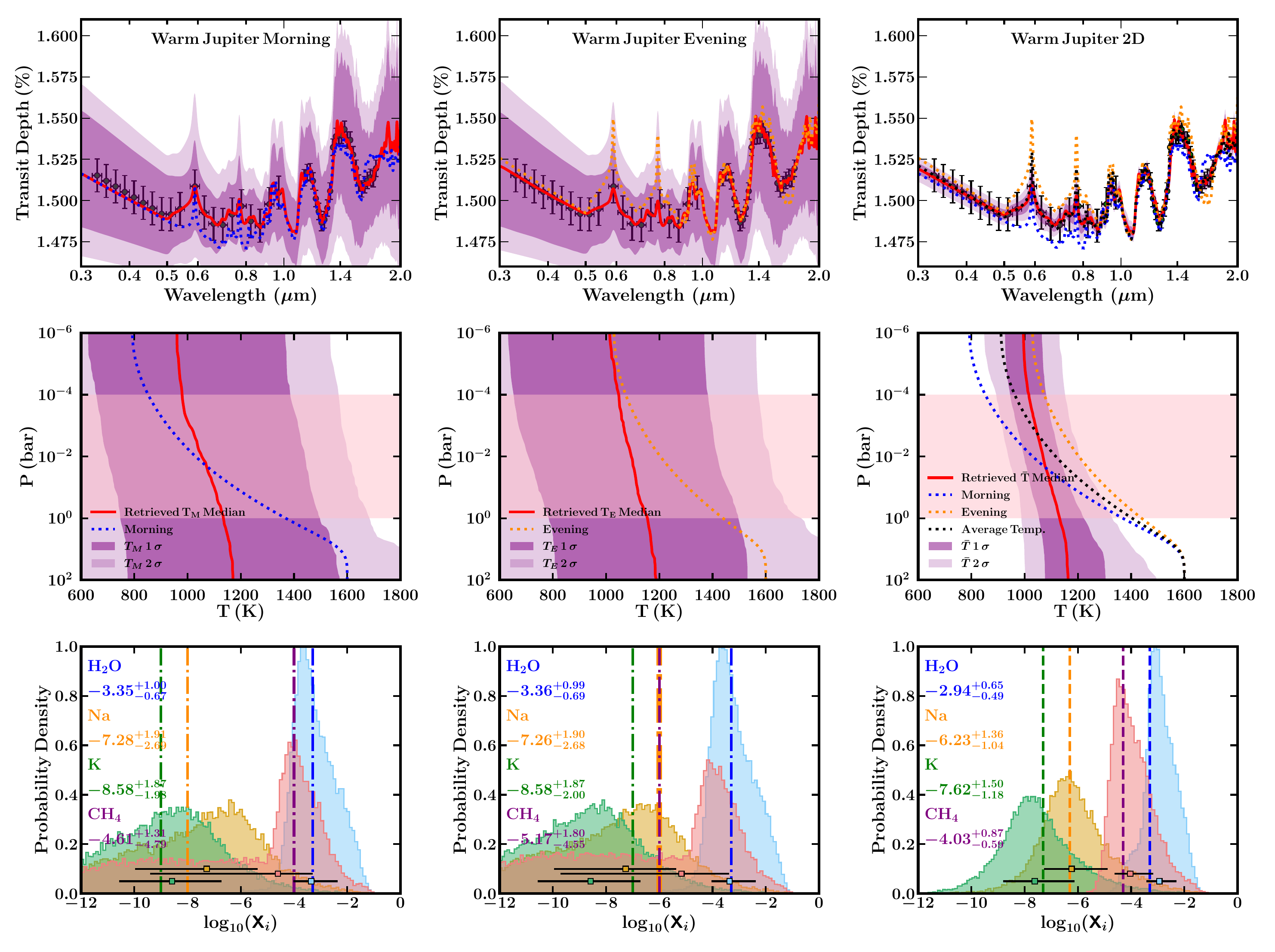}
\caption[]{Results from our 2D retrievals on synthetic data from a 2D model for the warm Jupiter case. The top row shows the morning (blue) and evening (orange) spectra that are linearly combined to produce a 2D model (black) and HST-STIS/HST-WFC3 synthetic observations (error bars). The retrieved median spectrum (red) and $1\sigma$ and $2\sigma$ confidence intervals (purple shaded regions) from the retrieval are shown. Middle: The true morning (blue) and evening (orange) $P$--$T$ profiles used to generate the 2D model and synthetic data, along with their average (black), are shown using dotted lines. The retrieved $P$--$T$ profile is shown using a red line (median) with purple shaded regions showing the $1\sigma$ and $2\sigma$ confidence intervals. Pink horizontal shaded region shows the estimated photospheric pressure range from \MGL. Bottom: Posterior distributions corresponding to the chemical abundances and error bars indicating the median and $1\sigma$ intervals (numerical value shown in the labels). The coloured dot-dashed lines indicate the input values for each terminator while the dashed lines show the average abundance of the morning and evening terminators. The left column shows the results for the morning terminator, the middle column shows the results for the evening terminator, and the right column shows the 2D results from the combination of the morning and evening terminator and the average abundances as in Figure \ref{fig:2d_grid}.} \label{fig:wj_2d_grid}
\end{figure*}

\begin{figure*}
\includegraphics[width=1.0\textwidth]{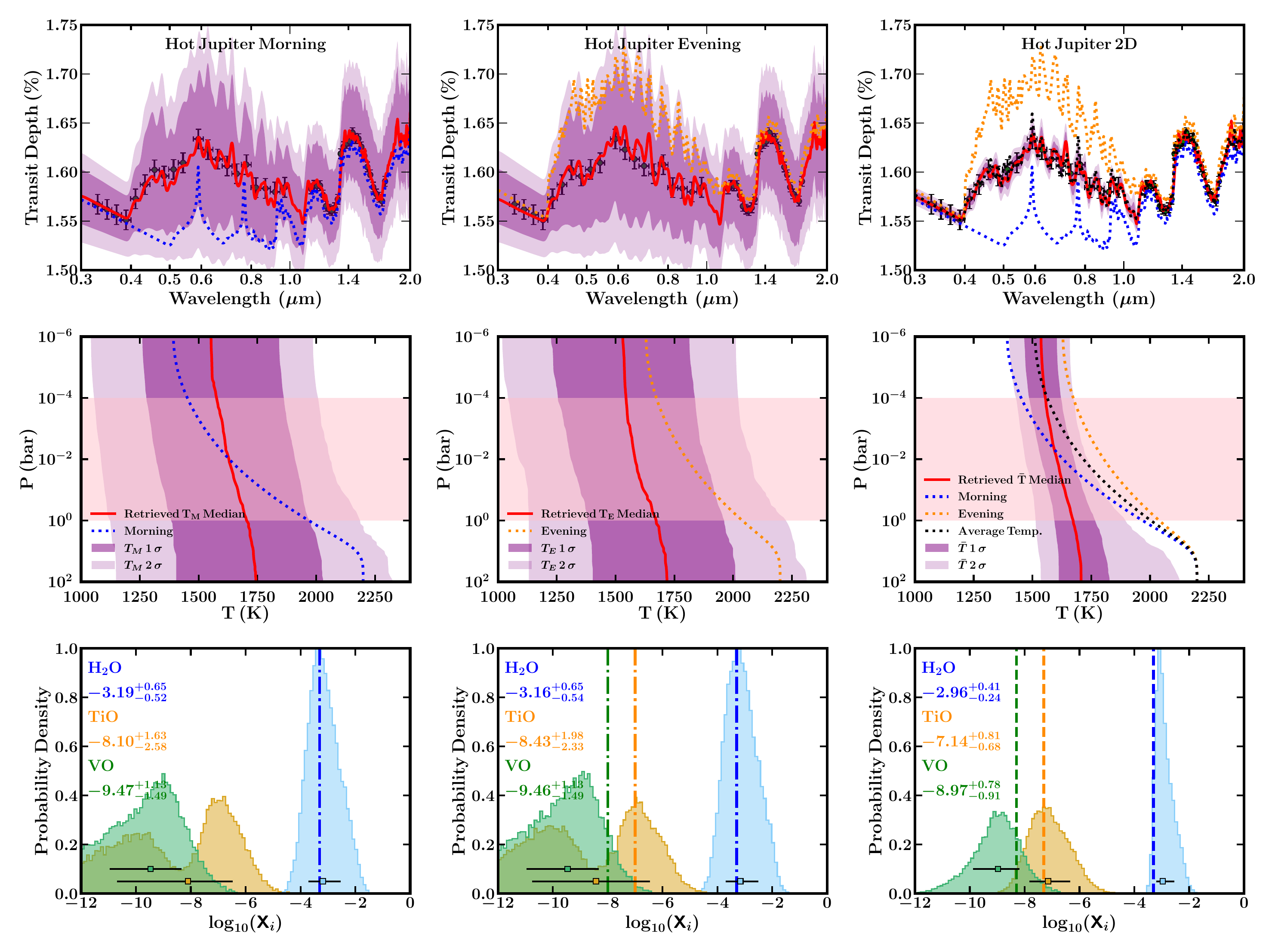}
\caption[]{As Figure \ref{fig:wj_2d_grid} but for the hot Jupiter case.} \label{fig:hj_2d_grid}
\end{figure*}

\begin{figure*}
\includegraphics[width=1.0\textwidth]{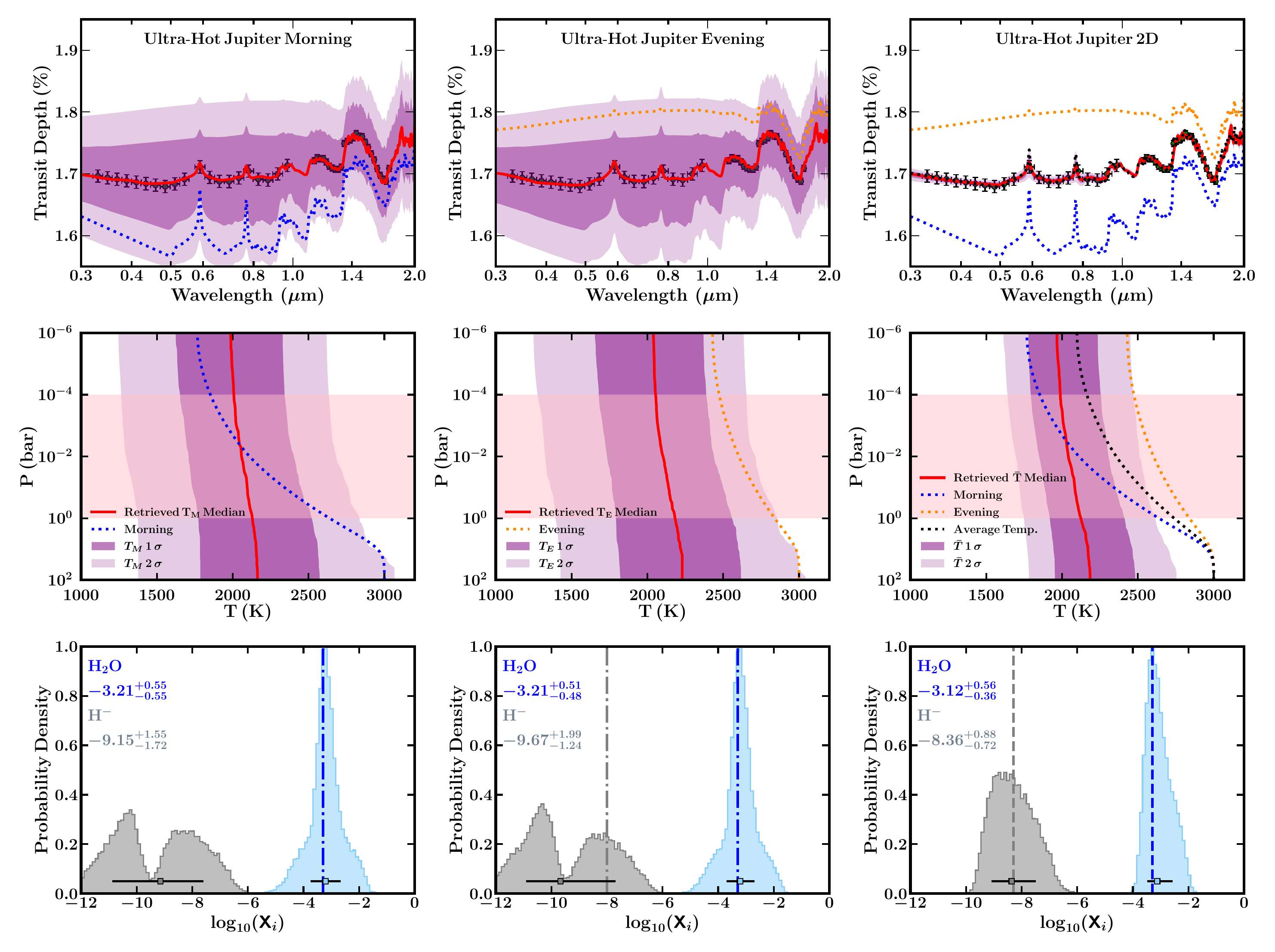}
\caption[]{As Figure \ref{fig:wj_2d_grid} but for the ultra-hot Jupiter case.} \label{fig:uhj_2d_grid}
\end{figure*}

Here we present the retrieved transmission spectra and posterior distributions for the 2D retrievals in Section \ref{sec:2D_retrievals}. Figures \ref{fig:wj_2d_grid}, \ref{fig:hj_2d_grid} and \ref{fig:uhj_2d_grid} show the retrieved spectra, $P$--$T$ profiles, and chemical abundances for the morning and evening terminators. Additionally, the third column of each figure shows the retrieved 2D spectrum, 2D $P$--$T$ profile, and average abundances as shown in Figure \ref{fig:2d_grid}.

\newpage

\bibliographystyle{aasjournal}
\bibliography{biblio}

\begin{thebibliography}{}
\expandafter\ifx\csname natexlab\endcsname\relax\def\natexlab#1{#1}\fi
\providecommand{\url}[1]{\href{#1}{#1}}
\providecommand{\dodoi}[1]{doi:~\href{http://doi.org/#1}{\nolinkurl{#1}}}
\providecommand{\doeprint}[1]{\href{http://ascl.net/#1}{\nolinkurl{http://ascl.net/#1}}}
\providecommand{\doarXiv}[1]{\href{https://arxiv.org/abs/#1}{\nolinkurl{https://arxiv.org/abs/#1}}}

\bibitem[{{Allard} {et~al.}(2016){Allard}, {Spiegelman}, \&
  {Kielkopf}}]{Allard2016}
{Allard}, N.~F., {Spiegelman}, F., \& {Kielkopf}, J.~F. 2016, \aap, 589, A21,
  \dodoi{10.1051/0004-6361/201628270}

\bibitem[{{Allard} {et~al.}(2019){Allard}, {Spiegelman}, {Leininger}, \&
  {Molliere}}]{Allard2019}
{Allard}, N.~F., {Spiegelman}, F., {Leininger}, T., \& {Molliere}, P. 2019,
  \aap, 628, A120, \dodoi{10.1051/0004-6361/201935593}

\bibitem[{{Arcangeli} {et~al.}(2021){Arcangeli}, {D{\'e}sert}, {Parmentier},
  {Tsai}, \& {Stevenson}}]{Arcangeli2021}
{Arcangeli}, J., {D{\'e}sert}, J.~M., {Parmentier}, V., {Tsai}, S.~M., \&
  {Stevenson}, K.~B. 2021, \aap, 646, A94, \dodoi{10.1051/0004-6361/202038865}

\bibitem[{{Barber} {et~al.}(2014){Barber}, {Strange}, {Hill}, {Polyansky},
  {Mellau}, {Yurchenko}, \& {Tennyson}}]{Barber2014}
{Barber}, R.~J., {Strange}, J.~K., {Hill}, C., {et~al.} 2014, \mnras, 437,
  1828, \dodoi{10.1093/mnras/stt2011}

\bibitem[{{Barstow}(2020)}]{Barstow2020a}
{Barstow}, J.~K. 2020, \mnras, 497, 4183, \dodoi{10.1093/mnras/staa2219}

\bibitem[{{Barstow} {et~al.}(2017){Barstow}, {Aigrain}, {Irwin}, \&
  {Sing}}]{Barstow2017}
{Barstow}, J.~K., {Aigrain}, S., {Irwin}, P.~G.~J., \& {Sing}, D.~K. 2017,
  \apj, 834, 50, \dodoi{10.3847/1538-4357/834/1/50}

\bibitem[{{Bauschlicher} {et~al.}(2001){Bauschlicher}, {Ram}, {Bernath},
  {Parsons}, \& {Galehouse}}]{Bauschlicher2001}
{Bauschlicher}, C.~W., {Ram}, R.~S., {Bernath}, P.~F., {Parsons}, C.~G., \&
  {Galehouse}, D. 2001, \jcp, 115, 1312, \dodoi{10.1063/1.1377892}

\bibitem[{{Benneke} \& {Seager}(2012)}]{Benneke2012}
{Benneke}, B., \& {Seager}, S. 2012, \apj, 753, 100,
  \dodoi{10.1088/0004-637X/753/2/100}

\bibitem[{{Benneke} {et~al.}(2019){Benneke}, {Knutson}, {Lothringer},
  {Crossfield}, {Moses}, {Morley}, {Kreidberg}, {Fulton}, {Dragomir}, {Howard},
  {Wong}, {D{\'e}sert}, {McCullough}, {Kempton}, {Fortney}, {Gilliland },
  {Deming}, \& {Kammer}}]{Benneke2019a}
{Benneke}, B., {Knutson}, H.~A., {Lothringer}, J., {et~al.} 2019, \natas, 3,
  813, \dodoi{10.1038/s41550-019-0800-5}

\bibitem[{{B{\'e}tr{\'e}mieux} \& {Swain}(2017)}]{Betremieux2017}
{B{\'e}tr{\'e}mieux}, Y., \& {Swain}, M.~R. 2017, \mnras, 467, 2834,
  \dodoi{10.1093/mnras/stx257}

\bibitem[{{Blecic} {et~al.}(2017){Blecic}, {Dobbs-Dixon}, \&
  {Greene}}]{Blecic2017}
{Blecic}, J., {Dobbs-Dixon}, I., \& {Greene}, T. 2017, \apj, 848, 127,
  \dodoi{10.3847/1538-4357/aa8171}

\bibitem[{{Brown}(2001)}]{Brown2001b}
{Brown}, T.~M. 2001, \apj, 553, 1006, \dodoi{10.1086/320950}

\bibitem[{{Brown} {et~al.}(2001){Brown}, {Charbonneau}, {Gilliland}, {Noyes},
  \& {Burrows}}]{Brown2001a}
{Brown}, T.~M., {Charbonneau}, D., {Gilliland}, R.~L., {Noyes}, R.~W., \&
  {Burrows}, A. 2001, \apj, 552, 699, \dodoi{10.1086/320580}

\bibitem[{{Buchner} {et~al.}(2014){Buchner}, {Georgakakis}, {Nandra}, {Hsu},
  {Rangel}, {Brightman}, {Merloni}, {Salvato}, {Donley}, \&
  {Kocevski}}]{Buchner2014}
{Buchner}, J., {Georgakakis}, A., {Nandra}, K., {et~al.} 2014, \aap, 564, A125,
  \dodoi{10.1051/0004-6361/201322971}

\bibitem[{{Burrows} {et~al.}(2010){Burrows}, {Rauscher}, {Spiegel}, \&
  {Menou}}]{Burrows2010}
{Burrows}, A., {Rauscher}, E., {Spiegel}, D.~S., \& {Menou}, K. 2010, \apj,
  719, 341, \dodoi{10.1088/0004-637X/719/1/341}

\bibitem[{{Caldas} {et~al.}(2019){Caldas}, {Leconte}, {Selsis}, {Waldmann},
  {Bord{\'e}}, {Rocchetto}, \& {Charnay}}]{Caldas2019}
{Caldas}, A., {Leconte}, J., {Selsis}, F., {et~al.} 2019, \aap, 623, A161,
  \dodoi{10.1051/0004-6361/201834384}

\bibitem[{{Charbonneau} {et~al.}(2002){Charbonneau}, {Brown}, {Noyes}, \&
  {Gilliland}}]{Charbonneau2002}
{Charbonneau}, D., {Brown}, T.~M., {Noyes}, R.~W., \& {Gilliland}, R.~L. 2002,
  \apj, 568, 377, \dodoi{10.1086/338770}

\bibitem[{{de Wit} \& {Seager}(2013)}]{DeWit2013}
{de Wit}, J., \& {Seager}, S. 2013, \sci, 342, 1473,
  \dodoi{10.1126/science.1245450}

\bibitem[{{Delrez} {et~al.}(2018){Delrez}, {Madhusudhan}, {Lendl}, {Gillon},
  {Anderson}, {Neveu-VanMalle}, {Bouchy}, {Burdanov}, {Collier-Cameron},
  {Demory}, {Hellier}, {Jehin}, {Magain}, {Maxted}, {Queloz}, {Smalley}, \&
  {Triaud}}]{Delrez2018}
{Delrez}, L., {Madhusudhan}, N., {Lendl}, M., {et~al.} 2018, \mnras, 474, 2334,
  \dodoi{10.1093/mnras/stx2896}

\bibitem[{{Deming} {et~al.}(2013){Deming}, {Wilkins}, {McCullough}, {Burrows},
  {Fortney}, {Agol}, {Dobbs-Dixon}, {Madhusudhan}, {Crouzet}, {Desert},
  {Gilliland}, {Haynes}, {Knutson}, {Line}, {Magic}, {Mand ell}, {Ranjan},
  {Charbonneau}, {Clampin}, {Seager}, \& {Showman}}]{Deming2013}
{Deming}, D., {Wilkins}, A., {McCullough}, P., {et~al.} 2013, \apj, 774, 95,
  \dodoi{10.1088/0004-637X/774/2/95}

\bibitem[{{Dulick} {et~al.}(2003){Dulick}, {Bauschlicher}, {Burrows}, {Sharp},
  {Ram}, \& {Bernath}}]{Dulick2003}
{Dulick}, M., {Bauschlicher}, C.~W., J., {Burrows}, A., {et~al.} 2003, \apj,
  594, 651, \dodoi{10.1086/376791}

\bibitem[{{Espinoza} \& {Jones}(2021)}]{Espinoza2021}
{Espinoza}, N., \& {Jones}, K. 2021, \aj, 162, 165,
  \dodoi{10.3847/1538-3881/ac134d}

\bibitem[{{Espinoza} {et~al.}(2019){Espinoza}, {Rackham}, {Jord{\'a}n}, {Apai},
  {L{\'o}pez-Morales}, {Osip}, {Grimm}, {Hoeijmakers}, {Wilson}, {Bixel},
  {McGruder}, {Rodler}, {Weaver}, {Lewis}, {Fortney}, \&
  {Fraine}}]{Espinoza2019}
{Espinoza}, N., {Rackham}, B.~V., {Jord{\'a}n}, A., {et~al.} 2019, \mnras, 482,
  2065, \dodoi{10.1093/mnras/sty2691}

\bibitem[{{Feng} {et~al.}(2020){Feng}, {Line}, \& {Fortney}}]{Feng2020}
{Feng}, Y.~K., {Line}, M.~R., \& {Fortney}, J.~J. 2020, \aj, 160, 137,
  \dodoi{10.3847/1538-3881/aba8f9}

\bibitem[{{Feng} {et~al.}(2016){Feng}, {Line}, {Fortney}, {Stevenson}, {Bean},
  {Kreidberg}, \& {Parmentier}}]{Feng2016}
{Feng}, Y.~K., {Line}, M.~R., {Fortney}, J.~J., {et~al.} 2016, \apj, 829, 52,
  \dodoi{10.3847/0004-637X/829/1/52}

\bibitem[{{Feng} {et~al.}(2018){Feng}, {Robinson}, {Fortney}, {Lupu}, {Marley},
  {Lewis}, {Macintosh}, \& {Line}}]{Feng2018}
{Feng}, Y.~K., {Robinson}, T.~D., {Fortney}, J.~J., {et~al.} 2018, \aj, 155,
  200, \dodoi{10.3847/1538-3881/aab95c}

\bibitem[{{Feroz} {et~al.}(2009){Feroz}, {Hobson}, \& {Bridges}}]{Feroz2009}
{Feroz}, F., {Hobson}, M.~P., \& {Bridges}, M. 2009, \mnras, 398, 1601,
  \dodoi{10.1111/j.1365-2966.2009.14548.x}

\bibitem[{{Feroz} {et~al.}(2013){Feroz}, {Hobson}, {Cameron}, \&
  {Pettitt}}]{Feroz2013}
{Feroz}, F., {Hobson}, M.~P., {Cameron}, E., \& {Pettitt}, A.~N. 2013, arXiv
  e-prints, arXiv:1306.2144.
\newblock \doarXiv{1306.2144}

\bibitem[{{Fortney} {et~al.}(2010){Fortney}, {Shabram}, {Showman}, {Lian},
  {Freedman}, {Marley}, \& {Lewis}}]{Fortney2010}
{Fortney}, J.~J., {Shabram}, M., {Showman}, A.~P., {et~al.} 2010, \apj, 709,
  1396, \dodoi{10.1088/0004-637X/709/2/1396}

\bibitem[{{Fortney} {et~al.}(2003){Fortney}, {Sudarsky}, {Hubeny}, {Cooper},
  {Hubbard}, {Burrows}, \& {Lunine}}]{Fortney2003}
{Fortney}, J.~J., {Sudarsky}, D., {Hubeny}, I., {et~al.} 2003, \apj, 589, 615,
  \dodoi{10.1086/374387}

\bibitem[{{Gandhi} \& {Madhusudhan}(2017)}]{Gandhi2017}
{Gandhi}, S., \& {Madhusudhan}, N. 2017, \mnras, 472, 2334,
  \dodoi{10.1093/mnras/stx1601}

\bibitem[{{Gandhi} \& {Madhusudhan}(2018)}]{Gandhi2018}
---. 2018, \mnras, 474, 271, \dodoi{10.1093/mnras/stx2748}

\bibitem[{{Gandhi} {et~al.}(2020{\natexlab{a}}){Gandhi}, {Madhusudhan}, \&
  {Mandell}}]{Gandhi2020b}
{Gandhi}, S., {Madhusudhan}, N., \& {Mandell}, A. 2020{\natexlab{a}}, \aj, 159,
  232, \dodoi{10.3847/1538-3881/ab845e}

\bibitem[{{Gandhi} {et~al.}(2020{\natexlab{b}}){Gandhi}, {Brogi}, {Yurchenko},
  {Tennyson}, {Coles}, {Webb}, {Birkby}, {Guilluy}, {Hawker}, {Madhusudhan},
  {Bonomo}, \& {Sozzetti}}]{Gandhi2020a}
{Gandhi}, S., {Brogi}, M., {Yurchenko}, S.~N., {et~al.} 2020{\natexlab{b}},
  \mnras, 495, 224, \dodoi{10.1093/mnras/staa981}

\bibitem[{{Gillon} {et~al.}(2012){Gillon}, {Triaud}, {Fortney}, {Demory},
  {Jehin}, {Lendl}, {Magain}, {Kabath}, {Queloz}, {Alonso}, {Anderson},
  {Collier Cameron}, {Fumel}, {Hebb}, {Hellier}, {Lanotte}, {Maxted},
  {Mowlavi}, \& {Smalley}}]{Gillon2012}
{Gillon}, M., {Triaud}, A.~H.~M.~J., {Fortney}, J.~J., {et~al.} 2012, \aap,
  542, A4, \dodoi{10.1051/0004-6361/201218817}

\bibitem[{{Gillon} {et~al.}(2014){Gillon}, {Anderson}, {Collier-Cameron},
  {Delrez}, {Hellier}, {Jehin}, {Lendl}, {Maxted}, {Pepe}, {Pollacco},
  {Queloz}, {S{\'e}gransan}, {Smith}, {Smalley}, {Southworth}, {Triaud},
  {Udry}, {Van Grootel}, \& {West}}]{Gillon2014}
{Gillon}, M., {Anderson}, D.~R., {Collier-Cameron}, A., {et~al.} 2014, \aap,
  562, L3, \dodoi{10.1051/0004-6361/201323014}

\bibitem[{{Goyal} {et~al.}(2018){Goyal}, {Mayne}, {Sing}, {Drummond},
  {Tremblin}, {Amundsen}, {Evans}, {Carter}, {Spake}, {Baraffe}, {Nikolov},
  {Manners}, {Chabrier}, \& {Hebrard}}]{Goyal2018}
{Goyal}, J.~M., {Mayne}, N., {Sing}, D.~K., {et~al.} 2018, \mnras, 474, 5158,
  \dodoi{10.1093/mnras/stx3015}

\bibitem[{{Greene} {et~al.}(2016){Greene}, {Line}, {Montero}, {Fortney},
  {Lustig-Yaeger}, \& {Luther}}]{Greene2016}
{Greene}, T.~P., {Line}, M.~R., {Montero}, C., {et~al.} 2016, \apj, 817, 17,
  \dodoi{10.3847/0004-637X/817/1/17}

\bibitem[{{Hargreaves} {et~al.}(2010){Hargreaves}, {Hinkle}, {Bauschlicher},
  {Wende}, {Seifahrt}, \& {Bernath}}]{Hargreaves2010}
{Hargreaves}, R.~J., {Hinkle}, K.~H., {Bauschlicher}, Charles~W., J., {et~al.}
  2010, \aj, 140, 919, \dodoi{10.1088/0004-6256/140/4/919}

\bibitem[{{Hellier} {et~al.}(2011){Hellier}, {Anderson}, {Collier Cameron},
  {Gillon}, {Jehin}, {Lendl}, {Maxted}, {Pepe}, {Pollacco}, {Queloz},
  {S{\'e}gransan}, {Smalley}, {Smith}, {Southworth}, {Triaud}, {Udry}, \&
  {West}}]{Hellier2011}
{Hellier}, C., {Anderson}, D.~R., {Collier Cameron}, A., {et~al.} 2011, \aap,
  535, L7, \dodoi{10.1051/0004-6361/201117081}

\bibitem[{{Heng} \& {Kitzmann}(2017)}]{Heng2017}
{Heng}, K., \& {Kitzmann}, D. 2017, \mnras, 470, 2972,
  \dodoi{10.1093/mnras/stx1453}

\bibitem[{{Hill} {et~al.}(2013){Hill}, {Yurchenko}, \& {Tennyson}}]{Hill2013}
{Hill}, C., {Yurchenko}, S.~N., \& {Tennyson}, J. 2013, \icarus, 226, 1673,
  \dodoi{10.1016/j.icarus.2012.07.028}

\bibitem[{{Iro} {et~al.}(2005){Iro}, {B{\'e}zard}, \& {Guillot}}]{Iro2005}
{Iro}, N., {B{\'e}zard}, B., \& {Guillot}, T. 2005, \aap, 436, 719,
  \dodoi{10.1051/0004-6361:20048344}

\bibitem[{{Iyer} \& {Line}(2020)}]{Iyer2020}
{Iyer}, A.~R., \& {Line}, M.~R. 2020, \apj, 889, 78,
  \dodoi{10.3847/1538-4357/ab612e}

\bibitem[{{John}(1988)}]{John1988}
{John}, T.~L. 1988, \aap, 193, 189

\bibitem[{{Kataria} {et~al.}(2015){Kataria}, {Showman}, {Fortney}, {Stevenson},
  {Line}, {Kreidberg}, {Bean}, \& {D{\'e}sert}}]{Kataria2015}
{Kataria}, T., {Showman}, A.~P., {Fortney}, J.~J., {et~al.} 2015, \apj, 801,
  86, \dodoi{10.1088/0004-637X/801/2/86}

\bibitem[{{Kataria} {et~al.}(2016){Kataria}, {Sing}, {Lewis}, {Visscher},
  {Showman}, {Fortney}, \& {Marley}}]{Kataria2016}
{Kataria}, T., {Sing}, D.~K., {Lewis}, N.~K., {et~al.} 2016, \apj, 821, 9,
  \dodoi{10.3847/0004-637X/821/1/9}

\bibitem[{{Kirk} {et~al.}(2021){Kirk}, {Rackham}, {MacDonald},
  {L{\'o}pez-Morales}, {Espinoza}, {Lendl}, {Wilson}, {Osip}, {Wheatley},
  {Skillen}, {Apai}, {Bixel}, {Gibson}, {Jord{\'a}n}, {Lewis}, {Louden},
  {McGruder}, {Nikolov}, {Rodler}, \& {Weaver}}]{Kirk2021}
{Kirk}, J., {Rackham}, B.~V., {MacDonald}, R.~J., {et~al.} 2021, \aj, 162, 34,
  \dodoi{10.3847/1538-3881/abfcd2}

\bibitem[{{Kreidberg} {et~al.}(2014{\natexlab{a}}){Kreidberg}, {Bean},
  {D{\'e}sert}, {Benneke}, {Deming}, {Stevenson}, {Seager}, {Berta-Thompson},
  {Seifahrt}, \& {Homeier}}]{Kreidberg2014a}
{Kreidberg}, L., {Bean}, J.~L., {D{\'e}sert}, J.-M., {et~al.}
  2014{\natexlab{a}}, \nat, 505, 69, \dodoi{10.1038/nature12888}

\bibitem[{{Kreidberg} {et~al.}(2014{\natexlab{b}}){Kreidberg}, {Bean},
  {D{\'e}sert}, {Line}, {Fortney}, {Madhusudhan}, {Stevenson}, {Showman},
  {Charbonneau}, {McCullough}, {Seager}, {Burrows}, {Henry}, {Williamson},
  {Kataria}, \& {Homeier}}]{Kreidberg2014b}
---. 2014{\natexlab{b}}, \apjl, 793, L27, \dodoi{10.1088/2041-8205/793/2/L27}

\bibitem[{{Kreidberg} {et~al.}(2018){Kreidberg}, {Line}, {Parmentier},
  {Stevenson}, {Louden}, {Bonnefoy}, {Faherty}, {Henry}, {Williamson},
  {Stassun}, {Beatty}, {Bean}, {Fortney}, {Showman}, {D{\'e}sert}, \&
  {Arcangeli}}]{Kreidberg2018}
{Kreidberg}, L., {Line}, M.~R., {Parmentier}, V., {et~al.} 2018, \aj, 156, 17,
  \dodoi{10.3847/1538-3881/aac3df}

\bibitem[{{Kurucz}(1992)}]{Kurucz1992}
{Kurucz}, R.~L. 1992, \rmxaa, 23

\bibitem[{{Lacy} \& {Burrows}(2020)}]{Lacy2020}
{Lacy}, B.~I., \& {Burrows}, A. 2020, \apj, 905, 131,
  \dodoi{10.3847/1538-4357/abc01c}

\bibitem[{{Lecavelier Des Etangs} {et~al.}(2008{\natexlab{a}}){Lecavelier Des
  Etangs}, {Pont}, {Vidal-Madjar}, \& {Sing}}]{Lecavelier2008a}
{Lecavelier Des Etangs}, A., {Pont}, F., {Vidal-Madjar}, A., \& {Sing}, D.
  2008{\natexlab{a}}, \aap, 481, L83, \dodoi{10.1051/0004-6361:200809388}

\bibitem[{{Lecavelier Des Etangs} {et~al.}(2008{\natexlab{b}}){Lecavelier Des
  Etangs}, {Vidal-Madjar}, {D{\'e}sert}, \& {Sing}}]{Lecavelier2008b}
{Lecavelier Des Etangs}, A., {Vidal-Madjar}, A., {D{\'e}sert}, J.-M., \&
  {Sing}, D. 2008{\natexlab{b}}, \aap, 485, 865,
  \dodoi{10.1051/0004-6361:200809704}

\bibitem[{{Lewis} {et~al.}(2014){Lewis}, {Showman}, {Fortney}, {Knutson}, \&
  {Marley}}]{Lewis2014}
{Lewis}, N.~K., {Showman}, A.~P., {Fortney}, J.~J., {Knutson}, H.~A., \&
  {Marley}, M.~S. 2014, \apj, 795, 150, \dodoi{10.1088/0004-637X/795/2/150}

\bibitem[{{Lewis} {et~al.}(2010){Lewis}, {Showman}, {Fortney}, {Marley},
  {Freedman}, \& {Lodders}}]{Lewis2010}
{Lewis}, N.~K., {Showman}, A.~P., {Fortney}, J.~J., {et~al.} 2010, \apj, 720,
  344, \dodoi{10.1088/0004-637X/720/1/344}

\bibitem[{{Line} \& {Parmentier}(2016)}]{Line2016a}
{Line}, M.~R., \& {Parmentier}, V. 2016, \apj, 820, 78,
  \dodoi{10.3847/0004-637X/820/1/78}

\bibitem[{{Line} {et~al.}(2013){Line}, {Wolf}, {Zhang}, {Knutson}, {Kammer},
  {Ellison}, {Deroo}, {Crisp}, \& {Yung}}]{Line2013a}
{Line}, M.~R., {Wolf}, A.~S., {Zhang}, X., {et~al.} 2013, \apj, 775, 137,
  \dodoi{10.1088/0004-637X/775/2/137}

\bibitem[{{MacDonald} {et~al.}(2020){MacDonald}, {Goyal}, \&
  {Lewis}}]{MacDonald2020}
{MacDonald}, R.~J., {Goyal}, J.~M., \& {Lewis}, N.~K. 2020, \apjl, 893, L43,
  \dodoi{10.3847/2041-8213/ab8238}

\bibitem[{{MacDonald} \& {Madhusudhan}(2017)}]{MacDonald2017a}
{MacDonald}, R.~J., \& {Madhusudhan}, N. 2017, \mnras, 469, 1979,
  \dodoi{10.1093/mnras/stx804}

\bibitem[{{Madhusudhan}(2018)}]{Madhusudhan2018}
{Madhusudhan}, N. 2018, in Handbook of Exoplanets, ed. H.~J. {Deeg} \& J.~A.
  {Belmonte} (Springer), 104, \dodoi{10.1007/978-3-319-55333-7_104}

\bibitem[{{Madhusudhan}(2019)}]{Madhusudhan2019}
{Madhusudhan}, N. 2019, \araa, 57, 617,
  \dodoi{10.1146/annurev-astro-081817-051846}

\bibitem[{{Madhusudhan} {et~al.}(2014){Madhusudhan}, {Crouzet}, {McCullough},
  {Deming}, \& {Hedges}}]{Madhusudhan2014a}
{Madhusudhan}, N., {Crouzet}, N., {McCullough}, P.~R., {Deming}, D., \&
  {Hedges}, C. 2014, \apjl, 791, L9, \dodoi{10.1088/2041-8205/791/1/L9}

\bibitem[{{Madhusudhan} \& {Seager}(2009)}]{Madhusudhan2009}
{Madhusudhan}, N., \& {Seager}, S. 2009, \apj, 707, 24,
  \dodoi{10.1088/0004-637X/707/1/24}

\bibitem[{{McKemmish} {et~al.}(2016){McKemmish}, {Yurchenko}, \&
  {Tennyson}}]{McKemmish2016}
{McKemmish}, L.~K., {Yurchenko}, S.~N., \& {Tennyson}, J. 2016, \mnras, 463,
  771, \dodoi{10.1093/mnras/stw1969}

\bibitem[{{Molli{\`e}re} {et~al.}(2019){Molli{\`e}re}, {Wardenier}, {van
  Boekel}, {Henning}, {Molaverdikhani}, \& {Snellen}}]{Molliere2019}
{Molli{\`e}re}, P., {Wardenier}, J.~P., {van Boekel}, R., {et~al.} 2019, \aap,
  627, A67, \dodoi{10.1051/0004-6361/201935470}

\bibitem[{{Moses} {et~al.}(2011){Moses}, {Visscher}, {Fortney}, {Showman},
  {Lewis}, {Griffith}, {Klippenstein}, {Shabram}, {Friedson}, {Marley}, \&
  {Freedman}}]{Moses2011}
{Moses}, J.~I., {Visscher}, C., {Fortney}, J.~J., {et~al.} 2011, \apj, 737, 15,
  \dodoi{10.1088/0004-637X/737/1/15}

\bibitem[{{Moses} {et~al.}(2013){Moses}, {Line}, {Visscher}, {Richardson},
  {Nettelmann}, {Fortney}, {Barman}, {Stevenson}, \& {Madhusudhan}}]{Moses2013}
{Moses}, J.~I., {Line}, M.~R., {Visscher}, C., {et~al.} 2013, \apj, 777, 34,
  \dodoi{10.1088/0004-637X/777/1/34}

\bibitem[{{Nikolov} {et~al.}(2018){Nikolov}, {Sing}, {Fortney}, {Goyal},
  {Drummond}, {Evans}, {Gibson}, {De Mooij}, {Rustamkulov}, {Wakeford},
  {Smalley}, {Burgasser}, {Hellier}, {Helling}, {Mayne}, {Madhusudhan},
  {Kataria}, {Baines}, {Carter}, {Ballester}, {Barstow}, {McCleery}, \&
  {Spake}}]{Nikolov2018}
{Nikolov}, N., {Sing}, D.~K., {Fortney}, J.~J., {et~al.} 2018, \nat, 557, 526,
  \dodoi{10.1038/s41586-018-0101-7}

\bibitem[{{Parmentier} {et~al.}(2016){Parmentier}, {Fortney}, {Showman},
  {Morley}, \& {Marley}}]{Parmentier2016}
{Parmentier}, V., {Fortney}, J.~J., {Showman}, A.~P., {Morley}, C., \&
  {Marley}, M.~S. 2016, \apj, 828, 22, \dodoi{10.3847/0004-637X/828/1/22}

\bibitem[{{Parmentier} \& {Guillot}(2014)}]{Parmentier2014}
{Parmentier}, V., \& {Guillot}, T. 2014, \aap, 562, A133,
  \dodoi{10.1051/0004-6361/201322342}

\bibitem[{{Parmentier} {et~al.}(2018){Parmentier}, {Line}, {Bean}, {Mansfield},
  {Kreidberg}, {Lupu}, {Visscher}, {D{\'e}sert}, {Fortney}, {Deleuil},
  {Arcangeli}, {Showman}, \& {Marley}}]{Parmentier2018}
{Parmentier}, V., {Line}, M.~R., {Bean}, J.~L., {et~al.} 2018, \aap, 617, A110,
  \dodoi{10.1051/0004-6361/201833059}

\bibitem[{{Patrascu} {et~al.}(2015){Patrascu}, {Yurchenko}, \&
  {Tennyson}}]{Patrascu2015}
{Patrascu}, A.~T., {Yurchenko}, S.~N., \& {Tennyson}, J. 2015, \mnras, 449,
  3613, \dodoi{10.1093/mnras/stv507}

\bibitem[{Pierrehumbert(2010)}]{Pierrehumbert2010}
Pierrehumbert, R.~T. 2010, Principles of planetary climate (Cambridge
  University Press)

\bibitem[{{Pinhas} {et~al.}(2019){Pinhas}, {Madhusudhan}, {Gandhi}, \&
  {MacDonald}}]{Pinhas2019}
{Pinhas}, A., {Madhusudhan}, N., {Gandhi}, S., \& {MacDonald}, R. 2019, \mnras,
  482, 1485, \dodoi{10.1093/mnras/sty2544}

\bibitem[{{Pinhas} {et~al.}(2018){Pinhas}, {Rackham}, {Madhusudhan}, \&
  {Apai}}]{Pinhas2018}
{Pinhas}, A., {Rackham}, B.~V., {Madhusudhan}, N., \& {Apai}, D. 2018, \mnras,
  480, 5314, \dodoi{10.1093/mnras/sty2209}

\bibitem[{{Pluriel} {et~al.}(2021){Pluriel}, {Leconte}, {Parmentier},
  {Zingales}, {Falco}, {Selsis}, \& {Borde}}]{Pluriel2021}
{Pluriel}, W., {Leconte}, J., {Parmentier}, V., {et~al.} 2021, arXiv e-prints,
  arXiv:2110.09080.
\newblock \doarXiv{2110.09080}

\bibitem[{{Pluriel} {et~al.}(2020){Pluriel}, {Zingales}, {Leconte}, \&
  {Parmentier}}]{Pluriel2020}
{Pluriel}, W., {Zingales}, T., {Leconte}, J., \& {Parmentier}, V. 2020, \aap,
  636, A66, \dodoi{10.1051/0004-6361/202037678}

\bibitem[{{Pont} {et~al.}(2008){Pont}, {Knutson}, {Gilliland}, {Moutou}, \&
  {Charbonneau}}]{Pont2008}
{Pont}, F., {Knutson}, H., {Gilliland}, R.~L., {Moutou}, C., \& {Charbonneau},
  D. 2008, \mnras, 385, 109, \dodoi{10.1111/j.1365-2966.2008.12852.x}

\bibitem[{{Rackham} {et~al.}(2018){Rackham}, {Apai}, \&
  {Giampapa}}]{Rackham2018}
{Rackham}, B.~V., {Apai}, D., \& {Giampapa}, M.~S. 2018, \apj, 853, 122,
  \dodoi{10.3847/1538-4357/aaa08c}

\bibitem[{{Rackham} {et~al.}(2019){Rackham}, {Apai}, \&
  {Giampapa}}]{Rackham2019}
---. 2019, \aj, 157, 96, \dodoi{10.3847/1538-3881/aaf892}

\bibitem[{{Richard} {et~al.}(2012){Richard}, {Gordon}, {Rothman}, {Abel},
  {Frommhold}, {Gustafsson}, {Hartmann}, {Hermans}, {Lafferty}, {Orton},
  {Smith}, \& {Tran}}]{Richard2012}
{Richard}, C., {Gordon}, I.~E., {Rothman}, L.~S., {et~al.} 2012, \jqsrt, 113,
  1276, \dodoi{10.1016/j.jqsrt.2011.11.004}

\bibitem[{{Rothman} {et~al.}(2010){Rothman}, {Gordon}, {Barber}, {Dothe},
  {Gamache}, {Goldman}, {Perevalov}, {Tashkun}, \& {Tennyson}}]{Rothman2010}
{Rothman}, L.~S., {Gordon}, I.~E., {Barber}, R.~J., {et~al.} 2010, \jqsrt, 111,
  2139, \dodoi{10.1016/j.jqsrt.2010.05.001}

\bibitem[{{Schwenke}(1998)}]{Schwenke1998}
{Schwenke}, D.~W. 1998, \fadi, 109, 321, \dodoi{10.1039/a800070k}

\bibitem[{{Seager}(2010)}]{Seager2010_book}
{Seager}, S. 2010, {Exoplanet Atmospheres: Physical Processes} (Princeton
  University Press)

\bibitem[{{Seager} \& {Sasselov}(1998)}]{Seager1998}
{Seager}, S., \& {Sasselov}, D.~D. 1998, \apjl, 502, L157,
  \dodoi{10.1086/311498}

\bibitem[{{Seager} \& {Sasselov}(2000)}]{Seager2000}
---. 2000, \apj, 537, 916, \dodoi{10.1086/309088}

\bibitem[{{Showman} {et~al.}(2009){Showman}, {Fortney}, {Lian}, {Marley},
  {Freedman}, {Knutson}, \& {Charbonneau}}]{Showman2009}
{Showman}, A.~P., {Fortney}, J.~J., {Lian}, Y., {et~al.} 2009, \apj, 699, 564,
  \dodoi{10.1088/0004-637X/699/1/564}

\bibitem[{{Sing}(2018)}]{Sing2018}
{Sing}, D.~K. 2018, in Astrophysics of Exoplanetary Atmospheres: 2nd Advanced
  School on Exoplanetary Science, ed. V.~{Bozza}, L.~{Mancini}, \&
  A.~{Sozzetti} (Springer).
\newblock \url{https://ui.adsabs.harvard.edu/abs/2018arXiv180407357S}

\bibitem[{{Sing} {et~al.}(2016){Sing}, {Fortney}, {Nikolov}, {Wakeford},
  {Kataria}, {Evans}, {Aigrain}, {Ballester}, {Burrows}, {Deming},
  {D{\'e}sert}, {Gibson}, {Henry}, {Huitson}, {Knutson}, {Lecavelier Des
  Etangs}, {Pont}, {Showman}, {Vidal-Madjar}, {Williamson}, \&
  {Wilson}}]{Sing2016}
{Sing}, D.~K., {Fortney}, J.~J., {Nikolov}, N., {et~al.} 2016, \nat, 529, 59,
  \dodoi{10.1038/nature16068}

\bibitem[{{Steinrueck} {et~al.}(2021){Steinrueck}, {Showman}, {Lavvas},
  {Koskinen}, {Tan}, \& {Zhang}}]{Steinrueck2021}
{Steinrueck}, M.~E., {Showman}, A.~P., {Lavvas}, P., {et~al.} 2021, \mnras,
  504, 2783, \dodoi{10.1093/mnras/stab1053}

\bibitem[{{Stevenson} {et~al.}(2017){Stevenson}, {Line}, {Bean}, {D{\'e}sert},
  {Fortney}, {Showman}, {Kataria}, {Kreidberg}, \& {Feng}}]{Stevenson2017}
{Stevenson}, K.~B., {Line}, M.~R., {Bean}, J.~L., {et~al.} 2017, \aj, 153, 68,
  \dodoi{10.3847/1538-3881/153/2/68}

\bibitem[{{Taylor} {et~al.}(2020){Taylor}, {Parmentier}, {Irwin}, {Aigrain},
  {Lee}, \& {Krissansen-Totton}}]{Taylor2020}
{Taylor}, J., {Parmentier}, V., {Irwin}, P. G.~J., {et~al.} 2020, \mnras, 493,
  4342, \dodoi{10.1093/mnras/staa552}

\bibitem[{{Torres} {et~al.}(2008){Torres}, {Winn}, \& {Holman}}]{Torres2008}
{Torres}, G., {Winn}, J.~N., \& {Holman}, M.~J. 2008, \apj, 677, 1324,
  \dodoi{10.1086/529429}

\bibitem[{{Venot} {et~al.}(2012){Venot}, {H{\'e}brard}, {Ag{\'u}ndez},
  {Dobrijevic}, {Selsis}, {Hersant}, {Iro}, \& {Bounaceur}}]{Venot2012}
{Venot}, O., {H{\'e}brard}, E., {Ag{\'u}ndez}, M., {et~al.} 2012, \aap, 546,
  A43, \dodoi{10.1051/0004-6361/201219310}

\bibitem[{{Venot} {et~al.}(2020){Venot}, {Parmentier}, {Blecic}, {Cubillos},
  {Waldmann}, {Changeat}, {Moses}, {Tremblin}, {Crouzet}, {Gao}, {Powell},
  {Lagage}, {Dobbs-Dixon}, {Steinrueck}, {Kreidberg}, {Batalha}, {Bean},
  {Stevenson}, {Casewell}, \& {Carone}}]{Venot2020}
{Venot}, O., {Parmentier}, V., {Blecic}, J., {et~al.} 2020, \apj, 890, 176,
  \dodoi{10.3847/1538-4357/ab6a94}

\bibitem[{{Weaver} {et~al.}(2020){Weaver}, {L{\'o}pez-Morales}, {Espinoza},
  {Rackham}, {Osip}, {Apai}, {Jord{\'a}n}, {Bixel}, {Lewis}, {Alam}, {Kirk},
  {McGruder}, {Rodler}, \& {Fienco}}]{Weaver2020}
{Weaver}, I.~C., {L{\'o}pez-Morales}, M., {Espinoza}, N., {et~al.} 2020, \aj,
  159, 13, \dodoi{10.3847/1538-3881/ab55da}

\bibitem[{{Welbanks} \& {Madhusudhan}(2019)}]{Welbanks2019a}
{Welbanks}, L., \& {Madhusudhan}, N. 2019, \aj, 157, 206,
  \dodoi{10.3847/1538-3881/ab14de}

\bibitem[{{Welbanks} \& {Madhusudhan}(2021)}]{Welbanks2021}
---. 2021, \apj, 913, 114, \dodoi{10.3847/1538-4357/abee94}

\bibitem[{{Welbanks} {et~al.}(2019){Welbanks}, {Madhusudhan}, {Allard},
  {Hubeny}, {Spiegelman}, \& {Leininger}}]{Welbanks2019b}
{Welbanks}, L., {Madhusudhan}, N., {Allard}, N.~F., {et~al.} 2019, \apjl, 887,
  L20, \dodoi{10.3847/2041-8213/ab5a89}

\bibitem[{{Wende} {et~al.}(2010){Wende}, {Reiners}, {Seifahrt}, \&
  {Bernath}}]{Wende2010}
{Wende}, S., {Reiners}, A., {Seifahrt}, A., \& {Bernath}, P.~F. 2010, \aap,
  523, A58, \dodoi{10.1051/0004-6361/201015220}

\bibitem[{{Yurchenko} \& {Tennyson}(2014)}]{Yurchenko2014}
{Yurchenko}, S.~N., \& {Tennyson}, J. 2014, \mnras, 440, 1649,
  \dodoi{10.1093/mnras/stu326}

\bibitem[{{Yurchenko} {et~al.}(2013){Yurchenko}, {Tennyson}, {Barber}, \&
  {Thiel}}]{Yurchenko2013}
{Yurchenko}, S.~N., {Tennyson}, J., {Barber}, R.~J., \& {Thiel}, W. 2013,
  Journal of Molecular Spectroscopy, 291, 69, \dodoi{10.1016/j.jms.2013.05.014}

\end{thebibliography}

\end{document}